\documentclass[aps,prd,twocolumn,twoside,superscriptaddress,floatfix, nofootinbib,longbibliography,10pt]{revtex4-1}
\usepackage{ifthen,amsmath,amssymb, bm}
\usepackage{graphicx}
\usepackage{multirow}
\usepackage{siunitx}
\usepackage{printlen}
\usepackage{layouts}
\usepackage{color}
\usepackage{svg}
\usepackage{float}
\usepackage{svg-extract}
\usepackage{hyperref}
\usepackage{cancel}
\usepackage{mathtools}
\usepackage{aas_macros}
\usepackage{afterpage}

\usepackage[normalem]{ulem}
\usepackage[scr=rsfso]{mathalfa} 
\DeclareMathOperator{\sinc}{sinc}
\newcommand{\vk} {\bm{k}}
\newcommand{\WIN}{{\sf \Pi}}
\renewcommand{\O} {\mathcal{O}}
\newcommand{\N} {\mathcal{N}}
\newcommand{\m}{m_\Lambda}
\newcommand{\I} {\xi}
\def\SNR{\textbf{\textsf{SNR}}}
\renewcommand{\i} {\texttt{i}}
\renewcommand{\L} {\textsf{L}}
\renewcommand{\j} {{\texttt{j}}}
\newcommand{\C} {{\mathcal{C}}}

\newcommand{\vu} {\bm{u}}
\newcommand{\KL} {{K_{\rm LIM}}}
\newcommand{\chimax}{\chi_{\sf max}}
\newcommand{\chimin}{\chi_{\sf min}}

\newcommand{\Kk} {{K_\kappa}}
\newcommand{\KLo} {{K^{\LO}_{\rm LIM}}}
\newcommand{\chip} {{\chi'}}
\newcommand{\chipp} {{\chi''}}
\newcommand{\vl} { {\bm{\ell}} }
\newcommand{\vb} { {\bm{b}} }
\newcommand{\vlp} { {\bm{\ell}'} }
\newcommand{\F} { {\HI} }
\newcommand{\V} { {\mathcal {V}} }
\let\originalLambda\Lambda
\renewcommand{\Lambda}{{\sf \originalLambda}}
\newcommand{\HI} { {\mathsf {Hi}_\Lambda} }
\newcommand{\LO} { {\mathsf {Lo}_\Lambda} }

\newcommand{\vm} { {\vL} }
\newcommand{\fV}{{\Gamma}}
\newcommand{\vL} { {\bm{L}} }
\newcommand{\prevv}{Moodley:2023lmu, Obuljen:2017jiy, Li:2018izh, Modi:2019hnu, Guandalin:2021sxw}

\newcommand{\vx} {\bm{x}}

\newcommand{\vn} {\widehat{\bm{n}}}
\newcommand{\vtheta} {\bm{\theta}}

\newcommand{\kpar}{{k_\parallel}}

\newcommand{\kparp}{{k_\parallel'}}
\newcommand{\kparpp}{{k_\parallel''}}
\newcommand{\qpar}{{q_\parallel}}

\newcommand{\kperp}{\vk_\perp}
\newcommand{\chib}{\overline\chi}
\newcommand{\beq} {\begin{equation}}
\newcommand{\eeq} {\end{equation}}
\newcommand{\bal} {\begin{aligned}}
\newcommand{\eal} {\end{aligned}}
\newcommand{\kF}{k_{\sf F}}
\newcommand{\lr}[3]{\biggr(\frac{#1}{#2}\biggr)^{#3}}

\DeclareRobustCommand{\Sec}[1]{Sec.~\ref{sec:#1}}

\DeclareRobustCommand{\App}[1]{App.~\ref{app:#1}}
\DeclareRobustCommand{\Apps}[2]{Apps.~\ref{app:#1} and \ref{app:#2}}
\DeclareRobustCommand{\Tab}[1]{Table~\ref{tab:#1}}

\DeclareRobustCommand{\Fig}[1]{Fig.~\ref{fig:#1}}
\DeclareRobustCommand{\Figs}[2]{Figs.~\ref{fig:#1} and \ref{fig:#2}}
\DeclareRobustCommand{\Eq}[1]{Eq.~(\ref{eq:#1})}
\DeclareRobustCommand{\Eqs}[2]{Eqs.~(\ref{eq:#1}) and (\ref{eq:#2})}
\DeclareRobustCommand{\Reff}[1]{Ref.~\cite{#1}}
\DeclareRobustCommand{\Refs}[1]{Refs.~\cite{#1}}

\definecolor{c1}{RGB}{249,65,68} 
\definecolor{c1}{RGB}{255,0,0} 
\definecolor{c4}{RGB}{255,111,114} 
\definecolor{c2}{RGB}{0,168,50} 
\definecolor{c3}{RGB}{39,125,161} 
\definecolor{c5}{RGB}{157,111,255} 
\definecolor{c6}{RGB}{251,105,255} 

\definecolor{mRed}{RGB}{230, 0, 50}
\colorlet{newtextColor}{mRed}

\begin{document}

\title{
Direct correlation of line intensity mapping and CMB lensing from {evolution along the} lightcone
}

\author{Delon Shen}
\email{delon@stanford.edu}
\affiliation{Department of Physics, Stanford University, Stanford, CA, USA 94305-4085}
\affiliation{Kavli Institute for Particle Astrophysics and Cosmology, 382 Via Pueblo Mall Stanford, CA 94305-4060, USA}
\affiliation{SLAC National Accelerator Laboratory 2575 Sand Hill Road Menlo Park, California 94025, USA}
\author{Nickolas Kokron}
\email{kokron@ias.edu}
\affiliation{School of Natural Sciences, Institute for Advanced Study, 1 Einstein Drive, Princeton, NJ, 08540, USA}
\affiliation{Department of Astrophysical Sciences, Princeton University, 4 Ivy Lane, Princeton, NJ, 08544, USA}
\author{Emmanuel Schaan}
\email{eschaan@stanford.edu}
\affiliation{Kavli Institute for Particle Astrophysics and Cosmology, 382 Via Pueblo Mall Stanford, CA 94305-4060, USA}
\affiliation{SLAC National Accelerator Laboratory 2575 Sand Hill Road Menlo Park, California 94025, USA}

\begin{abstract}
Line intensity mapping (LIM) promises to probe previously inaccessible corners of the faint and high-redshift universe.
{However, confusion with bright foregrounds is a major challenge for current-era pathfinder LIM experiments.}
Cross-correlation with cosmic microwave background (CMB) lensing is a promising avenue to enable the first LIM detections at high redshifts, a pristine probe of fundamental physics but sparsely populated by faint galaxies, and to further probe the connection between matter and spectral line emission, expanding our understanding of galaxies and the IGM.
Previous works have suggested that this direct correlation between LIM and CMB lensing is {effectively} impossible because smoothly varying modes in the intensity map are lost to bright foregrounds.
In this work, we analytically revisit the direct correlation of foreground-filtered line intensity mapping with CMB lensing, highlighting lightcone evolution's previously neglected yet unavoidable and crucial effects.
Indeed, {the growth of structure and evolution of line emission along the lightcone} breaks statistical translational invariance and thus induces mode coupling, even in linear theory, which enables the recovery of the smoothly varying modes lost to bright foregrounds.
We compute the effects of these lightcone evolution-induced mode couplings on the LIM$\times$CMB lensing cross-spectrum detectability, predicting that future wider-sky versions of \textsf{COMAP} and \textsf{CCAT} will be able to precisely measure this cross-correlation.
{Although we focus on the direct correlation of LIM with CMB lensing in this paper, our arguments generalize to the direct correlation of LIM with \textit{any projected field}.}
\end{abstract}

\maketitle

\begin{figure*}
    \centering
    \includegraphics{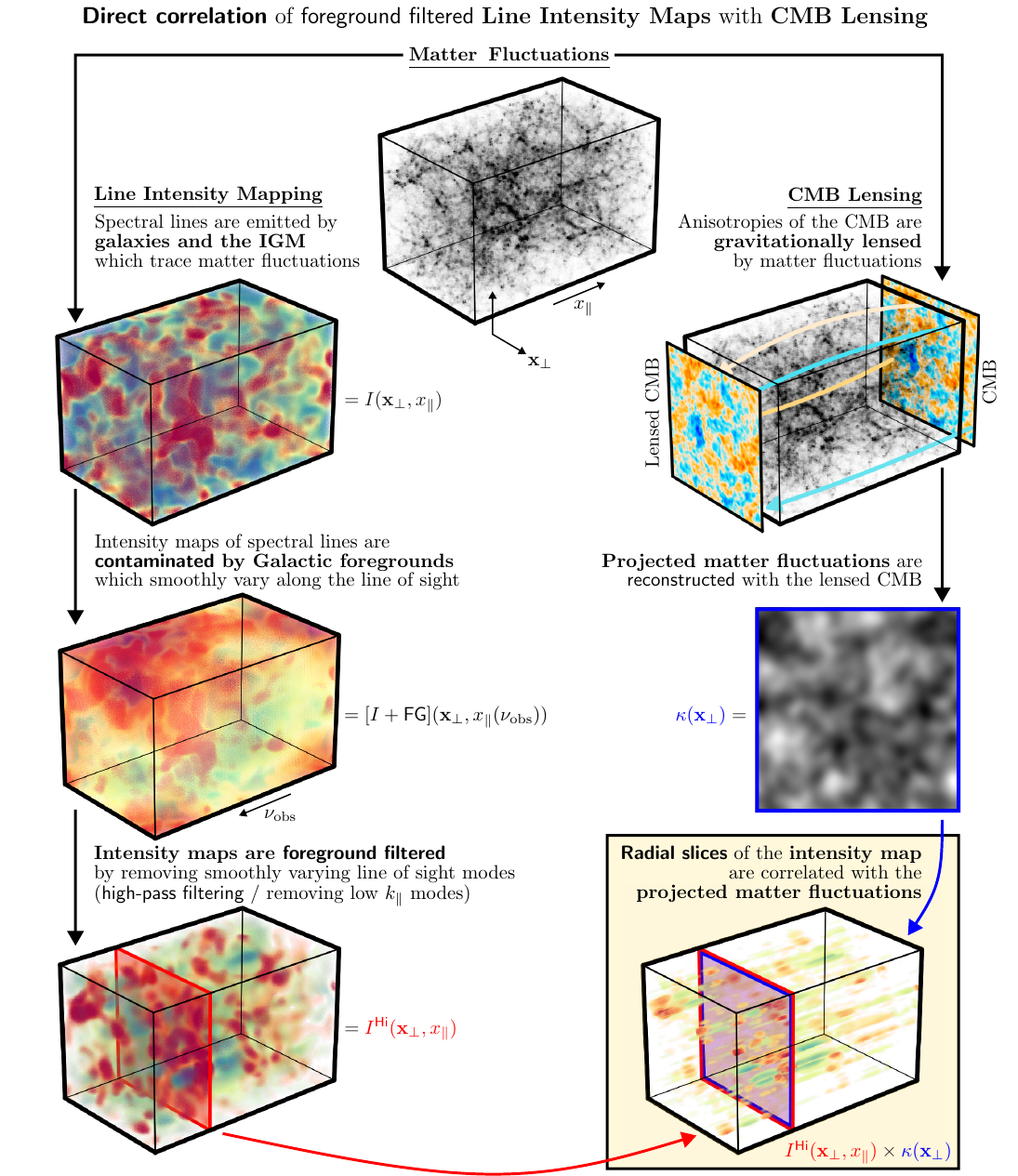}
    \caption{
    Visualization of the direct correlation between foreground-filtered line intensity maps (\Sec{LIM}) and CMB lensing (\Sec{cmb}) which we explore analytically in this paper (\Sec{spectra}) and argue is a promising observable for future LIM experiments (\Sec{SNR}).
    }
    \label{fig:summary}
\end{figure*}

\section{Introduction}

Observations of the faint and high redshift universe provide crucial information about astrophysics and cosmology.
Such observations extend our understanding of the cosmic star formation and reionization history, valuable inputs for our understanding of galaxy evolution (\Refs{McQuinn:2015icp, 2016ARAA..54..761S, Ouchi:2020zce, Robertson:2021ljt}).
Simultaneously, they probe a growing number of perturbative cosmological modes, pristine probes of fundamental physics (\Refs{Sailer:2021yzm, Ferraro:2022cmj, Cabass:2022epm, Ebina:2024ojt}).
In recent years, line intensity mapping (LIM) has emerged as a powerful technique to efficiently map out these previously inaccessible corners of the universe (\Reff{Bernal:2022jap}).

LIM measures the aggregate emission of atomic and molecular spectral lines from galaxies and the intergalactic medium (IGM).
By focusing on specific spectral lines and observing at multiple frequencies, LIM gains access to tomographic line-of-sight information.
Additionally, because LIM does not seek to resolve individual sources, it can map out the universe on large scales more rapidly than galaxy surveys (\Reff{Karkare:2022bai}).

In the current pathfinder era of LIM experiments, detecting faint or high-redshift signals from observed intensity maps alone faces many challenges.
Indeed, the first detections of such signals likely must come from cross-correlations (\Refs{Alvarez:2005sa,Adshead:2007ij,Pen:2008fw,2010arXiv1007.3709C, 2013ApJ...763L..20M, Anderson:2017ert, Pullen:2017ogs, COMAP:2018svn, Li:2020pre, 2021arXiv211105354S, Bernal:2022jap,CHIME:2022kvg,eBOSS:2021ebm,Cunnington:2022uzo, Chung:2022lpr, CHIME:2023til, Roy:2024kzc, 2024ApJ...965....7D}). 
Cross-correlation with weak lensing of the cosmic microwave background (CMB) is a promising path to make the first detection of LIM at high redshifts, where galaxy surveys may be too sparse.
At the same time, as a direct probe of the matter density field, CMB lensing allows us to probe the connection between matter and spectral line emission.
In this paper, we analytically explore this direct correlation of LIM with CMB lensing (visualized in \Fig{summary}).

While the direct correlation of LIM with CMB lensing is a promising path to probe the faint and high redshift universe, several previous works (\Refs{\prevv}) have suggested that this direct correlation is impossible in practice.
{Indeed}, the LIM signal of interest in observed intensity maps are contaminated by spectrally smooth and bright foregrounds (\Refs{Shaw:2014khi, Yue:2015sua, 2018MNRAS.479..275S, 2018MNRAS.479.4041S,Chung:2023ncd}) whose compositions varies from line-to-line but whose origin {generally} are primarily Galactic\footnote{Other contributions to the foregrounds include interloper line emissions, {radio-continuum point sources}, and the spectrally smooth {extragalactic background light (e.g. cosmic infrared background)}.
However, unlike the Galactic contributions, these sources contain a cosmological signal, making them a qualitatively different foreground component. 
We will focus primarily on Galactic foregrounds in this work.
}.
Filtering out these spectrally smooth foregrounds renders the data insensitive to slowly varying modes.
{On the other hand, because the projection kernel of CMB lensing is a smooth function of cosmic distance, it only retains the modes that vary slowly along the line-of-sight.
Naively, foreground-filtered LIM and CMB lensing would therefore lack overlap in Fourier space and thus could not be directly correlated.}
While a few numerical studies focusing on specific foreground removal methods have re-examined these claims (\Reff{Sangka:2024vfg, Marins:2025hzm}), 
a general theoretical understanding of this direct correlation between foreground-filtered line intensity maps and CMB lensing has yet to be established.

In this paper, we revisit this direct correlation from a purely analytical perspective, highlighting lightcone evolution's previously neglected yet crucial role.
By breaking statistical translational invariance along the line-of-sight, {lightcone evolution}, {specifically the growth of structure and evolution of line emission along the lightcone,} unavoidably induces mode couplings in LIM observations.
These mode couplings enable the recovery of slowly varying modes along the line-of-sight direction otherwise lost to bright foregrounds, even in \textit{linear theory}.
This approach is thus complementary to methods that invoke non-linear gravitational evolution to recover slowly varying modes like tidal and lensing reconstruction (\Refs{Zhu:2016esh, Li:2018izh, Schaan:2018yeh, Karacayli:2019iyd, Foreman:2018gnv, Liu:2019awk, Darwish:2020prn}).
We predict that the effects of lightcone evolution alone enable this cross-correlation to be detectable.
{In particular, we compute in this paper that this cross-correlation will be precisely measured by wider-sky versions of \textsf{COMAP} and \textsf{CCAT}.
Additionally, based on our calculations for \textsf{CHIME} in this paper, we infer that future 21cm intensity mapping experiments may also be able to precisely measure this cross-correlation.}
Finally, while we focus on the direct correlation with CMB lensing in this paper, our arguments generalize to direct correlation with any projected field like {tomographic galaxy clustering, galaxy shear, CIB, tSZ, and kSZ}.

This paper is organized as follows.
We begin in \Sec{summary} by briefly summarizing our main results and presenting an illuminating toy model to provide intuition.
In \Sec{models}, {we review the} modeling of CMB lensing and line intensity mapping. 
We restrict this modeling to linear theory to highlight the effects of lightcone evolution.
{In \Sec{spectra}, we predict the flat-sky angular spectra using analytical approximations beyond Limber, which are required for modeling foreground filtered intensity maps (see \App{limber_snr}).}
\Sec{SNR} presents our main results on the detectability of direct correlations between foreground-filtered LIM and CMB lensing.
We conclude in \Sec{conclusion}.

\subsection{Notation \& conventions}

Throughout this work we assume a fixed flat-$\originalLambda$CDM cosmology corresponding to the Planck 2018 (\Reff{Planck:2018vyg}) \texttt{TT,TE,EE+lowE+lensing} best fit.
We also utilize the shorthand notation for $n-$dimensional real and Fourier space integrals
\begin{equation}
    \int_{\vx} \equiv \int d\vx,\ \int_{\vk} \equiv \int \frac{d\vk}{(2\pi)^n},
\end{equation}
and follow the usual Fourier convention for infinite volumes
\begin{align}
\nonumber    f(\vx) = \int_{\vk} e^{i\vk\cdot\vx}f(\vk),\ f(\vk) &=\int_{\vx} e^{-i\vk\cdot\vx}f(\vx),\\
    \Rightarrow (2\pi)^n\delta^{(D)}(\vk+\vk') &= \int_{\vx} e^{\pm i (\vk+\vk')\cdot\vx},
\end{align}
and finite volumes $V$ 
\begin{align}
\nonumber    f(\vx) = \frac 1 V \sum_{\vk}e^{i\vk\cdot\vx}f_{\vk},\ f_{\vk} &= \int_{\vx \in V} e^{-i\vk\cdot\vx}f(\vx)\\
    \Rightarrow V\delta^{(K)}_{\vk, -\vk'} &= \int_{\vx \in V} e^{\pm i (\vk+\vk')\cdot \vx},
\end{align}
where $\delta^{(D)}$ is the Dirac delta and $\delta^{(K)}$ is the Kronecker delta function.
The smallest Fourier space element accessible (the fundamental) in the finite volume case is $\kF^n = (2\pi)^n/V$.

For a statistically homogeneous and isotropic field $f$, the power spectrum $P^f$ is defined as 
\begin{equation}
    \langle f(\vk) f(\vk')\rangle = (2\pi)^n\delta^{(D)}(\vk  + \vk') P^f(k).
\end{equation}
{Similarly for statistically isotropic fields $f$, the angular spectrum $C_\ell$ is defined as
\begin{equation}
    \langle f_\vl f_{\vl'}\rangle = (2\pi)^2 \delta^{(D)}(\vl+\vl') C_\ell.
\end{equation}
}
Throughout this work, we will utilize the shorthand notation for the two-point functions of statistically {symmetric} fields that drops the {relevant} momentum-conserving Dirac delta:
\begin{align}
    \nonumber
    \langle f(\vk)f(\vk')\rangle'  &= P^f(k),\\
    \langle f_\vl f_{\vlp}\rangle' &= C_\ell.
\end{align}

\section{Summary and intuition of main results}
\label{sec:summary}
The key result of this work is that the effects of lightcone evolution
allow future LIM experiments like wider-sky versions of \textsf{COMAP} and \textsf{CCAT} to precisely measure the direct correlation of LIM with CMB lensing, {unlike what was found in (\Refs{\prevv})}.
Indeed, lightcone evolution breaks statistical translational invariance along the line-of-sight and thus induces couplings between long- and short-wavelength modes, \textit{even in linear theory}.
Thus, long-wavelength modes lost to bright foregrounds can be recovered.
Before diving into our full calculation, we discuss the intuition behind our main results with an illuminating toy model.

\subsection{Toy model: the importance of lightcone evolution in the direct correlation}
\label{sec:toy}

\begin{figure}[t]
    \centering
    \includegraphics{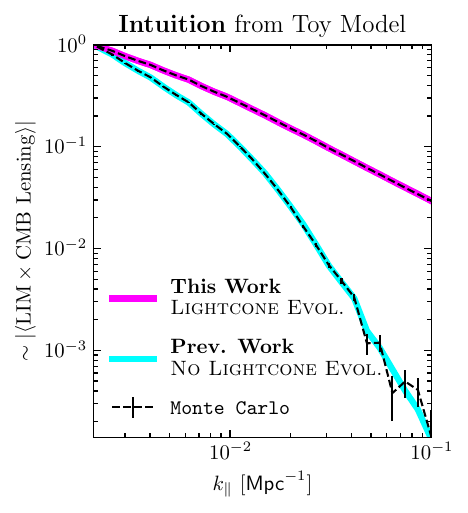}
    \caption{
    The toy model (\Sec{toy}) highlights how lightcone evolution (magenta), inescapable in observation but previously neglected (cyan), distributes information from long-wavelength modes of the linear matter density field into all modes of the line intensity map (\Eq{Itoy}), significantly affecting the direct correlation of LIM with CMB lensing which is sensitive only to long-wavelength modes of the linear matter density field (\Eq{ktoy}).
    This has dramatic consequences for the detectability of this cross-spectrum (\Fig{toySNR}).
    {The computation of all curves are described in \App{toy} and} normalized by the same factor such that the curve accounting for lightcone evolution (magenta) starts at $1$ at the minimum $k$ value plotted.
    }
    \label{fig:toyIk}
\end{figure}

\begin{figure}[b]
    \centering
    \includegraphics{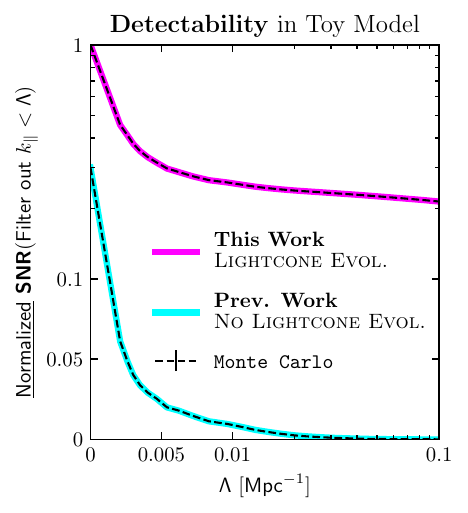}
    \caption{
    Lightcone evolution both (1) increases the detectability of the unfiltered LIM$\times$CMB lensing cross-spectrum and (2) decreases the suppression of detectability due to foreground-filtering {(discarding $\kpar<\Lambda$ modes)} {by distributing information from long--wavelength modes of the linear matter density field into all modes of the line intensity map} (see also \Fig{toyIk}).
    {The computation of all curves are described in \App{toy} and} normalized by the same factor such that the curve accounting for lightcone evolution (magenta) starts at $1$ when $\Lambda=0$.
    }
    \label{fig:toySNR}
\end{figure}

{In this toy model, we consider a single line-of-sight.}
Let $\chi$ be the distance along the line-of-sight and $\kpar$ be its Fourier conjugate.
We define the linear matter density contrast $\delta_m^0$ as a 1-D Gaussian random field with power
\begin{equation}
\langle \delta_m^0(\kpar)\delta_m^0(\kparp)\rangle = (2\pi)\delta^{(D)}(\kpar + \kparp) P(\kpar). \label{eq:toy_deltam}
\end{equation}
Long- and short-wavelength modes of the linear matter density field must be decoupled due to statistical translational invariance along the line-of-sight.

CMB lensing is sensitive only to slowly varying modes of the linear matter density field.
In particular, the CMB lensing field in our single angular pixel is a projection of all matter fluctuations between us and the surface of CMB photon last-scattering with a slowly varying kernel $\Kk(\chi)$ (discussed in \Sec{cmb}):
\begin{align}
     \kappa &= \int_\chi K_\kappa(\chi) \delta_m^0(\chi)= \int_\qpar K_\kappa(\qpar) {\delta^{0}_m}(-\qpar).
    \label{eq:ktoy}
\end{align}
Because $\Kk$ is slowly varying in real-space, it's Fourier transform is concentrated at low-$\qpar$.
Thus, $\Kk$ selects only low-$\qpar$ modes of $\delta_m^0$ in \Eq{ktoy}, rendering CMB lensing sensitive only to the slowly varying modes of linear matter density field.

If lightcone evolution is neglected when modeling line intensity maps like in \Refs{\prevv}, then filtering out smoothly varying foregrounds simultaneously filters out smoothly varying modes of the linear matter density field in the intensity map:
\begin{align}
\nonumber    \textbf{Prev. Work }(\textsc{No }&\textsc{Lightcone Evolution})\\
\nonumber    I(\chi) &= \KL \delta_m^0(\chi)\\
    \Rightarrow I(\kpar) &= \KL \delta_m^0(\kpar),\label{eq:ItoynLC}
\end{align}
where $\KL$ is some constant that relates the LIM to the linear matter density field.
The absence of lightcone evolution means that the Fourier modes of the intensity map are the same as the Fourier modes of the linear matter density field.
Since the CMB lensing kernel selects only slowly varying modes of the linear matter density field, the cross-correlation of LIM with CMB lensing $\langle I(\kpar)\kappa\rangle$ is suppressed by four orders of magnitude at $\kpar=0.1 /$Mpc (\Fig{toyIk}).

However, lightcone evolution is inescapable in observations and, crucially, distributes information from slowly varying modes of the linear matter density field into all modes of the line intensity map:
\begin{align}
\nonumber    \textbf{This Work }(&\textsc{Lightcone Evolution})\\
\nonumber    I(\chi) &= \KL(\chi) \delta_m^0(\chi)\\
    \Rightarrow I(\kpar) &= \int_\qpar \KL(\kpar-\qpar)\delta_m^0(\qpar),
    \label{eq:Itoy} 
\end{align}
where $\KL(\chi)$ is the LIM evolution kernel (\Sec{LIM} and \App{luminosity}).
The lightcone evolution kernel induces a convolution in \Eq{Itoy}, rendering each mode of the line intensity map a linear combination of all modes of the linear matter density field.
In particular, short-wavelength modes of the line intensity map contain contributions from long-wavelength modes of the linear matter density field and {the amplitude of these contributions increases with more evolution}.
Thus, filtering out the smoothly varying modes of the intensity map {does not completely remove} the smoothly varying modes of the linear matter density field.
The cross-correlation of LIM with CMB lensing $\langle I(\kparp)\kappa\rangle$ with lightcone evolution is then two orders of magnitude larger at $\kpar=0.1 /$Mpc relative to the case without lightcone evolution (\Fig{toyIk}).

We show in \Fig{toySNR} that lightcone evolution renders the direct correlation between line intensity maps and CMB lensing a fine observable whose detectability is not hopelessly suppressed by foreground filtering {(discarding $I(\kpar)$ for $\kpar <$ some cutoff $\Lambda$)} as previously claimed.
Indeed, we see that accounting for lightcone evolution
\begin{enumerate}
    \item increases the detectability of a direct correlation between \textit{unfiltered} line intensity maps and CMB lensing by a significant factor; and 
    \item decreases the suppression of detectability due to foreground filtering from exponential to a $O(1)$ factor.
\end{enumerate}
Again, lightcone evolution is inescapable in any observation of LIM and independent of any model of structure formation, cosmology, or astrophysics.
So in this paper, to highlight the critical effects of lightcone evolution, we restrict our modeling of structure formation to linear theory.

{For the sake of emphasizing the effects of lightcone evolution, this toy model neglects many complexities present in 3-D observations.
In the remainder of this paper, we will restore these complexities by
\begin{itemize}
    \item empirically {modeling of line luminosities from halo and galaxy properties} (\Eq{I0}, \App{luminosity}),
    \item introducing noise in both LIM (\Eq{eI0}, \App{experiments}) and CMB lensing (\Eq{NK}),
    \item accounting for the {limited redshift sensitivity} of individual LIM experiments, and
    \item characterizing the interaction of lightcone evolution induced line-of-sight mode coupling and angular spectra in the flat-sky approximation (\Sec{spectra}).
\end{itemize}
}

\subsubsection*{A quick aside: projecting the line intensity map}

This toy model suggests a \textit{projection} of the line intensity map may be directly correlated with CMB lensing without fear of suppression due to foreground filtering if appropriate projection weights are chosen.
Indeed, for a projection of the LIM  with normalized weights $w(\chi)$:
\begin{equation}
    I = \int_\chi w(\chi) I(\chi),
    \label{eq:toy_proj}
\end{equation}
we find an auspicious looking cross-spectrum with CMB lensing:
\begin{equation}
    \langle I \kappa\rangle = \int_\qpar \Kk(\qpar) P(\qpar) \int_\kpar w^*(\kpar) \KL(\kpar - \qpar).
\end{equation}
Of course a naive choice of constant $w(\chi)$ results in $w(k)$~$\sim$~$\delta^{(D)}(k)$, leading to any amount of foreground filtering annihilating the cross-spectrum.
More generally however, determining the weights which maximize this cross-spectrum when the line intensity map is filtered seems to be a well defined optimization problem whose solution could be of interest for future analyses.
We defer exploration of this to future work.

\begin{table*}[]
\renewcommand{\arraystretch}{1.15}
    \centering
    \begin{tabular}{r|ccccc}
    \hline
    \hline
         & & & & &\\
         \textbf{Experiment}
            & \textsf{\textbf{CHIME}}
            & \textsf{\textbf{HETDEX}} 
            & \textsf{\textbf{COMAP}} 
            & \textsf{\textbf{CCAT}}
            & \textsf{\textbf{SPHEREx}} \\
         {Line }
            & HI(21cm)
            & Ly-$\alpha$ 
            & CO(1$\rightarrow$0) 
            & [CII] 
            & Ly-$\alpha$\\
        &&&&&\\
        \hline
        $\nu_{\rm rest}$ 
            & 1420.406 MHz
            & 2456.43 THz
            & 115.27 GHz
            & 1900.5 GHz
            & 2456.43 THz\\
        $\nu_{\rm obs}$ 
            & 400-800 MHz
            & 545-857 THz
            & 26-34 GHz
            & 210-420 GHz
            & 270-400 THz
            \\
        $z_{\rm obs}$
            & 0.8 - 2.5
            & 1.9 - 3.5
            & 2.4 - 3.4
            & 3.5 - 8.1
            & 5.2 - 8\\
        $\mathcal R$
            & 1700
            & 800
            & 800
            & 100
            & 41
            \\
        \hline
        $\Omega_{\rm field}\ \rm [deg^2]$ 
            & 31000 
            & 540 
            & 12 
            & 8 
            & 200\\
        $\sqrt{\Omega_{\rm pixel}}$
            & \ang{;40;}
            & \ang{;;3}
            & $\ang{;4.5;}/\sqrt{8 \ln 2}$
            & $\ang{;;30}/\sqrt{8\ln 2}$
            & \ang{;;6}\\
        \hline
        $P^{\epsilon_I}$
            & \Eq{CHIME}
            & \Eq{HETDEX}
            & \Eq{COMAP}
            & \Eq{CCAT}
            & \Eq{SPHEREx}
            \\
        \hline
        \hline
    \end{tabular}
    \caption{Our assumed experimental configuration for {\sf CHIME} (\Refs{CHIME:2022dwe, CHIME:2022kvg}), {\sf HETDEX} (\Refs{Hill:2008mv,2016ASPC..507..393H,Gebhardt:2021vfo, Cheng:2018hox}), {\sf COMAP} (\Refs{Li:2015gqa, Cleary:2021dsp,COMAP:2021qdn,COMAP:2021pxy,COMAP:2021sqw, COMAP:2021lae}), {\sf CCAT} (\Refs{2020JLTP..199.1089C, Sato-Polito:2020cil}), and {\sf SPHEREx} (\Refs{SPHEREx:2014bgr, SPHEREx:2016vbo, SPHEREx:2018xfm, Cheng:2018hox}).
    For each experiment we report (1) rest frequencies of the targeted spectral line $\nu_{\rm rest}$, (2) frequencies and corresponding redshifts observed by an experiment $\nu_{\rm obs}/z_{\rm obs}$, (3) spectral resolving powers $\mathcal R = \nu_{\rm obs}/\delta\nu$ where $\delta\nu$ is the spectral resolution, (4) sky areas observed $\Omega_{\rm field}$, (5) angular sensitivities $\Omega_{\rm pixel}$, and (6) references to summaries of our instrumental noise models described in \App{experiments}.
    {The factor of $\sqrt{8\ln 2}$ for \textsf{COMAP} and \textsf{CCAT} are beam attenuation factors due to assuming a Gaussian profile for these experiments.}
    For \textsf{CHIME}, we assume $\Omega_{\rm pixel} = \ang{;40;}\times\ang{;40;}$ which is only a rough estimate.
    A more precise calculation with detailed beam modeling show actual \textsf{CHIME} observations have more resolution than our assumed $\Omega_{\rm pixel}$.
    }
    \label{tab:experiments}
\end{table*}

\begin{figure}
    \centering
    \includegraphics{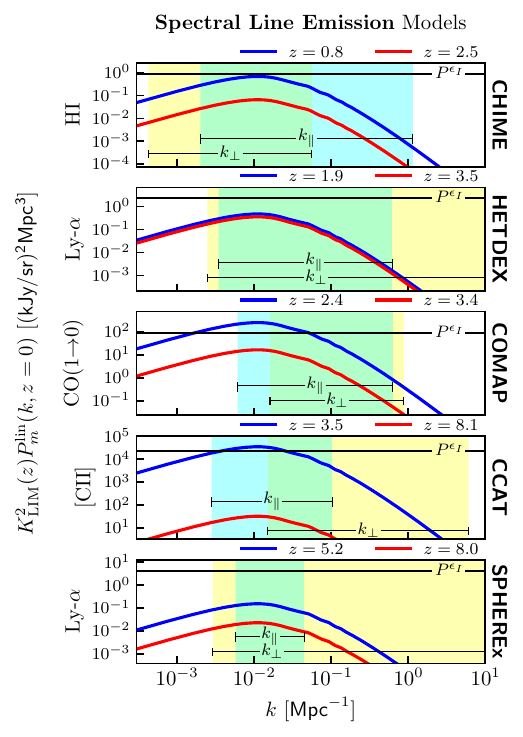}
    \caption{Our models for the cosmological signal in line intensity mapping (\Sec{lim_cosmo} and \App{luminosity}) from linear theory compared to the instrumental noise power (\Sec{lim_noise}) whose computations are described in \App{experiments} and summaries are referenced in \Tab{experiments}.
    For each experiment we also display the range of wave-number sensitivity parallel (cyan) and perpendicular (yellow) to the line of sight which are derived from assumed angular and line-of-sight resolutions reported in \Tab{experiments}.
    The cosmological signal for each experiment is plotted at the minimum (blue) and maximum (red) accessible redshift.
    }
    \label{fig:lim-model}
\end{figure}

\section{Review of Line Intensity Mapping \& CMB Lensing Modeling}
\label{sec:models}
In this section we review the modeling of line intensity mapping and CMB lensing.
The busy reader may want to skip ahead to \Tab{unify} where we summarize this modeling.
We restrict our modeling to \textbf{linear theory} in this paper to highlight the impact of lightcone evolution on the detectability of a direct correlation between foreground-filtered line intensity maps and CMB lensing.
\Apps{luminosity}{experiments} contain additional details of our modeling of particular lines and experiments considered in this paper (\Tab{experiments}).

\subsection{CMB lensing}
\label{sec:cmb}
The CMB lensing map is a projection of the linear matter density field with projection kernel $W_\kappa(\chi)$ (\Reff{Lewis:2006fu}):
\begin{equation}
W_\kappa(\chi) = \begin{cases} 
\displaystyle\frac{3H_0^2 \Omega_{m,0}}{2a} \frac{\chi(\chi_*-\chi)}{\chi_*} & \chi \leq \chi_* \\
0 & \chi > \chi_*,
\end{cases}
\label{eq:Wk}
\end{equation}
where $\chi_*=\chi(z_*\approx 1100)$ is the comoving radial distance to the surface of CMB photon last-scattering.
Let $\vtheta$ be the angular position in the plane of the sky around some line of sight $\hat{\mathbf{n}  }$. 
The CMB lensing map $\kappa(\vtheta)$ is then

\begin{align}
\kappa(\vtheta) &= \int d\chi\ \underbrace{W_\kappa(\chi) D(\chi)}_{\equiv \Kk(\chi)}\delta_m^0(\chi \vn, \chi \vtheta)
    \label{eq:k0}
\end{align}
where $D(\chi)$ is the linear growth factor and $\delta_m^0$ is the linear matter field at redshift $z=0$. 
For future convenience, we also defined the kernel for CMB lensing $K_\kappa(\chi)=W_\kappa(\chi)D(\chi)$.
For present day instrumental noise levels, reconstruction of the CMB lensing $\kappa$ map is primarily done with quadratic estimators \cite{Hu:2001kj}.
Such quadratic estimators lead to a noise bias in the CMB lensing spectrum, the dominant of which is called the $N^{(0)}$ bias, which we discuss more in \Sec{SNR}.

\subsection{Line intensity mapping}
\label{sec:LIM}

Our modeling of LIM is summarized in \Fig{lim-model} where we show the cosmological signal (\Sec{lim_cosmo}) for the experiments we consider (\Tab{experiments}) in comparison to the instrumental noise power (\Sec{lim_noise}).
\subsubsection{Cosmological signal in LIM}
\label{sec:lim_cosmo}
The observed spectral intensity $I(\chi)$ is related to the underlying luminosity density of the source by
\begin{equation}
    I(\chi) = \frac c {4\pi \nu_{\rm rest} H(\chi)}\rho_L(\chi).
\end{equation}
Moving forward we will work in units where $c=1$.
Lines are sourced by galaxies and IGM within halos which are a biased tracer of the underlying matter field.
With a redshift dependent luminosity-halo mass relation $L(M_h,\chi)$, our models of which are described in \App{luminosity}, we predict the mean luminosity 
\begin{equation}
\langle \rho_L\rangle(\chi) = \int dM_h\ L(M_h,\chi) \frac{ dn}{dM_h}(M_h, \chi),
\end{equation}
and luminosity-weighted linear bias 
\begin{equation}
    b_L(\chi) = \frac 1 {\langle \rho_L \rangle}  \int dM_h\ L(M_h,\chi) b(M_h, \chi) \frac{dn}{dM_h}(M_h,\chi),
\end{equation}
to relate fluctuations in the intensity of lines to the fluctuations in the underlying matter field\footnote{{We write the intensity fluctuations as a function of the comoving distance throughout this paper. However, line intensity maps are generally measured as a function of frequency $\nu$. The two can be related through $d\chi/d\nu = c(1+z)/H\nu$.}}:
\begin{align}
\delta I&(\chi\vn,\chi\vtheta)=
    \frac {\langle \rho_L\rangle(\chi) b_L(\chi)} {4\pi \nu_{\rm rest} H(\chi)}
    \times
    D(\chi)\delta_m^0(\chi, \vtheta).
    \label{eq:deltaI}
\end{align}
We use the linear bias $b(M_h,\chi)$ presented in \Reff{2010ApJ...724..878T}
and halo mass function $dn/dM_h$ presented in \Reff{Tinker:2008ff}.
For succinctness, in the remainder of the paper we will call the intensity fluctuations $I$, dropping the $\delta$ in front.
We can write \Eq{deltaI} in a form that parallels \Eq{k0} with a Dirac delta projection kernel:
\begin{align}
I(\chi, \vtheta) &= \int_\chip \underbrace{W_{\rm LIM}(\chip;\chi)D(\chip)}_{\equiv \KL(\chip;\chi)}\delta_m^0(\chip , \vtheta),\label{eq:I0}
\end{align}
where we have defined
\begin{equation}
W_{\rm LIM}(\chip;\chi) = \frac {\langle \rho_L\rangle(\chip) b_L(\chip)} {4\pi \nu_{\rm rest} H(\chip)}  \times \WIN(\chip)\delta^{(D)}(\chip-\chi),
\label{eq:WI}
\end{equation}
and $\KL(\chip;\chi) = W_{\rm LIM}(\chip;\chi)D(\chip)$.
Here we also have included a window function $\WIN(\chi)$ that encapsulates the redshift ranges probed by the LIM experiment of interest.
For example, from \Tab{experiments}, for \textsf{CCAT} we have $z_{\rm obs}~=~3.5-8.1$. 
So the window function $\WIN(\chi)$ in this case is
\begin{equation}
    \WIN_{\sf CCAT}(\chi) = \begin{cases}
        1& \chi(z=3.5) \leq \chi \leq \chi(z=8.1)\\
        0& \textrm{else}.
    \end{cases}
\end{equation}
{In the remainder of this paper, we will refer to the minimum and maximum $\chi$ probed by a survey as $\chimin$ and $\chimax$ and the length of the observed window as $\L = \chimax - \chimin$.
The fundamental mode observable along a LOS is then $\kF = 2\pi / \L$.
}
We clarify the utility of rewriting intensity fluctuations with a Dirac delta projection kernel in \Sec{unify}.

\subsubsection{Instrumental noise in LIM}
\label{sec:lim_noise}

Instrumental noise hinders observations of cosmological signals in LIM. 
We model this instrumental noise as Gaussian white noise with power $P^{\epsilon_I}$ in $\vk$ space and independent from the matter density field:
\begin{equation}
    \epsilon^I(\vk) \sim \mathcal {\sf Gaussian}(0, P^{\epsilon_I}).\label{eq:eI_base}
\end{equation}
Our computation of $P^{\epsilon_I}$ for each experiment is detailed in \App{experiments}.
To parallel \Eqs{k0}{I0}, we can write the LIM instrumental noise map as
\begin{equation}
    \epsilon(\chi,\vtheta) = \int_\chip K_{\epsilon,{\rm LIM}}(\chip;\chi) \epsilon^I(\chip,\vtheta),
    \label{eq:eI0}
\end{equation}
where we have defined
\begin{equation}
    K_{\epsilon,{\rm LIM}}(\chip;\chi) = \WIN(\chip)\delta^{(D)}(\chip-\chi).
\end{equation}
Just like in \Eq{WI}, we have included a window function $\WIN(\chi)$ that encapsulates the {redshift ranges probed by the LIM experiment of interest}.

\subsubsection{Foreground filtering of LIM}
\label{sec:lim_filter}

Continuum foregrounds contaminate large-scale line-of-sight Fourier modes of LIMs and are typically removed through filtering.
We model foreground filtering as high-pass filtering the LIM along the line-of-sight direction, cutting off all large-scale line-of-sight Fourier modes with wavenumber $\kpar < \Lambda$:
\begin{align}
    I^\HI(\vk) = \F(\kpar) I(\vk),
    \label{eq:filter}
\end{align}
where we have defined 
\begin{equation}
        \F(\kpar) = \begin{cases}
        1& |\kpar| \geq \Lambda\\
        0& |\kpar| < \Lambda.
    \end{cases}
    \label{eq:filter_explicit}
\end{equation}
For convenience, the high-pass filtered LIM can be rewritten in terms of the difference of the unfiltered and low-pass filtered LIM:
\begin{equation}
    I^\HI(\vk) = I(\vk) - \underbrace{\LO(\kpar)I(\vk)}_{\equiv I^\LO(\vk)},
\end{equation}
where we have introduced the low-pass (top-hat) filter
\begin{equation}
        \LO(\kpar) = \begin{cases}
        0& |\kpar| \geq \Lambda\\
        1& |\kpar| < \Lambda.
    \end{cases}
\end{equation}
This low-pass filter, having compact support in Fourier space, has a more well-defined Fourier transform.
Since filtering is applied to the finite volume observed box along the LOS direction, we must use the \textit{discrete} Fourier transform of the low-pass kernel to account for the finite length of the box along the LOS\footnote{Indeed, we show in \App{lo} how using the continuous Fourier transform would cause our computation of the observable to become un-physically sensitive to the window function $\WIN(\chi)$.}, $\L$, instead of the typical $\sinc$ (see \App{lo} for more details):
\begin{equation}
    \LO(\chi) = \frac 1 \L\sum_{|\kpar|\leq\Lambda}e^{i\kpar\chi}.
    \label{eq:LOx_sum}
\end{equation}
In particular, if $m_\Lambda$ is a integer such that 
$\m\kF \leq \Lambda < (\m+1)\kF$ 
then the discrete Fourier transform of the low-pass filter is
\begin{equation}
    \LO(\chi) = \frac 1 \L \frac{\sin \left(\frac \pi \L (1+2\m)\chi \right)}{\sin\left(\frac \pi \L \chi \right)}.
    \label{eq:LOx}
\end{equation}

The product of the LIM and low-pass filter is a convolution in real space:
\begin{align}
\nonumber     I^\LO(\chi,\vtheta) &= \int_\chip \LO(\chi-\chip) I(\chip, \vtheta). \\
    &= \int_\chip \underbrace{W_{\rm LIM}^\LO(\chip;\chi)D(\chip)}_{\equiv \KLo(\chip;\chi)}\delta_m^0(\chip,\vtheta),
\end{align}
where in the second equality we have defined 
\begin{equation}
    W_{\rm LIM}^\LO(\chip;\chi) = \frac{\langle \rho_L\rangle(\chip)b_L(\chip)}{4\pi\nu_{\rm rest} H(\chip)} \times \WIN(\chip)
   {\LO(\chi-\chip)}, 
    \label{eq:WLIMLo}
\end{equation}
and written $I^\LO$ in a form that parallels Eqs~(\ref{eq:k0}), (\ref{eq:I0}), and (\ref{eq:eI0}).

Note that the introduction of foreground filtering induces additional line-of-sight correlations.
Furthermore, these induced correlations becoming increasingly localized\footnote{The induced correlations also become increasingly oscillatory as they become more localized, rendering the evaluation of spectra involving foreground-filtered intensity maps more challenging. We expand on this in \App{II}.} with the amount of foreground filtering applied.
This is to be expected as low-pass filtering (1) is a local operation in Fourier space resulting in non-local couplings in real space and (2) becomes less localized in Fourier space as $\Lambda$ increases resulting in greater localization of power in real space.
These line-of-sight correlations induced by foreground filtering are the crucial reason why in \Sec{spectra} we explore approximations of the flat-sky angular spectra which go beyond the Limber approximation (\Refs{1954ApJ...119..655L, 1973ApJ...185..413P,1992ApJ...388..272K}) by retaining information from line-of-sight correlations.
Indeed, we show in \App{limber_snr} that the Limber approximation exhibits concerning pathologies when in the presence of these induced correlations, and thus argue it is unsuitable for foreground-filtered line intensity maps.

High-pass filtering occurs at the level of data, so the instrumental noise is filtered as well.
Similarly to the cosmological signal in LIM, we model the high-pass filtered instrumental noise as the difference of the unfiltered noise and low-pass filtered noise
\begin{equation}
    \epsilon^{I,\LO}(\chi,\vtheta)=\int_\chip K_{\epsilon, \rm LIM}^\LO(\chip;\chi) \epsilon^I(\chip,\vtheta),
\end{equation}
where we have defined 
\begin{equation}
    K_{\epsilon,\rm LIM}^\LO(\chip;\chi) = \WIN(\chip){\LO(\chi-\chip)},
    \label{eq:KeLo}
\end{equation}
to once again write the low-pass filtered LIM instrumental noise in a form that parallels our previous definitions.

\begin{table*}
    \centering
    \begin{tabular}{r@{\hspace{5pt}}|@{\hspace{5pt}}l@{\hspace{5pt}}|@{\hspace{5pt}}l}
    \hline
    \hline
    & & \\
    \textbf{2-D Projected Field} & \textbf{Projection Kernel} & \textbf{3-D Unprojected Field} \\
    $\phi_\vl(\bullet)$ (\Sec{unify}) & $K_\phi(\chip;\bullet)$ & $\Phi(\vk')$\\
    & & \\
    \hline
    & & \\
    \textbf{CMB Lensing}  & 
    \multirow{2}{*}{\begin{minipage}{8cm}\raggedright
    $\displaystyle K_\kappa(\chip) = \frac{3H_0^2 \Omega_{m,0}}{2a(\chip)} \frac{\chip(\chi_*-\chip)}{\chi_*} \times D(\chip)$
    \end{minipage}} 
    & Linear Matter Field \\
    $\kappa_\vl$  (\Sec{cmb})& 
      & $\delta_m^0(\vk')$\\
    & & \\
    \hline
    & & \\
    \textbf{Line Intensity Map} &  
    \multirow{2}{*}{\begin{minipage}{8cm}\raggedright
    $
    \displaystyle K_{\rm LIM}(\chip;\chi) = 
    \frac {\langle \rho _L \rangle (\chip)b_L(\chip)} {4\pi \nu_{\rm line} H(\chip)}  
    \times
    D(\chip)
     \WIN(\chip)
    \delta^{(D)}(\chip-\chi)
    $
    \end{minipage}}
    & Linear Matter Field\\
    $I_\vl(\chi)$  (\Sec{lim_cosmo}) & 
    & $\delta_m^0(\vk')$\\
    & & \\
    \textbf{\textsf{Low-Pass Filtered} LIM} & 
    \multirow{2}{*}{\begin{minipage}{8.9cm}\raggedright
    $\displaystyle K_{\rm LIM}^\LO(\chip;\chi)= 
    \frac {\langle \rho_L\rangle(\chip) b_L(\chip)} {4\pi \nu_{\rm rest} H(\chip)}
    \times
    D(\chip)
    \WIN(\chip)
    {\LO(\chi-\chip)}
    $
    \end{minipage}}
    &
    Linear Matter Field\\
    $I_\vl^\LO(\chi)$  (\Sec{lim_filter})&
    & $\delta_m^0(\vk')$\\
    & & \\
    \textbf{\textsf{High-Pass Filtered} LIM} &
    \multirow{2}{*}{\begin{minipage}{8cm}\raggedright
    $\displaystyle K_{\rm LIM}^\HI(\chip;\chi)=K_{\rm LIM}(\chip;\chi) - K^\LO_{\rm LIM}(\chip;\chi)$
    \end{minipage}}
    &
    Linear Matter Field\\
    $I_\vl^\HI(\chi)$  (\Sec{lim_filter})& 
    & $\delta_m^0(\vk')$\\
    & & \\
    \hline
    & & \\
    \textbf{\textsf{Instrumental} LIM {Noise}}  & 
    \multirow{2}{*}{\begin{minipage}{8cm}\raggedright
    $\displaystyle K_{\epsilon,\rm LIM}(\chip;\chi) = 
     \WIN(\chip)
    {\delta^{(D)}(\chip-\chi)}$
    \end{minipage}}
    & 
    Gaussian White Noise\\
    $\epsilon^I_\vl(\chi)$ (\Sec{lim_noise}) &
    & $\epsilon^I(\vk')$ (\Eq{eI_base})\\
    & & \\
    \textbf{\textsf{Low Pass Filtered} LIM Noise}  & 
    \multirow{2}{*}{\begin{minipage}{8cm}\raggedright
    $\displaystyle K_{\epsilon,\rm LIM}^\LO(\chip;\chi)=
    \WIN(\chip)
    {\LO(\chi-\chip)}
    $
    \end{minipage}}
    & Gaussian White Noise\\    
    $\epsilon^{I,\LO}_\vl(\chi)$ (\Sec{lim_filter})& 
    & $\epsilon^I(\vk')$ (\Eq{eI_base})\\
    & & \\
    \textbf{\textsf{High-Pass Filtered} LIM Noise}  & 
    \multirow{2}{*}{\begin{minipage}{8cm}\raggedright
    $\displaystyle K_{\epsilon,\rm LIM}^\HI(\chip;\chi)= K_{\epsilon,\rm LIM}(\chip;\chi) - K_{\epsilon,\rm LIM}^\LO (\chip;\chi)
    $
    \end{minipage}}
    & Gaussian White Noise\\    
    $\epsilon^{I,\HI}_\vl(\chi)$ (\Sec{lim_filter})& 
    & $\epsilon^I(\vk')$ (\Eq{eI_base})\\
    & & \\
    \hline
    \hline
\end{tabular}
    \caption{
    Summary of our models for CMB lensing (\Sec{cmb}), line intensity mapping (\Sec{lim_cosmo}), instrumental LIM noise (\Sec{lim_noise}), and foreground (or equivalently high-pass) filtering of line intensity maps (\Sec{lim_filter}) with the unifying notation introduced in \Sec{spectra}.
    These models will be used to examine the detectability of a direct correlation between foreground-filtered line intensity maps and CMB lensing in \Sec{SNR}.
    }
    \label{tab:unify}
\end{table*}

\section{Information from line-of-sight correlations in flat-sky angular spectra}
\label{sec:spectra}
In this section we first introduce notation which unifies our models of CMB lensing and line intensity mapping (summarized in \Tab{unify}).
Using this unifying notation, we discuss how to evaluate the approximate flat-sky angular spectra while retaining information from line-of-sight correlations following \Refs{Raccanelli:2023fle, Gao:2023rmo, Gao:2023tcd}.
This approximation goes beyond the usual Limber approximation (\Refs{1954ApJ...119..655L, 1973ApJ...185..413P,1992ApJ...388..272K}) which we argued in \Sec{lim_filter} and show quantitatively in \App{limber_snr} is important due to additional line-of-sight correlations induced by foreground filtering.
Finally, in \App{limber}, we show how the Limber approximation arises as a further approximation of the flat-sky approximation below.

\subsection{Unifying notation for flat-sky projected fields}
\label{sec:unify}
For our upcoming discussion of angular spectra in \Sec{sspectra}, it is convenient to introduce notation that unifies our modeling of CMB lensing and LIM.
Consider an arbitrary 3-D field $\Phi$ which is projected with kernel $K_\phi$ into another field $\phi$:
\begin{equation}
    \phi(\vtheta;\bullet) = \int_\chip  K_\phi(\chip;\bullet) \Phi(\chip,\vtheta).
\end{equation}
The symbol $\bullet$ in $\phi(\vtheta;\bullet)$ and $K_\phi(\chip ; \bullet)$ encapsulates any external variables $\phi$ and $K_\phi$ may be dependent on.
For example, in the unfiltered LIM (\Eq{I0}), the $\bullet$ encapsulates the line-of-sight slice position $\chi$.
This way, $\phi$ can either denote a 2-D field like CMB lensing, or a 3-D field like the filtered LIM.

We will now re-express this field $\phi$ in a more convenient form for our flat-sky angular spectra.
First, we can expand the 3-D field $\Phi$ in terms of it's Fourier transform while imposing the flat-sky approximation that the position across the line of sight satisfies
$\vx_\perp=\chi\vtheta$:
\begin{equation}
    \phi(\vtheta;\bullet) = \int_\chip  K_\phi(\chip;\bullet) \int_{\kpar}e^{i\kpar \chip} \int_{\kperp} e^{i\kperp\cdot\chi\vtheta}\Phi(\vk).
\end{equation}
The projected field $\phi$ can then be Fourier transformed with respect to $\vtheta$, whose conjugate variable is $\vl$, yielding
\begin{equation}
    \phi_\vl(\bullet) = \int_\chip \frac{K_\phi(\chip;\bullet)}{\chip^2}\int_\kpar e^{i\kpar\chip}\Phi(\kpar, \kperp = \vl/\chip).
    \label{eq:gen}
\end{equation}
We use this form of $\phi$ to evaluate the flat-sky angular spectra.
A summary of our CMB lensing and LIM models (\Sec{models}) within this unifying notation can be found in \Tab{unify}.

\begin{widetext}
\subsection{Flat-sky angular spectra without discarding line-of-sight correlations}
\label{sec:sspectra}
Consider two arbitrary fields $\phi,\psi$ following the form of \Eq{gen}:
\begin{align}
    \phi_\vl(\bullet) = \int_\chip \frac{K_\phi(\chip;\bullet) }{\chip^2}\int_\kparp e^{i\kparp \chip} \Phi(\kparp, \vl / \chip)\quad \textrm{ and }\quad
    \psi_\vm(\bullet) = \int_\chipp \frac{K_\psi(\chipp;\bullet)} {\chipp^2} \int_\kparpp e^{i\kparpp \chipp} \Psi(\kparpp, \vm / \chipp)   .
\end{align}
Assuming that the 3-D fields $\Phi,\Psi$ are statistically homogeneous, isotropic, and have cross spectrum $\langle \Phi\Psi\rangle' = P^{\Phi\Psi}$ leads to the cross-spectrum for $\phi,\psi$:
\begin{align}
\langle \phi_\vl(\bullet) \psi_\vm(\bullet) \rangle 
     = 
    \int_{\chip}\int_{\chipp}
    (2\pi)^2\delta^{(D)}\left(\frac{\vl}{\chip}  + \frac{\vm}{\chipp} \right)
    \frac{K_\phi(\chip ;\bullet)}{\chip^2}
    \frac{K_\psi(\chipp;\bullet)}{\chipp^2} 
    \int_\kparp e^{i\kparp(\chip - \chipp)}
    P^{\Phi\Psi}\left(k^2 = \kparp^2 + \frac{\ell}{\chip} \frac{L}{\chipp}\right)
,
    \label{eq:cross_full}
\end{align}
where we used the Dirac delta in \Eq{cross_full} to set $\ell^2 / \chip^2=\ell/\chip \times L/ \chipp$.
This off-diagonal Dirac delta which induces coupling between $\vl\ne\vm$ arises because the flat-sky approximation breaks the rotational invariance (isotropy) of the 3-D fields.
In particular, the flat-sky approximation replaces statistical rotational invariance with statistical translational invariance across the line-of-sight.
However, on small angular scales translational invariance across the line-of-sight can be approximately traded for rotational invariance.
Thus, there exists approximations for \Eq{cross_full} which restore the familiar statistically isotropic $\delta^{(D)}(\vl+\vm)$ structure of full-sky angular spectra.

The most commonly used approximation to evaluate \Eq{cross_full} is the Limber approximation (\Refs{1954ApJ...119..655L, 1973ApJ...185..413P,1992ApJ...388..272K}).
However, the Limber approximation discards information from correlations along the line-of-sight by approximating the innermost oscillatory integral in \Eq{cross_full} with the mean value of $P^{\Phi\Psi}$ along a given line of sight (see \App{limber} and \Reff{LoVerde:2008re} for more details).
However, as discussed in \Sec{lim_filter}, foreground filtering induces additional line-of-sight correlations which grow as more filtering is applied.
This calls into question whether it is truly safe to discard information from line-of-sight correlations.
To explore the detectability of a direct correlation between foreground-filtered line intensity maps and CMB lensing in a analytically robust way\footnote{{We believe an even more analytically robust approach would be to not make the flat-sky approximation and compute the full-sky angular spectrum but defer this to future work (see also \Reff{Kothari:2023keh}).}}, we decide to go beyond the Limber approximation by evaluating the flat-sky angular spectrum while retaining information from line-of-sight correlations.
Indeed, we show in \App{limber_snr} that the Limber approximations in the presence of foreground filtering exhibits concerning pathologies which leads us to argue against it's use for foreground filtered LIM.

Recently, \Refs{Raccanelli:2023fle, Gao:2023rmo, Gao:2023tcd} proposed a principled approximation for flat-sky angular spectra (\Eq{cross_full}) which retains information from line-of-sight correlations.
We now briefly review their results and also describe in \App{limber} the precise connection between this approximation and the Limber approximation.
To proceed, we first change variables from $\chip, \chipp$ to $\chib,\delta$ which are defined as
\begin{equation}
\begin{cases}
    \chib\phantom{\delta} = ({\chipp + \chip})/{2}\\
    \chib\delta = ({\chipp - \chip)}/{2}
\end{cases}
     \Leftrightarrow 
     \begin{cases}
     \chip\phantom{'} = \bar\chi(1-\delta)\\
     \chipp = \bar\chi(1+\delta).
     \end{cases}
\end{equation}
Since we have assumed that $0<\chip,\chipp<\infty$, the bounds of our new variables are $0 < \chib < \infty$ and $-1<\delta < 1$.
Rewriting the Dirac delta $\delta^{(D)}$ in \Eq{cross_full} in terms of $\chib, \delta$ yields
\begin{align}
\delta^{(D)}\left(\frac{\vl}{\chip}  + \frac{\vm}{\chipp} \right) 
    &= \bar\chi^2 (1-\delta^2)^2 \times \delta^{(D)}\left(\vl (1+\delta)+ \vm (1-\delta)\right) .
\end{align}
Instead of discarding information from line-of-sight correlations like in Limber, we make a softer approximation: line-of-sight correlations die off quickly.
In other words, we assume the dominant contributions to the angular spectrum come from terms where $|\delta|$ is small.
This allows us to expand the $\delta^{(D)}$ around a statistically isotropic approximation or, equivalently, taking ``momentum transfer" $\delta\times (\vl-\vm)$ to be our small expansion parameter:
\begin{equation}
    \label{eq:gaolimber}
    \delta^{(D)}\left(\frac{\vl}{\chip}  + \frac{\vm}{\chipp} \right) 
    =\bar\chi^2 (1-\delta^2)^2  \times \underbrace{\left(\delta^{(D)}(\vl + \vm) + \delta  \times (\vl - \vm)\cdot \frac{\partial \delta^{(D)}(\vl+\vm)}{\partial \vl} + \dots \right)}_{{\rm exp}\left\{\delta \times (\vl-\vm)\cdot \frac{\partial}{\partial \vl}\right\}\delta^{(D)}(\vl+\vm)}.
\end{equation}
Approximately restoring statistical isotropy in \Eq{cross_full} with the zero momentum transfer term of \Eq{gaolimber} then yields an approximation for the flat-sky angular cross-spectrum which retains some information from line-of-sight correlations (\Refs{Raccanelli:2023fle, Gao:2023rmo, Gao:2023tcd})\footnote{
In this expression, we have made use of the well-known geometric recalibration \cite{LoVerde:2008re,Gao:2023tcd} to replace $\ell^2$ with $\ell(\ell+1)$ in the argument of $P^{\Phi\Psi}$.
}:
\begin{align}
\langle \phi_\vl(\bullet) \psi_\vm(\bullet) \rangle' \approx 
    \int_{\chib}\ 2\chib\int_{-1}^1{d\delta} 
    \ 
    \frac{K_\phi(\chib(1-\delta);\bullet)K_\psi(\chib(1+\delta);\bullet)}
    {\chib^2}
    \underbrace{\int_\kparp e^{i\kparp\times 2\chib\delta}P^{\Phi\Psi}\left( k^2 = \kparp^2 + \frac{\ell(\ell+1)}{\bar\chi^2(1-\delta^2)}\right)}_{\equiv \I_\ell(\chib,\delta)}.
    \label{eq:gen_cross}
\end{align}
{The innermost integral $\I_\ell(\chib,\delta)$ dictates the contribution to the flat sky angular spectrum by correlations between slices separated by $2\chib\delta$ along the line-of-sight.
Thus we refer to $\I_\ell(\chib,\delta)$ as an angular spectrum of line-of-sight correlation functions\footnote{In \Refs{Gao:2023rmo, Gao:2023tcd}, $\xi_\ell(\chib,\delta)$ is referred to as the ``unequal-time angular power spectrum"}.
The line-of-sight correlation functions have support up to $\delta \sim 1/ \ell$ (\Fig{dkparp}) meaning that as we go to smaller angular scales (larger $\ell$), statistical translational invariance across the line of sight becomes a increasingly good approximation for statistical rotational invariance.
In the Limber approximation, this $\I_\ell(\chib,\delta)$ essentially reduces to a Dirac delta as we show in \App{limber}.}


\end{widetext}

\begin{figure*}
    \centering
    \includegraphics{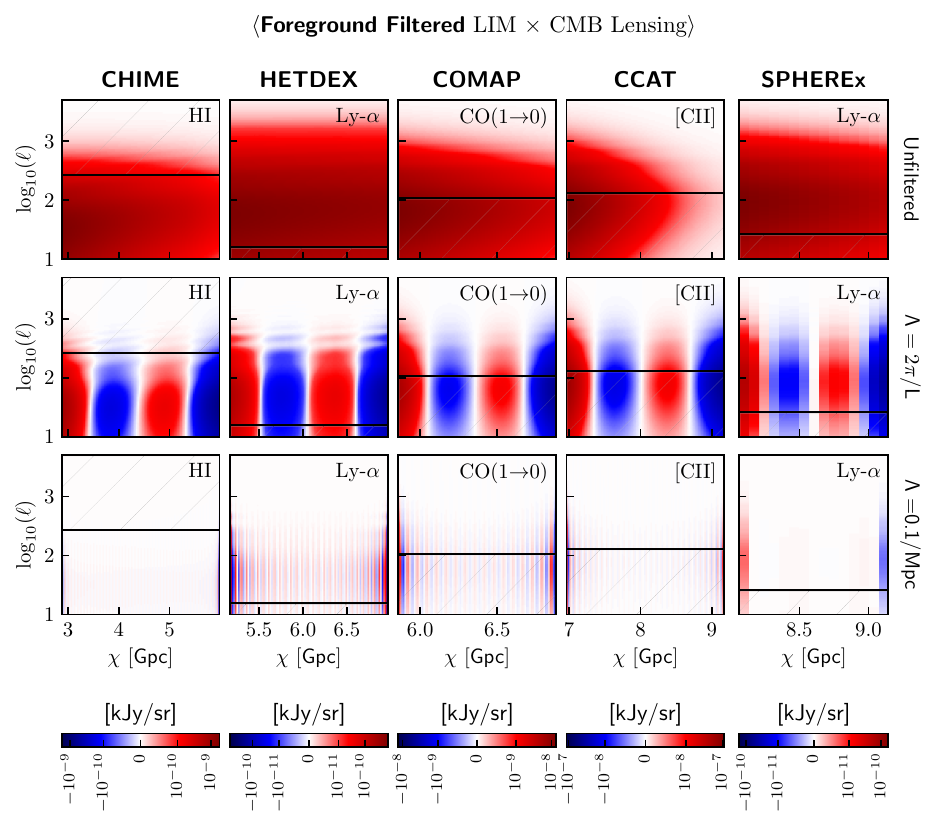}        
    \caption{The foreground-filtered LIM $\times$ CMB lensing cross-spectrum (\Eq{obs}) computed for each LIM experiment we consider (\Tab{experiments}) using our models described in \Sec{models} and summarized in \Tab{unify} for different amounts of filtering.
    The computation of this cross-spectrum is described in \App{Ik}.
    Hatched regions correspond to $\ell$ modes which are inaccessible given the smallest and largest angular scales observed by each experiment.
    }
    \label{fig:obs}
\end{figure*}

\section{Detectability of the direct correlation between foreground-filtered LIM and CMB lensing}
\label{sec:SNR}

In this section, we analytically compute the signal-to-noise ratio (\SNR) for the cross correlation of a high-pass filtered LIM and CMB lensing measured by the \textsf{Simons Observatory (SO)} (\Reff{SimonsObservatory:2018koc}) using our models of LIM, CMB lensing, LIM noise, and foreground filtering (described in \Sec{models} and summarized in \Tab{unify}) with the flat-sky angular spectrum approximation we describe in \Sec{spectra} which retains information from line-of-sight correlations.
The emission lines (\App{luminosity}) and experiments (\App{experiments}) we consider specifically in this section are
\begin{itemize}
    \item \textbf{HI} 21-cm emission (\App{HI}) measured by \textsf{{CHIME}} (\App{CHIME}, \Refs{CHIME:2022dwe, CHIME:2022kvg}),
    \item \textbf{Ly-$\boldsymbol\alpha$} emission (\App{Lya}) measured by \textsf{{HETDEX}} (\App{HETDEX}, \Refs{Hill:2008mv, 2016ASPC..507..393H, Gebhardt:2021vfo}) and \textsf{{SPHEREx}}\footnote{{We believe it is likely that \textsf{SPHEREx} H$\alpha$ may be a better candidate to cross correlate with CMB lensing than Ly-$\alpha$. Indeed, the redshift range sensitivity of \textsf{SPHEREx} H$\alpha$ ($z\sim 0.2-5$) both covers the peak of the CMB lensing kernel and a more active period of star formation relative to \textsf{SPHEREx} Ly-$\alpha$. We expect both these factors to boost the \textsf{SPHEREx} H$\alpha$ $\times$ CMB lensing cross spectrum relative to \textsf{SPHEREx} Ly-$\alpha$ but defer detailed exploration to future work.}} (\App{SPHEREx}, \Refs{SPHEREx:2014bgr, SPHEREx:2016vbo, SPHEREx:2018xfm}),
    \item \textbf{CO(1$\rightarrow$0)} rotational transition (\App{CO}) measured by \textsf{{COMAP}} (\App{COMAP}, \Refs{Li:2015gqa, Cleary:2021dsp,COMAP:2021qdn,COMAP:2021pxy,COMAP:2021sqw, COMAP:2021lae}), and
    \item \textbf{[CII]} emission (\App{CII}) measured by \textsf{{CCAT}} (\App{CCAT}, \Refs{2020JLTP..199.1089C}).
\end{itemize}
A summary of assumed experimental configurations can be found in \Tab{experiments}.

For future convenience, we define the following quantities: 
\begin{align}
    \mathcal O^{\kappa}_\ell &\equiv \langle \kappa_\vl \kappa_{-\vl}\rangle'&\textrm{(\App{KK})}\label{eq:Ok}\\
    \N^\kappa_\ell &\equiv [N^{(0)}\textrm{ bias}]& \textrm{(\Reff{Kesden:2003cc})}\label{eq:NK}\\
    \O^{\HI,\kappa}_\ell(\chi) &\equiv \langle I_\vl^\HI(\chi) \kappa_{-\vl}\rangle'&\textrm{(\App{Ik})}\label{eq:obs}\\
    \N^\HI_\ell(\chi,\chi') &\equiv \langle \epsilon^\HI_\vl(\chi) \epsilon^\HI_{-\vl}(\chi')\rangle'&\textrm{(\App{comp_ee})}\label{eq:NHI}\\
    \mathcal O^{\HI}_\ell(\chi, \chi') &\equiv \langle I_\vl^\HI(\chi) I_{-\vl}^\HI(\chi')\rangle'.&\textrm{(\App{II})}\label{eq:OHI}
\end{align}
We defer numerical evaluation details to \App{explicit} as enumerated in the above definitions. 
{In addition, we also publicly release all code used to numerically evaluate these quantities at \href{https://github.com/DelonShen/LIMxCMBL/blob/main/README.md}{\texttt{github.com/DelonShen/LIMxCMBL}} (\Reff{github}).}
The only term that we do not detail the computation of in this paper is the noise contribution to the CMB lensing spectrum, $\N^\kappa$ (\Eq{NK}), which we approximate with the so-called $N^{(0)}$ bias (see e.g. Eq. (23) of \Reff{Kesden:2003cc}) and compute with \texttt{LensQuEst} (\Reff{Schaan:2018tup}).
Our computation of the primary observable, the direct correlation of foreground-filtered line intensity maps with CMB lensing, $\O^{\HI,\kappa}$ (\Eq{obs}), is visualized in \Fig{obs} for all experiments we consider at various amounts of foreground filtering.
Figures showing other quantities can be found in their corresponding appendices.

\subsection{Estimator of the LIM$\times$CMB lensing cross-spectrum}
For a LIM experiment which observes a sky area $\Omega_{\rm field}$, the simplest unbiased estimator for the LIM$\times$CMB lensing cross-spectrum, $\O^{\HI,\kappa}_\ell(\chi)$, in a bin of $\chi$ with width $\delta\chi$ and annuli of $\vl$ with radial width $\delta\ell$ is
\begin{equation}
    \hat{\mathcal O}_{\ell}^{\HI,\kappa}(\chi) = \frac 1 {N_\ell}\sum_{[\ell_\pm]} \frac 1 {N_\chi} \sum_{[\chi_\pm]}\frac{I_{\vl'}^\HI(\chip) \kappa_{-\vl'} }{\Omega_{\rm field}},
    \label{eq:estimator_int}
\end{equation}
where the summation shorthand denote summing over all samples in the bin and $N_\ell,N_\chi$ denotes the number of samples in the $\ell$ and $\chi$ bin respectively.
Shortly we will need $N_\ell$ which we can compute with the size of the angular fundamental $\ell_F^2$ determined by $\Omega_{\rm field}$:
\begin{equation}
    N_\ell = \frac{2\pi\ell\delta \ell}{(\ell_F)^2} = \Omega_{\rm field}\times\frac{\ell \delta \ell}{2\pi}.
\end{equation}
Note that \Eq{estimator_int} can be naturally translated into integrals:
\begin{equation}
    \hat{\mathcal O}_{\ell}^{\HI,\kappa}(\chi) = \frac 1 {2\pi\ell\delta\ell}\int_{[\ell^\pm]} \frac 1 {\delta\chi} \int_{[\chi^\pm]}\frac{I_\vl^\HI(\chip) \kappa_{-\vl} }{\Omega_{\rm field}},
\end{equation}
where we have introduced shorthand notation to denote integration over these $\ell$ and $\chi$ bins:
\begin{align}
    \int_{[\ell^\pm]}(\cdots) &=  \int_0^{2\pi}d\theta \int_{\ell-\delta \ell/2}^{\ell+\delta\ell/2} \ell d\ell\times(\cdots)\\
    \int_{[\chi^\pm]} (\cdots)&= \int_{\chi-\delta\chi/2}^{\chi+\delta\chi/2}d\chip\times(\cdots).
\end{align}

\subsection{Covariance \& SNR of the LIM$\times$CMB lensing cross-spectrum}
\label{sec:snr_calc}

In our flat sky approximation (\Sec{spectra}), different angular scales $\ell\ne \ell'$ are uncorrelated, but different redshift slices $\chi\ne\chip$ are correlated.
Indeed, for a fixed $\ell$, the covariance between two $\chi$ bins, which we will index by integers $\i,\j$, of our cross-spectrum estimator (\Eq{estimator_int}) is
\begin{align}
\C_{\i\j} = \frac 1 {N_\ell} \biggr\{\O_{\ell,\i}^{\HI,\kappa}\O_{\ell,\j}^{\HI,\kappa}
   +(\O^\HI_{\ell,\i\j} + \N^\HI_{\ell,\i\j})(\O^\kappa_\ell+\N^\kappa_\ell)&\biggr\},
   \label{eq:cov}
\end{align}
where we will assume that everything varies slowly as a function of angular scale $\ell$ (but crucially not $\chi$, see \Fig{obs}), that the LIM instrumental noise and CMB lensing noise are uncorrelated with the underlying cosmological signal, and have introduced the shorthand notation for theory curves binned in $\chi$ with width $\delta \chi$:
\begin{align}
    f_{\i} &= \frac 1 {\delta\chi}\int_{[\chi^\pm_\i]}f(\chi'_\i)\\
    f_{\i\j} &= \frac 1 {(\delta\chi)^2}\int_{[\chi^\pm_\i]}\int_{[\chi^\pm_\j]}f(\chi'_\i, \chi'_\j).
    \label{eq:fij}
\end{align}

For \textsf{CHIME, HETDEX, COMAP,} and \textsf{CCAT}, we use {100} uniformly sized $\chi$ bins and for \textsf{SPHEREx}, we use 15 uniform $\chi$ bins from the minimum to maximum $\chi$ probed by each experiment.
Our assumed spectral resolving power reported in \Tab{experiments} sets the smallest bin width allowable for each experiment and our $\chi$-binning choices are consistent with our assumed spectral resolving power.

The squared signal-to-noise ratio ($\SNR^2$) in an $\ell$-bin of width $\delta \ell$ can then be computed as
\begin{equation}
    \delta \SNR^2_\ell = \sum_{\i,\j} \O^{\HI,\kappa}_{\ell,\i}(\C^{-1})_{\i\j} \O^{\HI,\kappa}_{\ell,\j}.
    \label{eq:dsnr}
\end{equation}
{The numerical evaluation of $\delta\SNR^2_\ell$ requires some care due to the singular value structure of the covariance matrix $\C$ which we discuss more in \App{SNR}.}
The total $\SNR^2$ for an experiment then is
\begin{equation}
    \SNR^2 = \sum_\ell \delta{\SNR^2_\ell}. 
\end{equation}
With this formalism in place, we are now ready to examine the detectability of a direct correlation between foreground-filtered line intensity maps and CMB lensing.

\begin{figure}
    \includegraphics{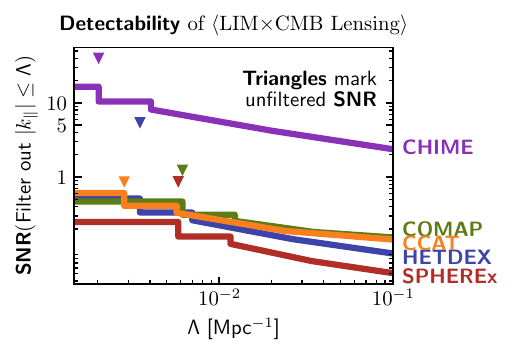}
    \caption{
    Foreground filtering suppresses by only an $O(1)$ factor the detectability of a direct correlation between line intensity maps and CMB lensing, though it is unlikely this direct correlation will be detectable with current era LIM experiments (colored lines, \Tab{experiments}) when correlated with CMB lensing measured by the \textsf{Simons Observatory} (\Reff{SimonsObservatory:2018koc}).
    However, we show in \Fig{SNR-aspir} this direct correlation is imminently detectable by future LIM experiments, principally because the suppression of \SNR{} due to foreground filtering is not catastrophic as previously claimed in the literature.
    Triangles mark the $\SNR$ when no filtering is applied to the LIM.
    Though the curves should change only in discrete steps at integer multiples of the fundamental, we only show these discrete steps at $\kF$ and $2\kF$.
    }
    \label{fig:SNR-SO}
    
\end{figure}
\begin{figure}
    \includegraphics{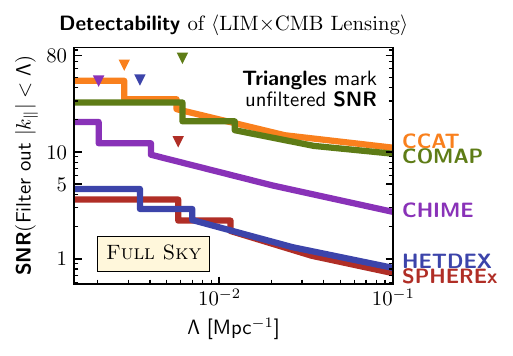}
    \caption{
    The direct correlation of LIM with CMB lensing is likely detectable with future LIM experiments (colored lines), which we approximate as current era experiments (\Tab{experiments}) that observe the full sky, when correlated with CMB lensing measured by the \textsf{Simons Observatory} (\Reff{SimonsObservatory:2018koc}).
    In particular we predict that wider-field verions of \textsf{COMAP} and \textsf{CCAT} will be able to precisely measure the LIM$\times$CMB lensing cross-spectrum {($\SNR \geq 10$ at $\Lambda = 0.1\ {\sf Mpc}^{-1}$)}.
    Furthermore, as we show in \Fig{Ofield_detect}, these experiments need only observe a smaller fraction of the sky to detect the direct correlation.
    Triangles mark the $\SNR$ when no filtering is applied to the LIM.
    Though the curves should change only in discrete steps at integer multiples of the fundamental, we only show these discrete steps at $\kF$ and $2\kF$.
    }
    \label{fig:SNR-aspir}
\end{figure}

\begin{figure}
    \includegraphics{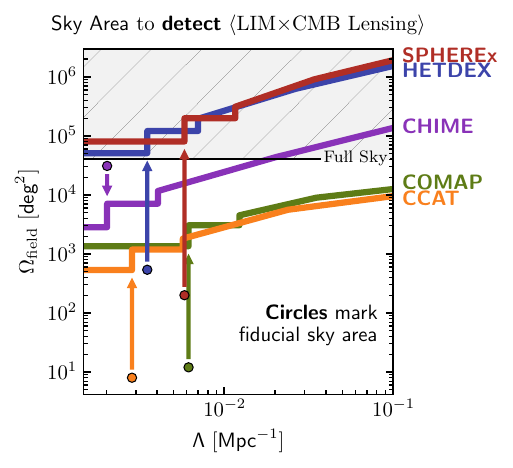}
    \caption{
    We showed in \Fig{SNR-aspir} that if current experiments could observe the full sky, our proxy for future experiments, a detection the direct correlation would be likely.
    Here, we show that the direct correlation of LIM with \textsf{Simons Observatory} (\Reff{SimonsObservatory:2018koc}) CMB lensing is detectable ($\SNR\geq 5$) by future LIM experiments even when only a fraction of the sky is observed.
    The circles mark the fiducial sky area each experiment observes (\Tab{experiments}).
    Though the curves should change only in discrete steps at integer multiples of the fundamental, we only show these discrete steps at $\kF$ and $2\kF$.
    }
    \label{fig:Ofield_detect}
\end{figure}

\begin{figure}
    \centering
    \includegraphics{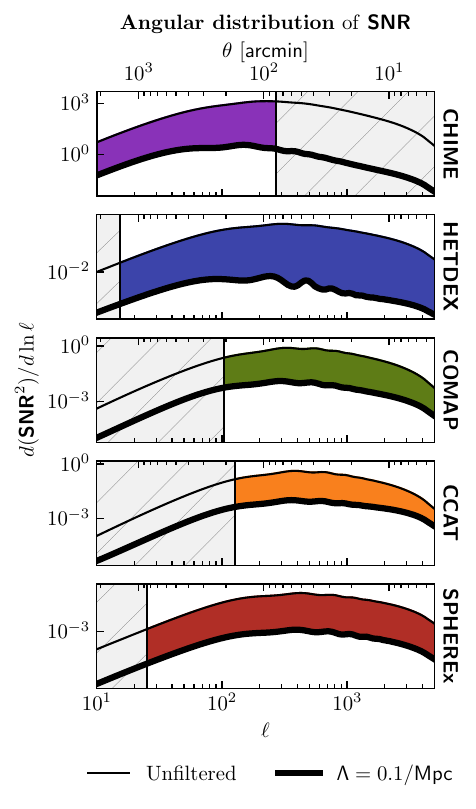}
    \caption{
    The largest contributions to the detectability of a direct correlation between LIM $\times$ CMB lensing come from features of angular scale $\theta\sim \ang{;10;}-\ang{;100;}$, or equivalently $\ell\sim 100-1000$, for the current and upcoming LIM experiments we consider (\Tab{experiments}) when correlated with CMB lensing measured by the \textsf{Simons Observatory} (\Reff{SimonsObservatory:2018koc}).
    Because the largest contribution to \SNR{} come from angular scales which the CMB lensing power spectrum measured by \textsf{SO} will be signal dominated (\Fig{ClKK}), improvements in CMB lensing measurements will not greatly improve the detectability of a direct correlation as we show in \Fig{SNR_N0}.
    The gray hatched regions correspond to $\ell$ modes which are inaccessible given the smallest and largest angular scales observed by each experiment and the solid color block between each pair of curves indicates that all intermediate amounts of filtering fall between the two curves shown.
    }
    \label{fig:dSNR_dlnl}
\end{figure}

\begin{figure}
    \includegraphics{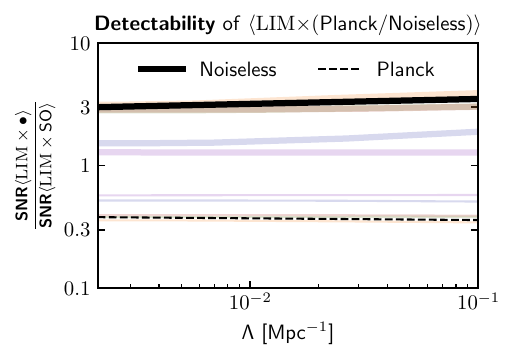}
    \caption{
    The detectability of the direct correlation between foreground-filtered LIM and CMB lensing changes by less than an order of magnitude when replacing the expected noise in the CMB lensing spectrum measured by \textsf{Simons Observatory} (\Reff{SimonsObservatory:2018koc}) with either (1) a perfectly noiseless measurement of the  CMB lensing spectrum (solid black) or (2) \textsf{Planck 2018} noise (dashed black, \Reff{Planck:2018lbu}).
    Indeed, (1) is to be expected from \Fig{dSNR_dlnl} as the the largest contribution to $\SNR{}$ come from angular scales $\ell$ where \textsf{SO} is already signal dominated (\Fig{ClKK}).
    Thus, improvements towards detecting the LIM $\times$ CMB lensing cross-spectrum will be driven primarily by improvements in LIM experiments in the near future.
    Light lines show the relevant ratios of \SNR{} for all LIM experiments and black lines pick out the general trend followed by all experiments except for {\sf CHIME} (light purple) {and {\sf HETDEX} (light blue)}.
    }
    \label{fig:SNR_N0}
\end{figure}

\subsection{Detectability for current and future LIM experiments}
\label{sec:detect}

Encouragingly, \Fig{SNR-SO} shows that foreground filtering suppresses the detectability of a direct correlation between LIM and \textsf{SO} CMB lensing only by a $O(1)$ factor\footnote{{A key exception to this is \textsf{HETDEX} which traces Ly-$\alpha$ emission across a period where observed line emission does not evolve strongly}}.
However, we find that a detection of this direct correlation may be unlikely for current era-experiments.

Because the suppression of \SNR{} due to foreground filtering is mild, we predict (\Fig{SNR-aspir}) that future experiments like wider-sky area version of \textsf{COMAP} and \textsf{CCAT}\footnote{In particular, we assume a fixed instrumental noise level $P^{\epsilon_I}$ but increase the observing area.
Of course, the noise level scales with amount of sky area observed (\App{experiments}) unless other change in experimental design can compensate.} will be able to precisely measure this direct correlation of LIM with CMB lensing\footnote{%
{Specifically, we predict they will have a $\SNR\geq 10$ at $\Lambda~=~0.1~{\sf Mpc}^{-1}$. This choice of reference $\Lambda$ is motivated by the fact that (1) optical- and mm-wavelength observations are generally less dominated by foregrounds than 21cm observations and (2) in the detection of 21cm emission by \textsf{CHIME} in \Reff{CHIME:2023til}, foregrounds required a $\Lambda\sim 0.1\ {\sf Mpc}^{-1}$.}
}.
Indeed, we show in \Fig{Ofield_detect} that these experiments can detect the direct correlation while observing small fractions of the full-sky.

{Of particular note in \Fig{Ofield_detect} is \textsf{CHIME} whose fiducial sky area we predict is sufficient to have ${\SNR}\geq 5$ at {$\Lambda = 2\pi/\L$}.
{Becaue of this,} we also infer\footnote{Note that \cite{CHIME:2023til} was required to use $\Lambda =0.13\ {\sf Mpc}^{-1}$ which is beyond what we compute.} that future 21cm intensity mapping experiments such as \textsf{CHORD} (\Reff{CHIME:2023til}), \textsf{HIRAX} (\Reff{Crichton:2021hlc}), and \textsf{SKA} (\Reff{SKA:2018ckk}) {may be able to detect this direct correlation} though defer precise computations of prospects for these experiments to future work.
}

Our ability to detect this direct cross spectrum comes primarily from angular features of size $\ang{;10;}-\ang{;100;}$, or equivalently $\ell\sim 100 - 1000$ (\Fig{dSNR_dlnl}).
Because of this, LIM experiments will be the limiting factor in detecting this direct correlation because the angular scales $\ell$ which contribute the most to \SNR{} are angular scales where measurements of the CMB lensing spectrum by {\sf SO} will be signal dominated.
Indeed, we show in \Fig{SNR_N0} that the difference in $\SNR{}$ between a perfectly noiseless reconstruction of the CMB lensing spectrum, a level of noise expected for \textsf{SO}, and the level of noise in \textsf{Planck 2018} (\Reff{Planck:2018lbu}) all differ by less than a order of magnitude.

{While no approximation beyond what was described in \Sec{spectra} have been made in the computation of \SNR{} in this section, features like the extreme delicacy of computing the LIM cosmic variance term (\App{II}) motivate the exploration in \App{approx} of two possible approximations which could simplify the \SNR{} computation carried out in this section:}
\begin{enumerate}
    \item  \textbf{Noise dominated approximation} (\App{noise_dom}, \textbf{recommended}): Here we examine the validity of approximating the covariance \Eq{cov} as noise dominated (\Eq{noise_dom}), thus discarding the cosmic variance ($\O^{\HI,\kappa}$ and $\O^\HI$) terms.
    Indeed the 3-D power spectrum of current and possibly future era LIM experiments are expected to be noise dominated\footnote{That is to say, for future era LIM experiments, it is possible instrumental noise level may still be the dominant contribution over the cosmological signal to the 3-D power spectrum. However, a wider area is able to be surveyed, thus dramatically increasing the number of modes accessible.} as we saw in \Fig{lim-model}.
    We show in \App{noise_dom} that this renders the noise dominated approximation a good approximation when computing the detectability of a direct correlation.
    Furthermore, the cosmic variance terms are more difficult to compute numerically (\App{II}), especially for large amounts of filtering which induce strong oscillatory off-diagonal terms (\Fig{corr_II}).
    Because of these factors, we {{recommend}} the use of the noise dominated approximation for future studies.
    \item \textbf{Limber approximation} (\App{limber_snr}, \textbf{not recommended}): We examine the validity of using the Limber approximation to compute the $\SNR{}$ instead of the approximation we use which retains information from line-of-sight correlations.
    As we have discussed throughout this paper, foreground filtering induces line-of-sight correlations which grow in strength as more filtering is applied.
    We show in \App{limber_snr} that while the Limber approximation and our approximation agree when no foreground filtering is present, the Limber approximation significantly deviates from our own when foreground filtering is applied.
    Counterintuitively,
    contributions to the \SNR{} by small angular scales are the first to deviate as more filtering is applied, the reason for this we defer to \App{limber_snr}.
    As many pathfinder experiments are restricted to probing small sky areas, this (counterintuitive) inability of the Limber approximation to correctly capture small angular scales in the presence of foreground filtering becomes a dangerous pathology.
    Because of these factors, we {{do not recommend}} the use of the Limber approximation for future studies.
\end{enumerate}

\section{Conclusion}
\label{sec:conclusion}
The direct correlation of line intensity mapping with CMB lensing is uniquely suited to probe the largely unmapped faint and high-redshift universe.
In this paper, we have established a theoretical understanding of this direct correlation, highlighting the critical role played by lightcone evolution.
We analytically show that foreground filtering suppresses the detectability (\SNR) of this direct correlation only mildly as opposed to exponentially as previously assumed (\Refs{\prevv}).
As a result, we predict future wider-sky versions of \textsf{COMAP} and \textsf{CCAT} should be able to precisely measure this direct correlation.
{Furthermore, based on our calculations for \textsf{CHIME} in this paper, we infer that future 21cm intensity mapping experiments may also be able to precisely measure this cross-correlation.}

There are several interesting avenues for future exploration.
While we focus on CMB lensing in this paper, our analytical framework generalizes to the direct correlation of foreground-filtered LIM with \textit{any projected field}.
Indeed, the direct correlation of LIM with other observables like the cosmic infrared background (\Reff{Zhou:2022gmu}), cosmic shear (\Refs{Chung:2022lpr, Sangka:2024vfg}), and kinematic Sunyaev–Zel'dovich effect (\Reff{Li:2018izh}) all hold promise.
{Our analytical framework also is restricted to the flat-sky approximation.
A analogous exploration without approximations using full-sky angular spectra may lead to additional insights into this direct correlation (see also \Reff{Kothari:2023keh}).}
In this paper, we have also restricted our exploration to linear theory in order to highlight the inescapable observational effects of lightcone evolution.
In doing so, we have neglected the impact of non-linear gravitational evolution and redshift-space distortions (\Refs{Zhu:2016esh, Li:2018izh, Karacayli:2019iyd, Foreman:2018gnv, Liu:2019awk, Darwish:2020prn, Zang:2022qoj}), whose interaction with lightcone evolution in this direct correlation is yet to be precisely explored.
{We have also only considered a restricted set of experiments and lines in this paper.
Thus, prospects of directly correlating CMB lensing with many other promising lines and experiments, like \textsf{SPHEREx} H$\alpha$, \textsf{CHORD} (\Reff{CHIME:2023til}), or \textsf{MeerKLASS} (\Reff{MeerKLASS:2017vgf}) among many others, are yet to be characterized.}
Finally, we have restricted our exploration to the detectability of this direct correlation, leaving open the exciting direction of exploring what physics this  observable of the faint and high-redshift universe can reveal.
All these we defer to future work.


\section*{Acknowledgments}
We wish to thank Dhayaa Anbajagane, Jos\'e Luis Bernal, Tzu-Ching Chang, Dongwoo Chung, Simon Foreman, Zucheng Gao, Dylan Jow, Adam Lidz, Aaron Parsons, Tristan Pinsonneault-Marotte, Anthony Pullen, Guochao Sun, and Martin White for useful discussions.
This research made use of computational resources at SLAC National Accelerator Laboratory, a U.S. Department of Energy Office of Science laboratory, and the Sherlock cluster at the Stanford Research Computing Center (SRCC). 
We would like to thank Stanford University, SRCC, and SLAC for providing computational resources that contributed to these research results.
D.S. is additionally supported by the National Science Foundation Graduate Research Fellowship under Grant No. DGE-2146755. 
NK acknowledges support from the Fund for Natural Sciences of the Institute for Advanced Study. 
Parts of this work were completed under the support of NSF award AST-2108126.
This work received support from the U.S. Department of Energy under contract number DE-AC02-76SF00515 to SLAC National Accelerator Laboratory.


\appendix

\section{Discarding information from line-of-sight correlations yields the Limber approximation}
\label{app:limber}

In this appendix we show how the Limber approximation arises from an information-discarding limit of the angular spectrum presented in \Refs{Raccanelli:2023fle, Gao:2023rmo, Gao:2023tcd} which we discussed in \Sec{spectra}.
Consider the innermost integral of \Eq{gen_cross}:
\begin{equation}
    \int \frac{d\kparp}{2\pi}\  e^{i\kparp\times 2\chib\delta}P^{\Phi\Psi}\left(k^2= {\kparp^2 + \frac{\ell(\ell+1)}{\bar\chi^2(1-\delta^2)}} \right).
\end{equation}
For convenience, lets define the dimensionless variable $\alpha = 2\bar\chi \kparp$.
Assume that $P^{\Phi\Psi}$ is slowly varying as a function of its argument.
The integrand we have now is a slowly varying function $P^{\Phi\Psi}$ integrated against a kernel  $e^{i\alpha\delta}$ that oscillates rapidly as $\alpha\rightarrow\infty$.
Contributions to the integral from rapidly oscillating terms will average out to zero.
So the integral is only non-zero when 
$\alpha \ll 1\Leftrightarrow \kparp \ll 1 / \chib$. 
Expanding $P^{\Phi\Psi}$ to zero-th order in $\alpha$
thus simply amounts to setting $\kparp$ to zero.
The integral thus reduces to:
\begin{align}
\nonumber \int_\kparp e^{i\kparp\times 2\chib\delta}P^{\Phi\Psi}\left(k^2={\kparp^2 + \frac{\ell(\ell+1)}{\bar\chi^2(1-\delta^2)}} \right)\\
\approx P\left(k^2=\frac {{\ell(\ell+1)}} {\bar\chi^2({1-\delta^2})} \right) \delta^{(D)}(\delta) \times \frac 1 {2\chib}
\label{eq:limber_trick}
\end{align}
Inserting this back into \Eq{gen_cross} then yields the well known Limber approximation:
\begin{equation}
    \langle \phi_\vl(\bullet) \psi_\vm(\bullet) \rangle' \approx 
    \int_{\chi}
    \frac{K_\phi(\chi;\bullet)K_\psi(\chi;\bullet)}{\chi^2} 
    P \left( \frac {\ell+1/2} \chi\right),
    \label{eq:Limber}
\end{equation}
where we have used the fact that $\sqrt{\ell(\ell+1)}\approx \ell + 1/2 + O(1/\ell)$.
Crucially, note that in \Eq{limber_trick}, we have a Dirac delta which forces the parameter $\delta$ to be zero.
This can be put in contrast to our assumption in \Sec{spectra} that $\delta$ is small (line-of-sight correlations die off quickly) while the Limber approximation fixes $\delta=0$ (discards line-of-sight correlation).

\begin{figure*}
    \centering
    \includegraphics{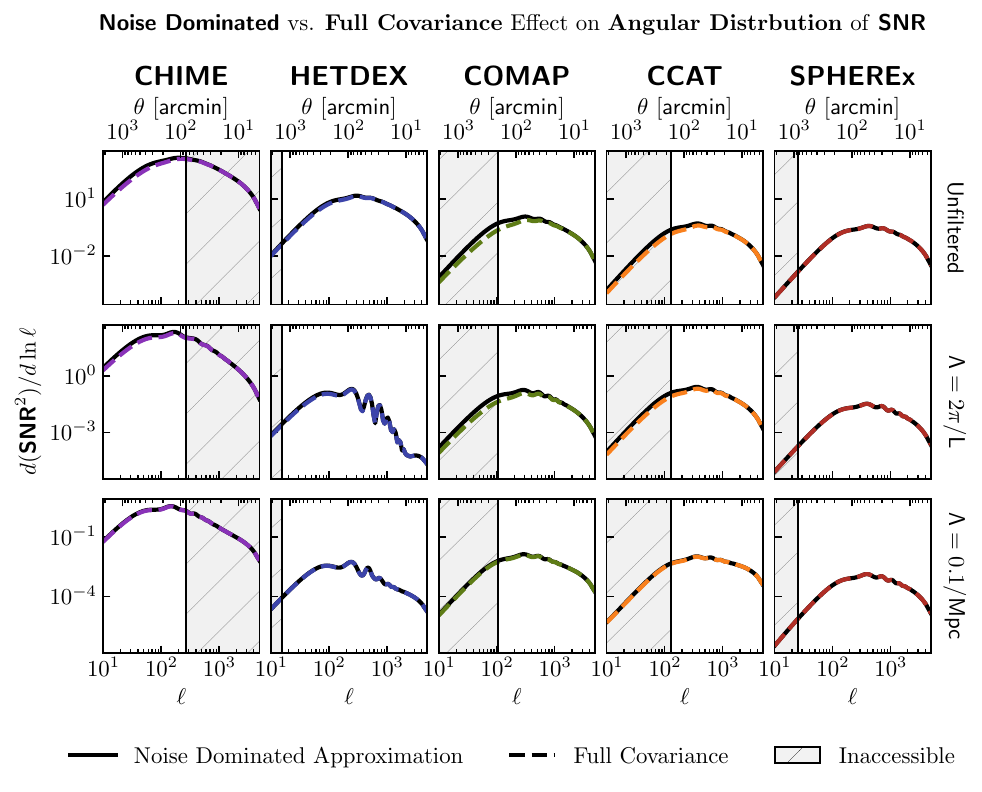}
    \caption{Contributions to the \SNR{} from large angular scales initially deviate between the noise dominated approximation and full covariance but converge towards each other as more filtering is applied.
    This initial deviation is likely due to the unfiltered instrumental noise covariance containing no $\chi\ne \chip$ correlations (\Eq{eIeI}) in our Gaussian white noise approximation while on large angular scales the cosmological signal intrinsically contain $\chi\ne\chip$ correlations.
    So, even if the instrumental noise dominates the variance ($\chi=\chip$), it misses these intrinsic $\chi\ne\chip$ covariances which play a role in the \SNR{} resulting in the deviations seen.
    Additionally, the cosmological signal on small angular scales have almost no $\chi\ne\chip$ covariance and are thus completely dominated by the instrumental noise, explaining the lack of deviation seen for high-$\ell$.
    Foreground filtering induces line-of-sight correlations whose strength is in proportion to the variance in the unfiltered case.
    So, the induced $\chi\ne\chip$ correlations in the instrumental noise dominate both the intrinsic and induced $\chi\ne\chip$ correlations in large angular scale cosmological modes and continue to dominate over the $\chi\ne\chip$ correlations of small angular scale cosmological modes.
    This explains the trends seen in \Fig{noise-dom-SNR}.
    }
    \label{fig:noise-dom-dsnr}
\end{figure*}

\begin{figure*}
    \centering
    \includegraphics{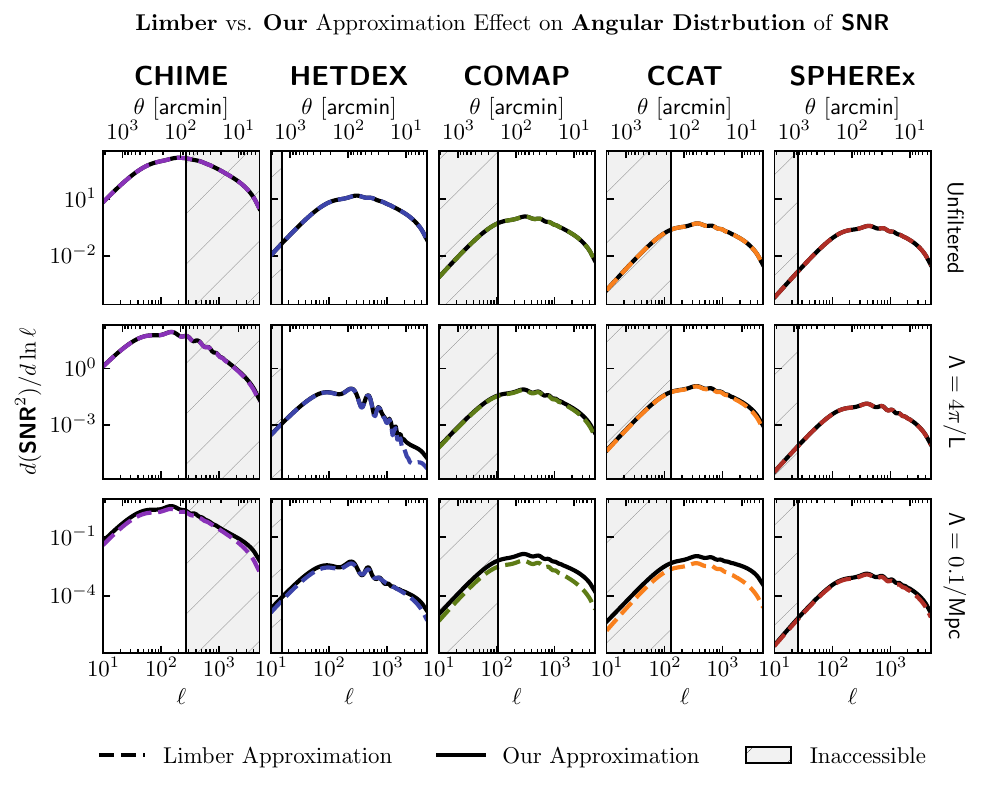}
    \caption{
    The contribution to \SNR{} from small angular scales (high-$\ell$) modes are the first to differ between the Limber and our approximation when foreground filtering is applied. 
    As more filtering is applied, the \SNR{} contribution from increasingly large angular modes deviate between the Limber and our approximation.
    {We speculate that} this is due to small angular scale cosmological modes having very little line-of-sight correlations. 
    Thus, the impact of foreground filtering induced line-of-sight correlations on the \SNR{} are felt immediately.
    Our approximations accounts for these line-of-sight correlations when computing the \SNR{} while the Limber approximation does not.
    Furthermore, large angular scale cosmological modes do inherently possess some line-of-sight correlations which may initially dominate over foreground filtering induced line-of-sight correlations.
    Thus the impact of filtering on the contribution to \SNR{} from large angular scales is delayed until foreground filtering induced line-of-sight correlations dominate over intrinsic line-of-sight correlations.
    This explains the trends seen in \Fig{limber-SNR}
    }
    \label{fig:limber-dsnr}
\end{figure*}

\begin{figure}
    \centering
    \includegraphics{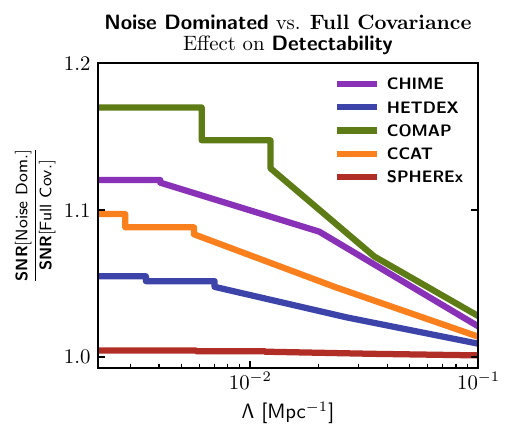}
    \caption{The instrumental noise power is expected to dominate over the cosmological signal in current era LIM experiments (\Fig{lim-model}) meaning that computing the detectability of the LIM$\times$CMB lensing cross-spectrum with a approximate covariance that assumes line intensity maps are noise dominated (\Eq{noise_dom}) results in minor errors which decrease as more foregound filtering is applied.
    Furthermore, the noise dominated approximation of the covariance is relatively cheap to compute (\App{comp_ee}) in comparison to the full covariance (\App{II}) making it a attractive option for future studies.
    We compare the angular ($\ell$) dependence of contribution to \SNR{} between the full and noise-dominated covariance in \Fig{noise-dom-dsnr} to better understand the trends seen in this figure.
    }
    \label{fig:noise-dom-SNR}
\end{figure}

\begin{figure}
    \centering
    \includegraphics{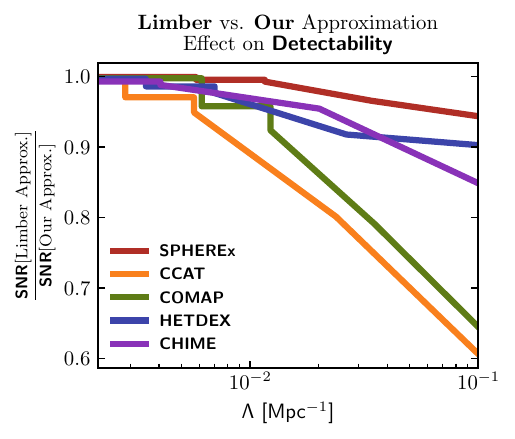}
    \caption{Foreground filtering induces line-of-sight correlations which contain information that the Limber approximation discards.
    Thus using the Limber approximation causes the signal-to-noise ratio of the LIM $\times$ CMB lensing cross-spectrum to be reduced in comparison to our calculation (\Sec{spectra} and \App{Ik}) which retains information from line-of-sight correlations. 
    Here we use the noise-dominated approximation (\App{noise_dom}) for the covariance which we show is a good approximation in \Fig{noise-dom-dsnr}.
    We compare the angular ($\ell$) dependence of contribution to \SNR{} between the Limber and our approximation in \Fig{limber-dsnr} to better understand the trends seen in this figure.
    }
    \label{fig:limber-SNR}
\end{figure}

\section{Validity of various approximations for foreground-filtered LIM}
\label{app:approx}

In this section we examine the validity of the noise dominated covariance approximation (\App{noise_dom}) and Limber approximation (\App{limber_snr}) when computing the detectability of the LIM $\times$ CMB lensing cross-spectrum.
By examining the angular contributions to \SNR{} in these approximations (\Figs{noise-dom-dsnr}{limber-dsnr}) we  are able to qualitatively understand the nature of deviations these approximations exhibit in total \SNR{} (\Figs{noise-dom-SNR}{limber-SNR}).
To summarize this appendix:
\begin{itemize}
    \item We \textbf{{recommend}} the use of the \textbf{noise dominated approximation} for current and possibly future era LIM experiments whose 3-D power spectrum are instrumental noise dominated (\Fig{lim-model}).
    \item We \textbf{{do not recommend}} the use of \textbf{the Limber approximation} for foreground-filtered LIM.
\end{itemize}

\subsection{Noise dominated covariance}
\label{app:noise_dom}

For current and upcoming LIM experiments, instrumental noise is expected to dominate over the cosmological signal as we saw in \Fig{lim-model}.
So, approximating the covariance $\C_{\i\j}$ as noise dominated, e.g.
\begin{equation}
    (\textsf{Noise Dominated})\ \C_{\i\j} \approx \frac 1 {N_\ell} \biggr\{ \N^\HI_{\ell,\i\j} (\O^\kappa_\ell + \N^\kappa_\ell )\biggr\},
    \label{eq:noise_dom}
\end{equation}
may be sufficiently accurate for future studies.
Indeed, computing the noise component of the LIM covariance (\App{comp_ee}) is much more efficient than computing than the cosmic variance terms(\App{II}).
Thus, it is advantageous if a sufficiently accurate prediction for the \SNR{} is achievable without needing to compute the cosmic variance component of the LIM covariance.

In \Fig{noise-dom-SNR} we see that the noise dominated approximation results in a mildly overoptimistic prediction for the \SNR{} in the unfiltered case, as expected, but an increasingly good prediction as more filtering is applied.
This is at most a $20\%$ overestimate of the \SNR{}, 
making it acceptable for forecasts.

The origin of the trend seen in \Fig{noise-dom-SNR} {is consistent with} \Fig{noise-dom-dsnr}, where we find that the contribution to \SNR{} from large angular scales (low-$\ell$) initially deviate between the noise dominated approximation and full covariance before settling to similar values as more filtering is applied. Simultaneously, the noise dominated approximation is always a good approximation for small angular scales (high-$\ell$).
{
The difference in low- and high-$\ell$ behavior can be understood through the following.
\begin{enumerate}
    \item {\sf Unfiltered}, high-$\ell$: The cosmic variance contribution to the unfiltered covariance on small angular scales (high-$\ell$) has very little intrinsic $\chi\ne\chip$ correlation (\Fig{corr_II}) making it essentially diagonal.
    The instrumental noise contribution to the unfiltered covariance is also diagonal (\Eq{eIeI}) and dominates over the nearly diagonal cosmic variance contribution (\Figs{var_eIeI}{var_II}).
    Thus, the noise dominated approximation is a fine approximation for small angular scales when no foreground filtering is applied
    \item {\sf Filtered}, high-$\ell$: Foreground filtering induces $\chi\ne\chip$ correlations in both the instrumental noise (\Fig{corr_eIeI}) and cosmic variance (\Fig{corr_II}) contribution to the covariance. 
    The strength of these induced correlations is in proportion to the variance in the unfiltered case.
    Thus, the induced $\chi\ne\chip$ correlations in the instrumental noise contribution dominate the induced $\chi\ne\chip$ correlations in the cosmic variance contribution.
    Furthermore, instrumental noise remains the dominant contribution to the diagonal.
    Because of this, the noise dominated approximation remains a fine approximation for small angular scales as foreground filtering is applied.
    \item {\sf Unfiltered}, low-$\ell$: The cosmic variance contribution to the unfiltered covariance on large angular scales (low-$\ell$) has some intrinsic $\chi\ne\chip$ correlations (\Fig{corr_II}) unlike in (1).
    However, the instrumental noise contribution to the unfiltered covariance is still diagonal (\Eq{eIeI}).
    So even if instrumental noise is the dominant contribution to the diagonal of the low-$\ell$ covariance, it cannot affect the $\chi\ne\chip$ correlations of the cosmic variance contribution.
    This is why the \SNR{} computed in the noise dominated approximation exhibits minor deviations for large angular scales when no foreground filtering is applied.
    \item {\sf Filtered}, high-$\ell$: Foreground filtering induces $\chi\ne\chip$ correlations in both the instrumental noise (\Fig{corr_eIeI}) and cosmic variance (\Fig{corr_II}) contribution to the covariance.
    The strength of these induced correlations is in proportion to the variance in the unfiltered case just like in (2).
    Thus, the induced $\chi\ne\chip$ correlations in the instrumental noise contribution (a) dominates over the foreground filtering induced $\chi\ne\chip$ correlations {and} (b) increasingly dominates the intrinsic $\chi\ne\chip$ correlations of the cosmic variance contribution mentioned in (3).
    Furthermore, instrumental noise remains the dominant contribution to the diagonal.
    Because of this, the noise dominated approximation becomes a increasingly fine approximation for large angular scales as foreground filtering is applied.
\end{enumerate}
We note that in the presence of non-white noise across frequency bands, $\chi\ne \chip$ correlations would exist in the instrumental noise even when no filtering is applied.
These would likely dominate over the intrinsic $\chi\ne\chip$ correlations of the cosmic variance contribution mentioned in (3). 
Thus, the noise-dominated approximation should become a fine approximation for all scales and amounts of filtering with a more careful treatment of instrumental noise.
}

{To summarize: because
\begin{enumerate}
    \item the noise dominated approximation is a good approximation at the level of $\SNR{}$ for experiments where the 3-D power spectrum is instrumental noise dominated,
    \item current and possibly future experiments will be instrumental noise dominated, and
    \item the noise contribution to the covariance is computationally efficient to compute,
\end{enumerate}
we \textbf{recommend} the noise dominated approximation for exploring foreground-filtered LIM experiments.}

\subsection{The Limber approximation}
\label{app:limber_snr}

As we discussed in \App{limber}, a core simplifying assumption of the Limber approximation is that information from line-of-sight correlations contributes negligibly to projected angular spectra.
While this simplifying assumption is reasonable for many current observables of interest, it is not sufficient when considering line intensity maps which have been foreground-filtered.
As we discussed in \Sec{lim_filter} and our toy model, \Sec{toy}, foreground filtering induces line-of-sight correlations which become increasingly strong as additional filtering is applied.
Whether it is still justified to use the Limber approximation and discard line-of-sight correlations after filtering has induced line-of-sight correlations is the question we answer in this appendix.

We see in \Fig{limber-SNR} that ignoring these line-of-sight correlations in the Limber approximation {underestimates} the \SNR{} by more than a factor of $2$ in the most extreme case where a large amount of filtering is applied.
{Here we use the noise-dominated approximation for the covariance (\Eq{noise_dom}) because (1) we found it to be a good approximation at the level of \SNR{} for current-era experiments (\App{noise_dom}) (2) the instrumental noise contribution to the covariance is the same in the Limber and our approximation (\App{comp_ee}), and (3) the Limber approximation still retains the features described in \App{II} which made the LIM spectrum covariance extremely expensive to compute.
This additionally allows us to highlight the effect of the Limber approximation on the signal.}

The origin of the trends seen in \Fig{limber-SNR} can be understood in more detail with \Fig{limber-dsnr} where we find that the contribution to \SNR{} from small angular scales (high-$\ell$) are the first to deviate as more filtering is applied when comparing the Limber approximation and our approximation.
We {speculate} the origin of this is from the fact that, in the unfiltered case, the cosmological signal on small angular scales possess very little line-of-sight correlations (see e.g. \Fig{corr_II}), a key reason why the Limber approximation is so successful in the high-$\ell$ regime.
Thus, the impact of foreground filtering induced line-of-sight correlations on the \SNR{} contribution of small angular scales is felt immediately.
Our approximation retains information from line-of-sight correlations while the Limber approximation does not.
On the other hand, large angular scales do possess some intrinsic line-of-sight correlations, a key reason why the Limber approximation struggles in the low-$\ell$ regime, which initially dominate over the line-of-sight correlations induced by foreground filtering.
Thus, the effects of filtering on the \SNR{} contribution from large angular scales cannot be seen until filtering induced line-of-sight correlations grow larger than intrinsic line-of-sight correlations.
{We defer a full intuitive understanding of the trends seen in \Fig{limber-dsnr} to future work.}

Because the majority of LIM experiments in this era will be most sensitive to small angular scales which are (counterintuitively) most susceptible to the pathologies of the Limber approximation applied to foreground-filtered line intensity maps, we do not recommend the Limber approximation.

{To summarize: because 
\begin{enumerate}
    \item small angular scales are most susceptible to the pathologies of the Limber approximation applied to foreground-filtered line intensity maps,
    \item majority of LIM experiments in this era will be most sensitive to small angular scales, and 
    \item the LIM spectrum covariance remains extremely expensive to compute (\App{II}) even in the Limber approximation,
\end{enumerate}
we \textbf{do not recommend} the Limber approximation for exploring foreground-filtered LIM experiments.}

\section{Subtlety in analytical modeling of low-pass filter\label{app:lo}}
\begin{figure}
    \centering
    \includegraphics{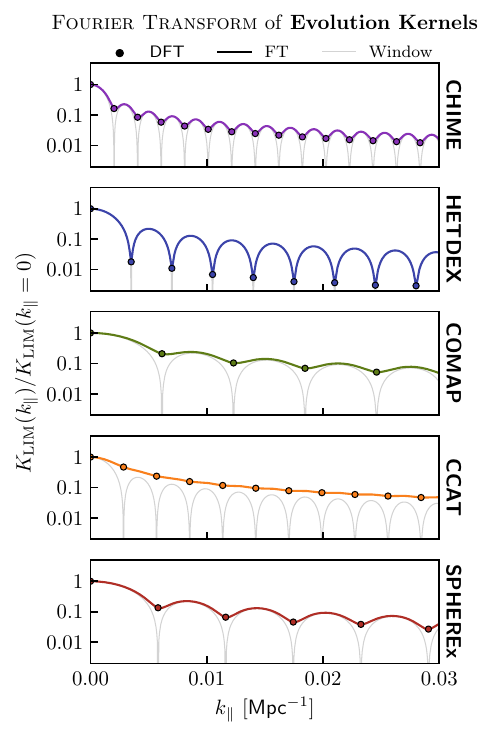}
    \caption{The Fourier transform of the evolution kernel from line emission evolution and growth of structure (colored dots) is modulated by the window function (grey and colored line) except at the modes accessible in a finite volume data cube, integer multiples of the fundamental $2\pi/ \L$. Because of this, to self consistently model the low-pass filtering without introducing corruption due to the window function $\WIN(\chi)$, we must use the discrete Fourier transform form of the real-space low pass filter \Eq{LOx} as opposed to the continuous Fourier transform form (\Eq{LoOld}) as discussed in \App{lo}.}
    \label{fig:kernel}
\end{figure}
\begin{figure}
    \centering
    \includegraphics{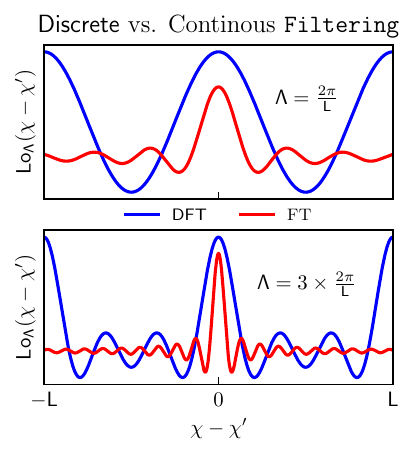}
    \caption{A comparison of the discrete (\Eq{LOx}, blue, used in this work) and continuous (\Eq{LoOld}, red) Fourier transform form of the real-space low pass filter discussed in \App{lo} for different amounts of filtering.}
    \label{fig:DFTvFT}
\end{figure}
There is a subtlety in analytically modeling low-pass filtering of the LIM (\Sec{lim_filter}).
In particular, the naive thing one might do is let $\LO(\chi)$ in \Eq{LOx} be the inverse continous Fourier transform of the low-pass filter:
\begin{align}
\nonumber\textsc{Continuous Fourier Transform}\\
    \LO(\chi) = \int_{-\Lambda}^{\Lambda} \frac{d\kpar}{2\pi} e^{i\kpar \chi}=\frac\Lambda \pi \sinc(\Lambda\chi).
    \label{eq:LoOld}
\end{align}
However this then causes inconsistencies in the modeling of the LIM signal.
In particular this implicitly introduces an additional un-physical source of lightcone evolution, the window function $\WIN(\chi)$, which in some cases is subdominant to the mode coupling due to the evolution of line luminosity (e.g. for {\sf CCAT} and {\sf COMAP}) while for others dominates over the mode coupling due to luminosity evolution (e.g. for {\sf HETDEX}).
This can be seen explicitly in \Fig{kernel} where the Fourier transform of th kernel, which dictates the spread of long-wavelength modes of the linear matter field $\delta_m^0$ into short-wavelength modes of the LIM (see \Sec{toy}), is modulated by the window function (grey). 
However, the Fourier transform at modes accessible in the finite box, integer multiples of the fundamental $2\pi/\L$, must be the same when either the continuous or discrete Fourier transform is used. 
These modes are the physically relevant ones since they are ones accessible in analyses of a finite volume data cube and correspond purely to the line luminosity evolution, un-corrupted by additional power from the LOS window function $\WIN(\chi)$. 
Because of this, to self consistently model the low-pass filtering of the LIM, we must use the discrete Fourier transform of the low pass filter implemented in \Eq{LOx}.
We visually show the difference between these two implementations of the low-pass filter in \Fig{DFTvFT}.

In a previous preprint version of this work, the \SNR's reported in e.g. \Fig{SNR-SO} differ minorly (e.g. for {\sf COMAP} and {\sf CCAT}) to, in the worst case, an order of magnitude (e.g. {\sf HETDEX}).
These results were primarily corrupted by the use of the continuous Fourier transform (\Eq{LoOld}) when analytically modeling low-pass filtering.
This then mixed the unphysical ``lightcone evolution" due to the LOS window function $\WIN(\chi)$ with the physical evolution of line emission and growth of structure along the LOS, significantly modifying the reported \SNR{} if the physical evolution was subdominant to the unphysical evolution (\Fig{kernel}).

\section{Numerical evaluation of the flat-sky angular spectra}
\label{app:explicit}

In this appendix we describe how we numerically evaluate the auto and cross-spectra for our models of CMB lensing and foreground-filtered line intensity maps summarized in \Tab{unify} with the approximation of the flat-sky angular spectrum described in \Sec{spectra}.
The structure of this appendix can be summarized in relation to \Eq{gen_cross}:
\begin{equation}
\langle \phi\psi \rangle' \approx 
    \overbracket{\int_{\chib} 2\chib
    \underbracket{\vphantom{\int_\kparp}
    \int_\delta
    \frac{K_\phi K_\psi}{\chib^2}
    }_{\textrm{\App{ddelta}}} \underbracket{\int_\kparp e^{i\kparp\times 2\chib\delta}P^{\Phi\Psi}(k)
    }_{\textrm{\App{dkparp}}}}
    ^{\textrm{Apps.~\ref{app:KK}, \ref{app:Ik}, \ref{app:comp_ee}, and \ref{app:II}}}.
\end{equation}
We also publicly release all code used to numerically evaluate flat-sky angular spectra computed in this paper at \href{https://github.com/DelonShen/LIMxCMBL/blob/main/README.md}{\texttt{github.com/DelonShen/LIMxCMBL}} (\Reff{github}).

\subsection{Angular spectrum of line-of-sight correlation functions}
\label{app:dkparp}

\begin{figure*}
    \centering
    \includegraphics{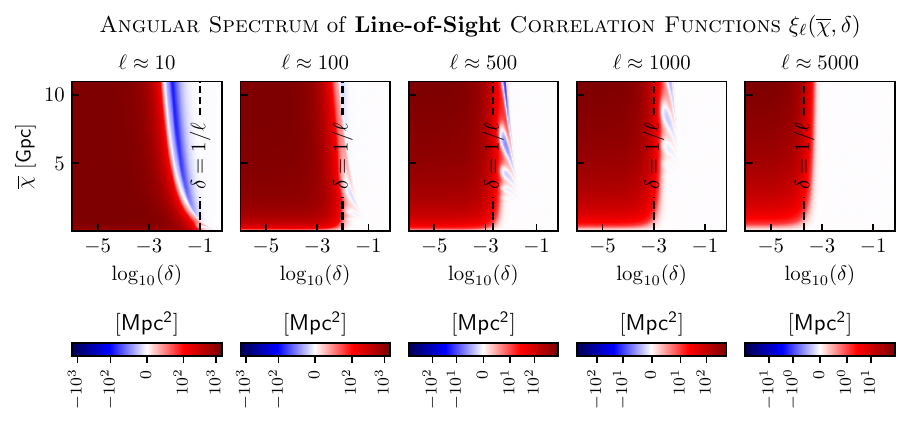}
    \caption{
    Our tabulation of the angular spectrum of line-of-sight correlation functions $\I_\ell(\chib,\delta)$ (\Eq{dkparp}) whose computation we describe in \App{dkparp}. 
    This is then used to compute angular spectra which we describe for generic tracers in \Sec{spectra} (specifically \Eq{gen_cross}) and detail numerical implementation for the relevant LIM and CMB lensing spectra in \App{explicit}.
    }
    \label{fig:dkparp}
\end{figure*}

In \Eq{gen_cross}, the innermost integral yields a line-of-sight correlation function $\I_\ell(\chib,\delta)$ which we defined as
\begin{equation}
    \I_\ell(\chib,\delta) = \int_\kparp e^{i\kparp\times 2\chib\delta}P^{\Phi\Psi}\left(k^2={ \kparp^2 + \frac{\ell(\ell+1)}{\bar\chi^2(1-\delta^2)}} \right).
\end{equation}
Since this inner-most integral depends only on cosmology and in this paper we consider a fixed cosmology, we only must tabulate this oscillatory integral once as a function of $(\ell,\chib,\delta\chi)$.
To evaluate the inner integral we utilize the $\kparp\rightarrow-\kparp$ symmetry of the real part of the integral and anti-symmetry of the imaginary part of the integral.
So $\I_\ell(\chib,\delta)$ can be simplified:
\begin{equation}
    \I_\ell(\chib,\delta) = 2\int_0^\infty d\kparp\times \cos(\kparp \times 2 \chib \delta)P^{\Phi\Psi}\left(k \right).
    \label{eq:dkparp}
\end{equation}
This integral can then be evaluated with adaptive quadrature where we set our relative error tolerance to $10^{-4}$.
We tabulate this integral within the region defined by
\begin{align}
\nonumber
    &\{10\leq \ell\leq 5000\}\\
\nonumber
    \times
    &\{10 \leq \chib \leq \chi(z=20) \}\\
    \times
    &\{10^{-6}\leq \delta \leq 0.7\},\label{eq:sampling}
\end{align}
and found that sampling $\ell$ uniformly in log space at 100 points, $\chib$ uniformly at $2^8$ points, and $\delta$ uniformly in log space at $2^7$ points was sufficiently dense in order for linear interpolation of our tabulation to not cause appreciable numerical artifacts for our study.
It should be noted that the evaluation of $\I_\ell(\chib,\delta)$ may be done more efficiently with the \texttt{FFTLog} algorithm as described in \Reff{Gao:2023tcd} but we found this additional efficiency unneeded for our study so we did not utilize this fact.
In \Fig{dkparp} we show our resulting $\I_\ell(\chib,\delta)$ used throughout this paper to compute angular spectra.

\subsection{Integral over line-of-sight correlations}
\label{app:ddelta}

In \Eq{gen_cross}, the bounds of the integral over line-of-sight correlations ($\int d\delta$) go from $-1$ to $1$. 
However, because $\delta$ appears somewhat symmetrically in the kernels $K_\phi,K_\psi$, we only need to integrate in the region $0<\delta < 1$.
Let $\uparrow$ denote $\chib(1+\delta)$ and $\downarrow$ denotes $\chib(1-\delta)$.
We can rewrite the $\int d\delta$ integral as
\begin{align}
\nonumber
    \int_{-1}^1 &d\delta\times \frac{K_\phi(\downarrow)K_\psi(\uparrow)}{\chib^2} \times \I_\ell(\chib,\delta)\\
    =     \int_{0}^1 & d\delta\times \frac{K_\phi(\downarrow)K_\psi(\uparrow)+K_\phi(\uparrow)K_\psi(\downarrow)}{\chib^2} \times \I_\ell(\chib,\delta).
\end{align}
Because $\int d\delta =\int d\ln\delta\times \delta$, small $\delta$ contributions vanish if $K_\phi,K_\psi$, and $\I_\ell(\chib,\delta)$ do not strongly vary strongly between $\delta=0$ and $\delta = 10^{-6}$.
We verify this is the case for our kernels for interest.
Furthermore, contributions from $\delta>0.7$ can safely be neglected since these values of $\delta$ physically correspond to very long range line-of-sight correlations which physically must be highly suppressed.
So, in practice, all of our $d\delta$ integrals are evaluated numerically in $\ln \delta$ space for $10^{-6} < \delta < 0.7$ with the trapezoidal rule using the samples described around \Eq{sampling}.

\begin{figure}
    \centering
    \includegraphics{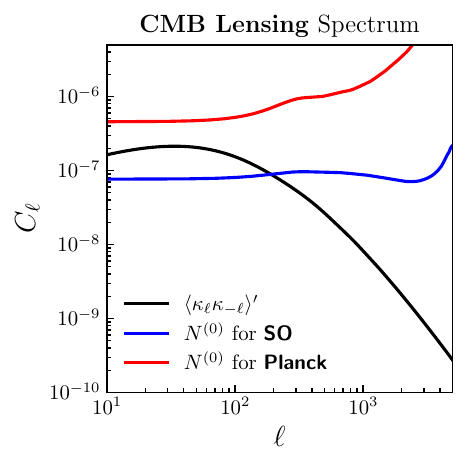}
    \caption{
    The CMB lensing spectrum whose computation we describe in \App{KK} as well as the expected $N^{(0)}$ noise bias for \textsf{Simons Observatory} (\Reff{SimonsObservatory:2018koc}) and \textsf{Planck} (\Reff{Planck:2018lbu}) which are computed with \texttt{LensQuEst} (\Reff{Schaan:2018tup}). Our fiducial forecasts in \Sec{SNR} utilize the \textsf{SO} $N^{(0)}$ curves while the \textsf{Planck} $N^{(0)}$ is utilized in \Fig{SNR_N0}.}
    \label{fig:ClKK}
\end{figure}
\subsection{CMB lensing spectrum}
\label{app:KK}

The most straightforward spectrum to compute for our study is the CMB lensing auto spectrum:
\begin{equation}
\langle \kappa_\vl \kappa_{-\vl} \rangle' = 
    \int_{\chib} \frac 2\chib\int_\delta
    {\Kk(\downarrow)\Kk(\uparrow)}
    \I_\ell(\chib,\delta).
\end{equation}
Thanks to the absence of any strong oscillatory features in the CMB lensing kernel $K_\kappa$, we evaluate this integral with the trapezoid rule using the $\chib$ and $\delta$ sampling described in \Eq{sampling}\footnote{Note that the sampling we described for $\chib$ in \Eq{sampling} does not go all the way to $\chi(z=z_{\rm source})$ while the CMB lensing kernel \Eq{Wk}  has support all the way to $\chi(z_{\rm source})$ meaning our computation of the CMB lensing spectrum is missing some high redshift contribution. In practice, our mistaken choice leads to a $<$10\% underestimate at high $\ell$. Crucially this ends up negligibly affecting our results due to the high $\ell$ errors being washed out by the dominant $N^{(0)}$ contribution (\Fig{ClKK}) when included in the covariance (\Eq{cov}) for our detectability calculations. This would affect the ``noiseless" curve of \Fig{SNR_N0} so for the ``noiseless" case specifically, we use the CMB lensing spectrum computed with the Limber approximation which is accurate at high-$\ell$. Additionally, the choice of sampling in \Eq{sampling} does not affect any other spectra because other kernels have compact support due to experimental window functions combined with $\I_\ell(\chib,\delta)$ suppressing large-$\delta$ contributions.}.
This convenient feature of the CMB lensing spectrum integral which allows us to use efficient methods and fixed sampling does not appear again for other angular spectra of interest which we will describe shortly.
In \Fig{ClKK} we plot our resulting CMB lensing spectrum as well as the expected $N^{(0)}$ bias, used as $\N_\ell^\kappa$ (\Eq{NK}) in the covariance \Eq{cov}, for the \textsf{Simons Observatory} (\Reff{SimonsObservatory:2018koc}) and \textsf{Planck} (\Reff{Planck:2018lbu}) which are computed with \texttt{LensQuEst} (\Reff{Schaan:2018tup}).

\subsection{Foreground-filtered LIM$\times$CMB lensing cross-spectrum}
\label{app:Ik}

The CMB lensing and line-intensity mapping cross-spectrum (\Eq{obs}, visualized in \Fig{obs}) can be decompose into two components: (1) the CMB lensing and unfiltered LIM cross-spectra and (2) the CMB lensing and low-pass filtered LIM cross spectra.
The computation of (1) can be simplified by inserting the Dirac deltas present in the unfiltered LIM kernel (\Eq{WI}):
\begin{align}
    \langle I_{\ell}(\chi) \kappa_{-\ell}\rangle' 
    &=2 \KL(\chi) \int_{\chib}\frac{\Kk(2\chib - \chi)}{\chib^2} \I_\ell\left(\chib, 1 - \frac \chi {\chib}\right).
\end{align}
The computation of (2) cannot be simplified and must be directly evaluated
\begin{equation}
   {
   \langle I_\ell^\LO(\chi) \kappa_{-\ell} \rangle' =
    \int_{\chib} \frac 2\chib\int_\delta 
    {\KLo(\downarrow;\chi)\Kk(\uparrow)}
    \I_\ell(\chib,\delta).
    }
\end{equation}
When computing $\langle I^\HI  \kappa \rangle $ we take the difference $\langle I\kappa \rangle - \langle I^\LO\kappa\rangle$ at the level of integrands, to avoid unnecessarily losing numerical accuracy. 
In other words, $\langle I^\HI\kappa\rangle$ is computed by evaluating
\begin{equation}
    \langle I^\HI\kappa\rangle\sim \int_{\chib}\int_\delta [\KL- \KLo]\Kk.
\end{equation}
We evaluate $d\delta$ integral as described in \App{ddelta}.
The $\chib$ and binning in $\chi$ integrals are computed with nested adaptive quadrature where we set our relative error tolerance to $10^{-3}$.
Contrary to the computation of $\langle\kappa\kappa\rangle$ described in \App{KK}, the use of adaptive quadrature for these integrals is preferred because $\KLo$ is an oscillatory kernel, especially oscillatory for aggressive filtering of the line intensity map.

\begin{figure*}
    \centering
    \includegraphics{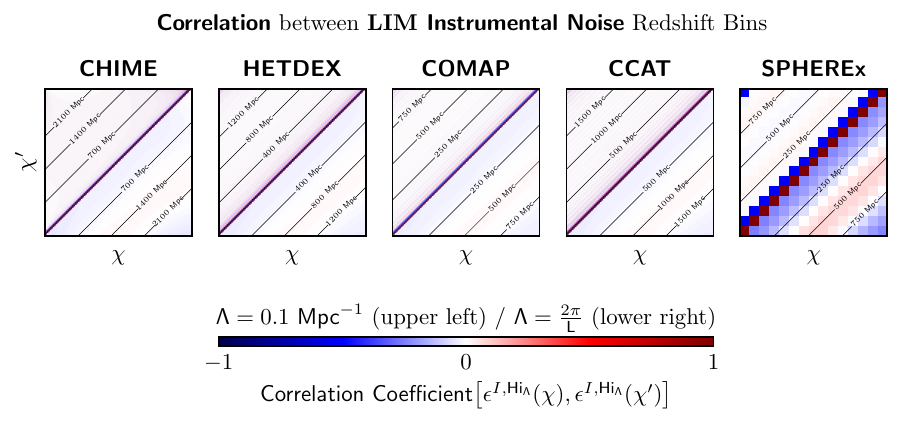}
    \caption{
    The magnitude of ``off-diagonal" correlation between redshift bins of the LIM instrumental noise (whose computation we describe in \App{comp_ee}) increases as more foreground filtering is applied. 
    Any filtering at all induces off-diagonal correlations that are not present in the unfiltered LIM instrumental noise map which is uncorrelated for $\chi\ne \chip$ (\Eq{eIeI}).
    Diagonal black lines correspond to contours of constant $|\chi-\chip|$.
    We plot the variance of LIM instrumental noise in \Fig{var_eIeI}
    }
    \label{fig:corr_eIeI}
\end{figure*}
\begin{figure*}
    \centering
    \includegraphics{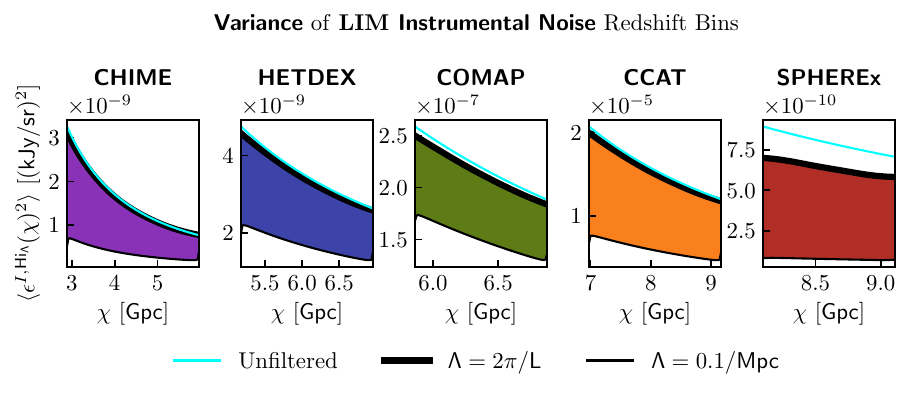}
    \caption{The variance of LIM instrumental noise in redshift bins (whose computation we describe in \App{comp_ee}) decreases in relation to the unfiltered (cyan) noise map as we go from less foreground filtering (thick black) to more foreground filtering (thin black).
    We plot correlation between different redshift bins of the LIM instrumental noise in \Fig{corr_eIeI}. 
    }
    \label{fig:var_eIeI}
\end{figure*}

\subsection{Instrumental noise contribution to foreground-filtered LIM spectrum covariance}
\label{app:comp_ee}

Since we model LIM instrumental noise (\Sec{lim_noise}) as Gaussian white noise, the instrumental noise contribution to the foreground-filtered LIM spectrum covariance for any amount of filtering can be evaluated analytically.
First consider the inner $\int d\kparp$ integral of \Eq{gen_cross}.
We described our evaluation of this integral for a cosmological spectrum in \App{dkparp}.
However, for instrumental noise we approximate the spectrum as constant. 
This then reduces the $\int d\kparp$ integral to a Dirac delta:
\begin{equation}
 \int \frac{d\kparp}{2\pi}  e^{i\kparp \times 2\chib \delta} P^{\epsilon_I} = \frac{P^{\epsilon_I}}{2\chib}\times \delta^{(D)}(\delta)\label{eq:dkparpeI}
\end{equation}
Note that this is essentially the same as the Limber approximation (\App{limber}).
We can then use this result to analytically compute the unfiltered LIM instrumental noise spectrum:
\begin{align}
\nonumber\langle \epsilon^{I}(\chi) \epsilon^{I}(\chip)\rangle' &=  P^{\epsilon_I}\int_0^\infty d\chib\frac{K_{\rm LIM}^\epsilon(\chib;\chi)K_{\rm LIM}^\epsilon(\chib;\chi')}{\chib^2}\\
&= \delta^{(D)}(\chi-\chi')\times \frac{P^{\epsilon,I}}{\chi^2}\times \WIN(\chi).
\label{eq:eIeI}
\end{align}
In the discrete case where we bin $\langle \epsilon^I \epsilon^I\rangle$ in $\chi,\chip$ with bin width $\Delta\chi$, {the unfiltered instrumental noise covariance becomes (in the shorthand notation introduced in \Sec{SNR})}
\begin{align}
{\langle \epsilon_\i^I\epsilon_\j^I\rangle =\delta^{(K)}_{\i\j} \times \frac{P^{\epsilon_I}}{(\delta\chi)^2}\left(\frac 1 {\chi_\i^-} - \frac 1 {\chi_\i^+} \right).}
\end{align}
Since we model the high-pass filtered LIM instrumental noise map as the difference of the unfiltered and low-pass filtered instrumental noise map, the high-pass filtered LIM instrumental noise spectrum includes both spectra of the unfiltered and low-pass filtered instrumental noise maps as well as their cross spectra:
\begin{equation}
    \langle \epsilon^{I,\HI} \epsilon^{I,\HI}\rangle  = \langle \epsilon^I \epsilon^I \rangle + \langle \epsilon^{I,\LO}\epsilon^{I,\LO} \rangle - \langle \epsilon^I \epsilon^{I,\LO}\rangle - \langle \epsilon^{I,\LO}\epsilon^I \rangle.\label{eq:ee_quad}
\end{equation}
Because of the $\delta^{(D)}(\delta)$ from \Eq{dkparpeI} and the $\delta^{(D)}$ in $K^\LO_{\epsilon,\rm LIM}$, the cross terms can also be evaluated analytically:
\begin{equation}
\langle \epsilon^{I}_\ell(\chi) \epsilon^{I,\LO}_{-\ell}(\chip)\rangle' =  \frac{P^{\epsilon,I}}{\chi^2}\times  \LO(\chi-\chip)\times \WIN(\chi).
\end{equation}
The low-pass filtered LIM instrumental noise spectrum requires the most work.
Applying our angular spectrum \Eq{gen_cross} to the low-pass filtered LIM noise kernel \Eq{KeLo} yields
\begin{align}
    \nonumber \langle \epsilon^{I,\LO}_{\ell}(\chi) \epsilon^{I,\LO}_{-\ell}(\chip)\rangle' =   P^{\epsilon_I}\int_{\chib}\frac{K_{\rm LIM}^{\epsilon,\LO}(\chib;\chi)K_{\rm LIM}^{\epsilon,\LO}(\chib;\chi')}{\chib^2}\\
    =  P^{\epsilon_I}\int_{\chimin}^{\chimax} d\chib\frac{{\LO(\chib-\chi)\LO(\chib-\chip)}}{\chib^2}.
    \label{eq:eLOieLOi_interm}
\end{align}
{Instead of evaluating the integral with the summed form (\Eq{LOx}), we take the product of the pre-summed version \Eq{LOx_sum} and evaluate the integral for each summand individually} in terms of efficiently computable sine- and cosine-integrals:
\begin{equation}
    {\rm Si}(x) = \int_0^x\frac{\sin t}{t}dt\qquad \textrm{and}\qquad {\rm Ci}(x) = \int_0^x\frac{\cos t}{t}dt.
\end{equation}
While it is possible to write down $\langle \epsilon^{I,\LO}\epsilon^{I,\LO}\rangle$ in terms of these special integrals, the explicit form is long and cumbersome.
We publicly share the Mathematica notebook which computes the explicit form of $\langle \epsilon^{I,\LO}\epsilon^{I,\LO}\rangle$ in terms of ${\rm Si}(x)$ and ${\rm Ci}(x)$ with the rest of our code in \Reff{github}.
Furthermore, we verify that this analytical form of the low-pass filtered LIM instrumental noise spectrum matches numerically evaluating the integral \Eq{eLOieLOi_interm} with adaptive quadrature where relative error tolerance is set to $10^{-3}$.
To bin the LIM noise spectrum in $\chi,\chip$, we once again use adaptive quadrature.
We visualize the resulting LIM noise spectrum in two figures:
\begin{enumerate}
    \item In \Fig{corr_eIeI} we show the \textit{correlation coefficient} $\rho$ between two redshift bins of the instrumental noise map:
    \begin{equation}
    \rho[\epsilon^{I,\HI}_{\tt i}, \epsilon^{I,\HI}_{\tt j}] = \frac{\langle \epsilon^{I,\HI}_{\tt i} \epsilon^{I,\HI}_{\tt j}\rangle}{\sqrt{\langle (\epsilon^{I,\HI}_{\tt i})^2\rangle \langle (\epsilon^{I,\HI}_{\tt j})^2\rangle}},
    \end{equation}
    when the noise map is highly filtered ($\Lambda=0.1\ {\sf Mpc}^{-1}$, upper left) and when the noise map is mildly filtered ($\Lambda=2\pi/\L$, lower right).
    Note that the instrumental noise map correlation matrix is diagonal if its unfiltered (\Eq{eIeI}).
    \item In \Fig{var_eIeI} we compare the variance of a instrumental noise redshift bin when the noise map is filtered (black) and when the noise map remains unfiltered (cyan).
\end{enumerate}

\begin{figure*}
    \centering
    \includegraphics{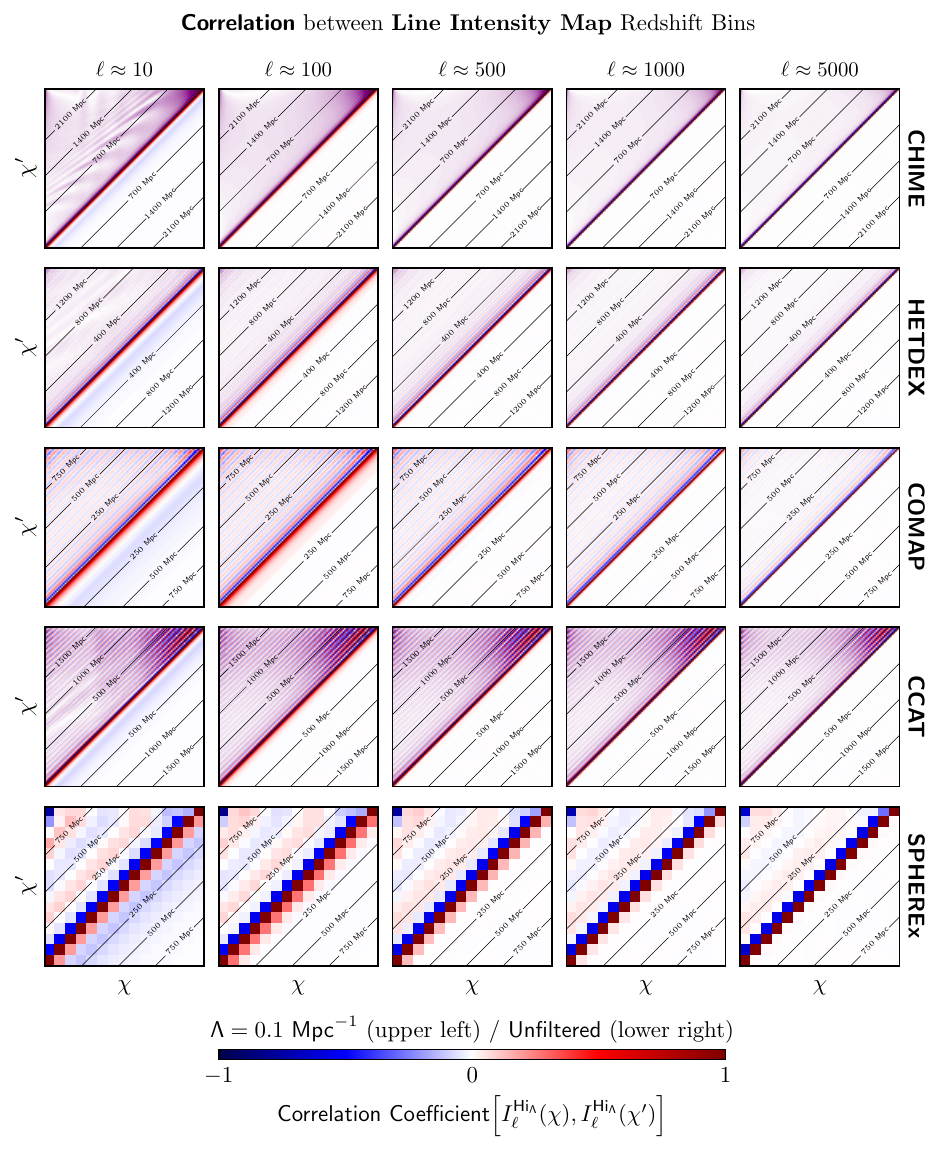}
    \caption{
    The magnitude of ``off-diagonal" correlation between redshift bins of the line intensity map (whose computation we describe in \App{II}) increases in relation to the unfiltered LIM (lower right) once foreground filtering is applied (upper left).
    The magnitude of off-diagonal correlations reduce as we consider smaller angular scales ($\ell$ increases).
    Diagonal black lines correspond to contours of constant $|\chi-\chip|$.
    We plot the variance of line intensity map slices in \Fig{var_II}
    }
    \label{fig:corr_II}
\end{figure*}
\begin{figure*}
    \centering
    \includegraphics{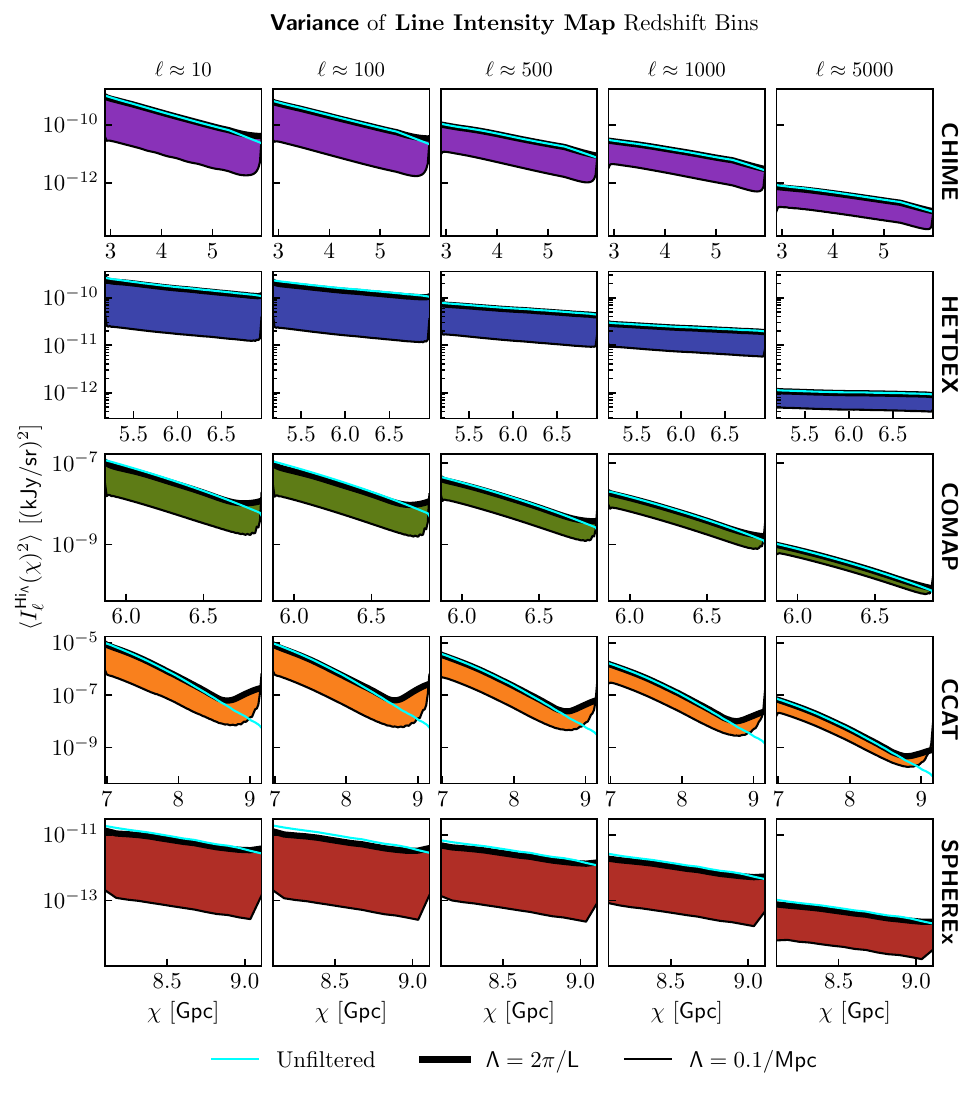}
    \caption{The variance of line intensity map redshift bins (whose computation we describe in \App{II}) 
    decreases in relation to the unfiltered (cyan) line intensity map as we go from less foreground filtering (thick black) to more foreground filtering (thin black).
    This difference in the variance decreases as we consider smaller angular scales ($\ell$ increases).
    We plot correlation between redshift bins of the line intensity map in \Fig{corr_II}.
    }
    \label{fig:var_II}
\end{figure*}

\subsection{Cosmic variance contribution to the foreground-filtered LIM spectrum covariance}
\label{app:II}

Of all spectra relevant for our work, the cosmic variance term in the foreground-filtered LIM spectrum covariance requires the most care and resources to compute.
In essence this is because of the $\LO$ functions present in $\KLo$ become increasingly oscillatory as filtering becomes more aggressive and we have two highly oscillatory $\LO$ coupled to eachother when computing the high-pass filtered LIM spectrum.
Though we compute the filtered LIM spectrum for all of our calculations in \Sec{SNR}, one does not necessarily need to compute this for near-term LIM experiments.
Indeed, many near-term LIM experiments are path finders where noise is expected to be large in comparison to the cosmological signal (\Fig{lim-model}).
So approximating the LIM $\times$ CMB lensing spectrum covariance \Eq{cov} as noise dominated is a valid approximation and yields reasonable results.
This is especially important since the computation of the filtered LIM instrumental noise term (\App{comp_ee}) can be done relatively efficiently wheras the computation of the cosmic variance term is extremely delicate and thus requires orders of magnitude more resources as we will describe shortly.
We expanded more on the validity of the noise dominated approximation in \App{noise_dom}. 

The unfiltered LIM spectrum can be simplified by using the $\delta^{(D)}$ present in $\KL(\chip;\chi)$ (\Eq{WI}):
\begin{align}
\nonumber
    \langle I_\ell(\chi) I_{-\ell}(\chip)\rangle' = \frac {\KL(\chi)\KL(\chip)} {\left( \left({\chi+\chip}\right)/2\right)^2} & \\
    \times \I_\ell\biggr(\chib = \frac{\chi+\chip}2, \delta = \frac{\chip-\chi}{\chi+\chip}& \biggr)
    .
\end{align}
Similar to the high-pass filtered LIM instrumental noise spectrum \Eq{ee_quad}, the high-pass filtered LIM spectrum includes both spectra of the unfiltered and low-pass filtered LIM as well as their cross spectra:
\begin{equation}
    \langle I^\HI I^\HI\rangle = \langle II\rangle + \langle I^\LO I^\LO\rangle - \langle II^\LO\rangle - \langle I^\LO I\rangle.\label{eq:comp_II}
\end{equation}
The cross-spectra simplifies slightly due to the $\delta^{(D)}$ in $\KL$ like for the unifiltered spectrum:
\begin{align}
\nonumber
    \langle I_\ell(\chi) &I_{-\ell}^\LO(\chip) \rangle' = \\
    &2\KL(\chi)\int_{\chib} \frac{\KLo(2\chib - \chi; \chip)}{\chib^2}
    \I_\ell\left(\chib, \frac {\chib-\chi} {\chib}\right).
\end{align}
The low-pass filtered LIM spectrum must be computed in full using the kernels $\KLo$ (\Eq{WLIMLo}) and \Eq{gen_cross}.
As described in \App{ddelta}, we are able to evaluate the $\int d\delta$ integral with the trapezoid rule even for cases of highly filtered intensity maps since it is possible to sample $\delta$ sufficiently densely.
However, the $\int d\chib$ integral as well as the additional two nested integral to bin $\langle I^\HI I^\HI\rangle$ in $\chi,\chip$ {(\Eq{fij})} must be evaluated with extreme care.
While the naive approach of using nested adaptive quadrature to compute the binned high-pass filtered LIM spectrum would in principle work, the computational cost grows exponentially with the number of dimensions to be integrated over.
We found that this curse of dimensionality renders evaluation of the filtered LIM spectrum intractable given the computational resources available to us.
Instead, we use quasi-Monte Carlo (QMC) integration, reviewed in detail by \Reff{practicalqmc}, to compute the filtered LIM spectrum.
The terms \Eq{comp_II} are combined at the level of integrands as opposed to integrating each term independently and combining the results.
The $\int_0^{\infty} d\chib$ is split up into 65 sub-intervals.
The first interval goes from $10\leq \chi \leq 0.9\times \chimin$ where $\chimin$ is the minimum $\chi$ a given experiment probes a given line. 
The edges of the next 63 sub-intervals are uniformly spaced between $0.9\times \chimin$ and $1.1\times \chimax$ where $\chimax$ is the maximum $\chi$ a given experiment probes a given line. 
The final sub-interval goes from $1.1\times\chimax$ to $\chi(z=20)$.
Each $\chib$ sub-interval is treated as an independent integral and evaluated independently.
To summarize, the rough form of our integral to compute the LIM spectrum for the $\i^{\rm th}$ $\chi$ bin and $\j^{\rm th}$ $\chip$ bin with bin width $\delta \chi$ is (in the shorthand notation introduced in \Sec{SNR})
\begin{align}
\nonumber
    \langle I^\HI_\i I^\HI_\j\rangle \sim {\frac 1 {(\delta \chi)^2}\int_{[\chi^\pm_\i]} \int_{[\chi^\pm_\j]}}\biggr[&\int_{10}^{0.9\chimin }+\int_{0.9\chimin}^{...} \\
    + \dots + \int_{...}^{1.1\chimax} +&\int_{1.1\chimax}^{\chi(z=20)}\biggr] d\chib.
\end{align}
We estimate $\langle I^\HI_{\i}I^\HI_{\j}\rangle$ only for bins $\j \geq \i$ due to the $\i\leftrightarrow\j$ symmetry.
Each $[\int d\chi \int d\chip \int d\chib]$ is estimated with $2^{17}$ points drawn from a Halton sequence.
This is repeated $2^3$ times to get a sense of the error of the integral estimate.
In particular, we check that the standard deviation within these $2^3$ repeated evaluations is of order .1\% of the actual estimate of the integral in this bin.
We then estimate the high-pass filtered LIM spectrum in each bin as the average of these $2^3$ estimates.

Due to minuscule numerical errors combined with the highly oscillatory nature of the covariance due to foreground filtering, our estimate for the high-pass filtered LIM spectrum is not a positive semi-definite matrix to within numerical precision.
Because we broadly trust our method for computing the spectrum, we assume the matrix we have computed $\langle I^\HI_{\sf i} I^\HI_{\sf j}\rangle$ is nearly the correct matrix and the lack of positive semi-definiteness is due purely to minor numerical issues and not a symptom of a more fundamental issue.
Thus we define our final estimate for $\langle I^\HI_{\sf i} I^\HI_{\sf j}\rangle$ as the positive semi-definite symmetric matrix which is closest to our numerical estimate with QMC.
In particular, we spectrally decompose our QMC estimated high-pass filtered LIM spectrum covariance with its eigenvalues $\lambda_i$ and a matrix $\sf Q$ whose $i$th column is our estimated matrix's $i$th eigenvector:
\begin{equation}
(\textrm{QMC esimate for }{\langle I^\HI_{\sf i} I^\HI_{\sf j}\rangle})= {\sf Q}_{\sf ia}\sf \lambda_a {\sf Q}_{aj}.
\end{equation}
The positive semi-definite symmetric matrix which is closest to 
this matrix under the Frobenius norm is then given by (\Reff{doi:10.1137/S0895479896302898})
\begin{align}
(\textrm{Final estimate for }\langle I^\HI_{\sf i} I^\HI_{\sf j}\rangle) = {\sf Q}_{ia} \tau_a {\sf Q}_{aj}&\\
\textrm{where} \quad \tau_a = \begin{cases}
        \lambda_a&\lambda_a \geq 0\\
        0&\lambda_a < 0.
    \end{cases}&
\end{align}
Put more simply, we zero out negative eigenvalues of our QMC estimated high-pass filtered LIM spectrum.
In practice this leads to minuscule changes numerically to the quasi-Monte Carlo estimated high-pass filtered LIM spectrum.
This final estimate is then used for all calculations carried out in \Sec{SNR}.
We visualize the resulting LIM spectrum in two figures:
\begin{enumerate}
    \item In \Fig{corr_II} we show the \textit{correlation coefficient} $\rho$ between two redshift bins of the LIM:
    \begin{equation}
    \rho[I^\HI_{\tt i}, I^\HI_{\tt j}] = \frac{\langle I^\HI_{\tt i} I^\HI_{\tt j}\rangle}{\sqrt{\langle (I^\HI_{\tt i})^2\rangle \langle (I^\HI_{\tt j})^2\rangle}},
    \end{equation}
    when the LIM is highly filtered ($\Lambda=0.1\ {\sf Mpc}^{-1}$, upper left) and when the LIM remains unfiltered (lower right) at several values of $\ell$.
    \item In \Fig{var_II} we compare the variance of a LIM redshift bin when the LIM is filtered (black) and when the LIM remains unfiltered (cyan) at several values of $\ell$.
\end{enumerate}

Despite the efficiency and favorable convergence rate of quasi-Monte Carlo integration, this integral would still take $O(\textrm{months})$ per experiment to compute on a CPU.
We bring this integral into the realm of something computable on a reasonable timescale by implementing the integrands on a GPU with \texttt{JAX} (\Reff{jax2018github}).
Doing this allows us to utilize a combination of Tesla A100's and NVIDIA L40S's to estimate the binned and filtered LIM spectrum in $O(\textrm{days})$ per experiment.
In total we required $O(10,000)$ GPU hours to compute the cosmic variance $\langle I^\HI I^\HI\rangle$ contribution to the covariance \Eq{cov} for all experiments/filtering of interest that is then used in our SNR calculations described in \Sec{SNR}.
Given the extreme computational cost of computing this spectrum, we would again like to emphasize that for LIM experiments whose 3-D power spectrum is instrumental noise dominated, approximating the covariance \Eq{cov} as noise dominated still yields fine results as we show in \App{noise_dom}.

\subsection{Numerical evaluation of the \SNR\label{app:SNR}}
Additional care must be taken when evaluating the \SNR{} as described in \Sec{SNR} due to the singular value structure of the covariance matrix.
In particular, while the covariance matrix is well behaved when unfiltered, it has a single null\footnote{
While analytically it may be intractable to see this in the full covariance, a practical limiting case where this can be seen analytically is for the noise-dominated approximation \Eq{noise_dom}.} (0) eigenvalue for $\m\geq 0$.
In other words, the rank of the covariance is equal to $n_{\rm bins}-1$.
Numerically this single null eigenvalue is exponentially difficult to realize in practice but can still be seen when computing the eigenvalues, or doing a singular value decomposition, of our covariance matrix.
In particular the spectrum will contain a single value that is roughly 8 orders of magnitude smaller than any other value. 
Because the inverse covariance matrix is needed when computing \SNR{}\footnote{More formally, a matrix with a null-eigenvalue is not invertible but practically we can interpret the \SNR{} computation with the given formally ill-behaved covariance as simply throwing away some data.}, this null eigenvalue causes numerical instabilities in the \SNR{}.
To handle this we prescribe the following recipe.
\begin{enumerate}
    \item Do a singular-value decomposition of the covariance matrix $\C$ in \Eq{dsnr}:
    \begin{equation}
        \C_{\i\j}= U_{\tt ia}\Sigma_{\tt ab}V_{\tt bj}
    \end{equation}
    where $\Sigma_{aa}$ are the singular value of $\C$.
    \item Let $\Sigma'$ be equal to $\Sigma$ except for the smallest singular value which is manually set to zero. 
    \item Recombine $\Sigma'$ with the original singular vectors $U,V$ to form a new covariance $\C'$:
    \begin{equation}
        \C_{\i\j}'= U_{\tt ia}\Sigma_{\tt ab}'V_{\tt bj}
    \end{equation}
    \item Compute the quadratic form in \Eq{dsnr} by first solving with least-squares for $x$ in the system
    \begin{equation}
        x_\i \C'_{\i\j}=  \O^{\HI,\kappa}_{\ell,\j},
    \end{equation}
    and then using $x$ to compute the $\SNR{}$:
    \begin{equation}
        \delta \SNR^2_\ell = \O^{\HI,\kappa}_{\ell,\i} x_\i.
    \end{equation}
\end{enumerate}

\section{Modeling of the luminosity-halo relation}
\label{app:luminosity}
In this appendix we describe our modeling of the halo mass-luminosity relation $L(M_h,\chi)$ used in \Sec{LIM}.
To briefly summarize this appendix:
\begin{itemize}
    \item \textbf{HI} (\App{HI}): 
    We relate neutral hydrogen mass $M_{\rm HI}$ to halo mass which then can be related to the neutral hydrogen 21-cm line luminosity.
    To model the halo mass-neutral hydrogen mass relation, we use the results of \Reff{Villaescusa-Navarro:2018vsg} which analyze the TNG100 magneto-hydrodynamic simulation.
    \item \textbf{Ly-$\boldsymbol\alpha$} (\App{Lya}) and \textbf{[CII]} (\App{CII}): 
    We relate halo mass to star-formation rate (SFR) and then use empirical relations to relate SFR to the [CII] and Ly-$\alpha$ emission line luminosity.
    To model the halo mass-SFR relation, we use the average relation between star-formation rate (SFR) and halos of mass $M_h$ at redshift $z$ derived from {UniverseMachine} (\Reff{2019MNRAS.488.3143B}).
    \item \textbf{CO(1$\rightarrow$0)} (\App{CO}): We use a direct scaling relation between halo-mass and the CO rotational transition line luminosity from \Refs{Li:2015gqa, COMAP:2021lae} that is tuned with COLDz \cite{Riechers:2018zjg}, UniverseMachine \cite{2019MNRAS.488.3143B}, and COPSS \cite{Keating:2016pka}.
\end{itemize}
In this appendix, log denotes a base-10 logarithm.

\subsection{HI 21-cm emission}
\label{app:HI}
To model the halo-luminosity relation for post-reionization HI 21-cm (spin-flip) transition, we first relate halo mass to neutral hydrogen mass $M_{\rm HI}$ with results derived from the TNG100 magneto-hydrodynamic simulation (\Reff{Villaescusa-Navarro:2018vsg}) and then relate neutral hydrogen mass to luminosity by assuming 
the spin temperature $T_S$,
defined by the relative abundance of neutral hydrogen atoms in the higher and lower energy hyperfine states $N_1/N_0\sim{\rm exp}\left\{-T_\star/T_S \right\}$,
is greater than the temperature of the background CMB and temperature associated with the photon released by the 21-cm transition, $T_\star\equiv h\nu_{21}/k_B$ where $h$ is Planck's constant and $k_B$ is the Boltzmann constant.
In particular this assumption about the spin temperature is satisfied if the following processes occurs frequently enough in the neutral hydrogen gas (\Refs{Field:1958rri, 1959ApJ...129..536F, 2012RPPh...75h6901P, 2016era..book.....C}):
\begin{enumerate}
    \item Scattering with Ly-$\alpha$ photons produced by nearby stars and galaxies which induce a 21-cm transition in neutral hydrogen atoms via an intermediary excited state\footnote{e.g. Consider a neutral hydrogen atom which starts in the lower hyperfine energy level of the ground state. After absorption of a Ly-$\alpha$ photon, the neutral hydrogen atom is excited into one of the central 2P hyperfine states. This atom can now emit a Ly-$\alpha$ photon and relax back to either hyperfine levels of the ground state. If this emission relaxes the atom to the higher hyperfine energy level of the ground state, then a spin-flip transition has occurred.}.
    \item Collisions with other hydrogen atoms, free electrons, and protons which excite the 21-cm transition.
\end{enumerate}
Under this assumption about the spin temperature, stimulated emissions almost exactly balance absorption (e.g. the attenuation coefficient is nearly zero) which allows us the write the luminosity purely in terms of spontaneous emission (\Reff{Bull:2014rha})\footnote{The factor of 3/4 in \Eq{LHI} occurs because, under the assumption we made that $T_S\gg T_\star$, the abundance of neutral hydrogen atoms in the higher energy hyperfine state $N_1={g_1}/{g_0}\times N_0$ (e.g. ${\rm exp}\{-T_\star/T_S\}\rightarrow 1$) where $g_1$ and $g_0$ are degeneracy factors. Since the higher energy hyperfine state is a triplet state, $g_1=3$ whereas the lower energy hyperfine state is a singlet so $g_0=1$. Thus $N_1=3N_0=3/4\times N$ where $N=M_{\rm HI}/m_p$ is the total number of neutral hydrogen atoms. 
}:
\begin{equation}
    L_{\rm HI}(M,z) = A_{10}\times h\nu_{21}\times \frac {3}{4} \frac{M_{\rm HI}(M,z)}{m_p}.
    \label{eq:LHI}
\end{equation}
$A_{10}\simeq 2.869\times 10^{-15}\ {\rm s}^{-1}$ is the Einstein spontaneous emission coefficient for the 21-cm hyperfine line of HI, $m_p$ is the mass of the proton, and $M_{\rm HI}(M,z)$ is modeled with the parameterization of \Refs{Villaescusa-Navarro:2018vsg, Sato-Polito:2022wiq}:
\begin{equation}
    M_{\rm HI}(M_h,z) = \frac{
    M_0(z)
    \left({M_h} /{M_{\rm min}(z)}\right)^{\alpha(z)
    }
    }{ {\rm exp}\left\{ \left( {M_{\rm min}(z)}/{M_h}\right)^{0.35}\right\} } .
\end{equation}
Informed by TNG100 we determine $M_0(z)$, $M_{\rm min}(z)$ and $\alpha(z)$ by linearly interpolating the best-fit values of these quantities reported at redshift $z=0,1,2,3,4,5$ in Table.~1 of \Reff{Villaescusa-Navarro:2018vsg} for Friends-of-Friends (FoF) halos (\Reff{Davis:1985rj}).
\subsection{Ly-${\alpha}$ emission}
\label{app:Lya}
To model the halo-luminosity relation for Ly-$\alpha$ emissions, we first relate halo mass and star-formation rate using the average relation between SFR and halos of mass $M_h$ at redshift $z$ derived from {UniverseMachine} (\Reff{2019MNRAS.488.3143B}), ${\rm SFR}(M_h,z)$, and then model the relation between SFR and luminosity with a linear Kennicutt-Schmidt like relation (\Reff{Kennicutt:1997ng}) following \Reff{COMAP:2018svn}:
\begin{equation}
    \frac{L_{\textrm{Ly-$\alpha$}}(M_h,z)}{\rm erg/s} = 1.6\times 10^{42} \left(\frac{\rm SFR(M_h,z)}{M_\odot/\rm yr} \right)f_{\rm esc}({\rm SFR}, z).
\end{equation}
The escape fraction $f_{\rm esc}$ captures the scattering and attenuation of the Ly-$\alpha$ luminosity.
Though $f_{\rm esc}$ is largely unconstrained, for $z\gtrsim 2$, which is the regime of interest for our studies, observations suggests two assumption can reasonably be made about the escape fraction's behavior:
\begin{enumerate}
    \item $f_{\rm esc}$ increases monotonically with redshift; and
    \item $f_{\rm esc}$ decreases with higher SFR.
\end{enumerate}
This motivates the parameterization of $f_{\rm esc}$ used in \Reff{COMAP:2018svn} which satisfies these two assumptions:
\begin{align}
    f_{\rm esc}(&{\rm SFR},z) = \left[\frac{\left(f_0 + \frac{1-f_0}{1 + ({\rm SFR / SFR_0})^\eta } \right)}{ (1+e^{-\xi(z-z_0)})^{\zeta}} \right]^2.
\end{align}
Informed by \Refs{2010ApJ...711..928C, 2012ApJ...749..106B, Sobral:2016toh, Gronwall:2007qd}, we match \Reff{COMAP:2018svn} and set 
$\xi=1.6$, 
$\zeta=1/4$, 
$\eta = 0.875$,
$z_0=3.125$,
$f_0=0.18$, 
and ${\rm SFR}_0=1.29\ M_\odot/\rm yr$.

\subsection{CO(1$\rightarrow$0) rotational transition}
\label{app:CO}
To model the halo-luminosity relation for CO rotational transitions from $J=1$ to $J=0$, we use the direct scaling relation between halo-mass and line luminosity used by \Refs{COMAP:2021lae,Li:2015gqa}:
\begin{equation}
    \frac{L_{\rm CO(1\rightarrow 0)}(M_h)}{L_\odot} = 4.9\times 10^{-5}\times 
    \frac C{(M_h / M)^A + (M_h/M)^B}.
\end{equation}
Informed by COLDz (\Reff{Riechers:2018zjg}), UniverseMachine (\Reff{2019MNRAS.488.3143B}), and COPSS (\Reff{Keating:2016pka}), we match \Reff{COMAP:2021lae} and set $A=-2.85$, $B=-0.42$, $\log C = 10.63$, and $\log M/M_\odot =12.3$.

\subsection{[CII] emission}
\label{app:CII}
To model the halo-luminosity relation for [CII] emission, we first relate halo mass and star-formation rate using the average relation between SFR and halos of mass $M_h$ at redshift $z$ derived from {UniverseMachine} (\Reff{2019MNRAS.488.3143B}), ${\rm SFR}(M_h,z)$, and then model the relation between SFR and luminosity with a power law following \Reff{Zhou:2022gmu}:
\begin{equation}
    \log \frac{L_{\rm [CII]}(M_h,z)}{L_\odot} = \alpha \log \frac{{\rm SFR}(M_h,z)}{M_\odot/{\rm yr}} + \beta.
\end{equation}
Informed by the semi-analytical model of \Reff{2021ApJ...911..132Y}, we match \Reff{Zhou:2022gmu} and set $\alpha = 1.26$ and $\beta=7.1$.

\section{Modeling of instrumental noise}
\label{app:experiments}
In this appendix we describe our modeling of the instrumental noise power $P^{\epsilon_I}$ described in \Eq{eI_base} in \Sec{lim_noise} for the experiments considered in this paper.
The experimental configurations we assume are detailed in \Tab{experiments} and summaries of our instrumental noise model for these experiments are referenced in the final line of this table.
We approximate the smallest 3-D volume element observable by an experiment as a cubic voxel whose depth is determined by the resolving power $\delta \nu$ and sky-area is determined by angular resolution $\Omega_{\rm pixel}$.
The volume of this volume element $\delta V$ can then be computed as
\begin{align}
\nonumber   \delta V(z) &= \overbracket{
   \vphantom{   \frac{c}{H(z)}\frac{1+z}{\mathcal R}}
   \chi^2(z)\frac{\Omega_{\rm pixel}}{\rm sr}
   }^{\rm Sky\ area}
   \overbracket{
   \frac{c}{H(z)}\frac{\delta\nu}{\nu_{\rm obs}}
   \frac{\nu_{\rm rest}}{\nu_{\rm obs}}
   }^{\rm Radial\ depth}\\
   &=  \underbracket{
   \vphantom{   \frac{c}{H(z)}\frac{1+z}{\mathcal R}}
   \frac{\delta\nu}{\nu_{\rm obs}}
   \frac{\nu_{\rm rest}}{\nu_{\rm obs}}
   \frac{\Omega_{\rm pixel}}{\rm sr}}_{\rm Experiment} 
   \underbracket{\left(\fV(z)\equiv \frac{c{\chi^2(z)}}{H(z)}\right)}_{\rm Cosmology},
   \label{eq:deltaV}
\end{align}
where in the second equality we separated experimental configuration from cosmology.
At best, experiments can observe only the average intensity within a volume element $\delta V$:
\begin{equation}
    I_{\sf obs} = \frac 1 {\delta V} \int_{\delta V}I(\vtheta,\chi)dV .
\end{equation}
Typically, our noise models first determine the power of fluctuations from instrumental noise on measurements of the binned 3-D intensity map, $P^{\epsilon_I}_{\delta V}$.
Assuming the instrumental noise can be modeled as a uncorrelated Gaussian random field then allows us to relate the power of fluctuations due to instrumental noise on the binned intensity map $I_{\sf obs}$ to the power of fluctuations due to instrumental noise on the continuous intensity map $I$ itself:
\begin{equation}
P^{\epsilon_I}_{\delta V} = \frac 1 {(\delta V)^2}\int_{\delta V} P^{\epsilon_I}\ dV = \frac{P^{\epsilon_I}}{\delta V}.
    \label{eq:sI_PeI}
\end{equation}
Throughout this section we report summaries of our instrumental noise models (listed in the final row of \Tab{experiments}) in terms of characteristic values for for each experiment which are only approximately correct but provide valuable intuition.
However, when actually computing values of $P^{\epsilon_I}$ we do not make use of these summarizing expressions and instead directly implement the computation described in the following sections which yield the summarizing expressions.
Additionally, while $P^{\epsilon_I}$ typically varies spatially, in this paper we conservatively choose to fix $P^{\epsilon_I}$ to its maximum value in the relevant region for each experiment.

In \Fig{lim-model}, we plot the instrumental noise powers computed in this section in comparison to the expected cosmological signal for each experiment.

\subsection{\textsf{CHIME}}
\label{app:CHIME}
Our model of $P^{\epsilon_I}$ for \textsf{CHIME} is informed by \Refs{Morales:2004ca, Furlanetto:2006jb, 2012ApJ...753...81P, 2017isra.book.....T, Bull:2014rha,CHIME:2022dwe, CHIME:2022kvg}.
In contrast to other experiments we consider which sample the intensity $I$ binned in real $(x,y,\nu_{\rm obs})$-space volume elements $\delta V$ (\Eq{deltaV}), interferometers sample the 2-D Fourier transform of intensity, the complex visibility $\mathcal V$, binned in $(u,v, \nu_{\rm obs})$-space volume elements $\delta U$ which we will define shortly.
Consider a pair of antennae separated by a baseline $\vb$.
This can be converted to angular coordinates $\vtheta$ based on the wavelength being observed:
\begin{equation}
    \frac{\theta_i}{\rm rad} \sim \frac{\lambda_{\rm obs}}{b_i}.
\end{equation}
The components of the corresponding vector $\vu$ in $(u,v)$ space are defined as the east-west ($u$) and north-south ($v$) baseline in inverse angular coordinates or, equivalently, in units of observed wavelength.
We approximate the effective collecting area for a single antenna in a cylindrical interferometer, $A_e$, following \Reff{CHIME:2022dwe}:
\begin{equation}
    \frac{A_e}{4.3\ \rm m^2}
    \approx
    \biggr(\frac{\eta}{0.7} \biggr)
    \biggr(\frac{w_{\rm cyl}}{20\ \rm m}\biggr)
    \biggr(\frac{d_{\rm ant}}{0.3048\ \rm m} \biggr)
    .
\end{equation}
$\eta$ is the aperture efficiency, $l_{\rm cyl}$ is the length along the cylinder axis that is instrumented with feeds, $w_{\rm cyl}$ is the cylinder width, and $N_{\rm ant}$ is the number of antenna per cylinder.
Combining $A_e$ with the observed wavelength then defines a effective beam\footnote{Note that because of the geometry of \textsf{CHIME}, this beam is anisotropic. However we follow \Refs{Bull:2014rha,CHIME:2022dwe} and approximate it as isotropic for simplicity.}:
\begin{equation}
    \frac{\Omega_{\rm beam} (\lambda_{\rm obs})}{157\ \rm deg^2}
    \approx
    \biggr(\frac{\lambda_{\rm obs}}{45\ \rm cm}\biggr)^{2}
    \biggr(\frac{A_e}{4.3\ \rm m^2}\biggr)^{-1}
    .
\end{equation}
Note that $\Omega_{\rm beam}/N_{\rm ant}$ is approximately the angular resolution of $\ang{;40;}$ for {\sf CHIME} reported in \Reff{CHIME:2022dwe} and \Tab{experiments} as expected.
We use this beam to define a area element of the $(u,v)$ plane, $\delta u^2=(2\pi)^2/\Omega_{\rm beam}$:
\begin{equation}
    \frac{\delta u^2(\lambda_{\rm obs})}{0.25\ \rm deg^{-2}} \approx
    \biggr( \frac{\Omega_{\rm beam} (\lambda_{\rm obs})}{157\ \rm deg^2} \biggr)^{-1},
\end{equation}
and assume that visibility measurements by \textsf{CHIME} are binned within a volume element $\delta U$ that we approximate as a cubic voxel whose depth is determined by the spectral resolution $\delta \nu$ and area is determined by $\delta u^2$.

To compute the instrumental noise power $P^{\epsilon_I}$ for \textsf{CHIME}, we begin by assuming that fluctuations due to instrumental noise in each visibility measurement by a pair of antenna is due to thermal fluctuations.
Furthermore we assume that the instrumental noise in each visibility measurement by a pair of antenna is independent.
Thus, combining $N$ visibility measurements beats down the noise like $1 / \sqrt{N}$.
Using the radiometer equation (\Reff{2016era..book.....C}) for each measurement and combining measurements while assuming independence, we can compute the power of fluctuation due to thermal instrumental noise in a volume element $\delta U$:
\begin{align}
\nonumber    \frac{P^{\epsilon_T}_{\delta U}}{3\ \rm \mu K^2}
   \approx 
    &
    \lr{T_{\rm system}}{55\ \rm K}{2} 
    \lr{N_{\delta U}}{2700}{-1}
    \lr{N_{\rm pol}}{2}{-1}
    \\
    &
    \lr{\tau_{\rm p}(\lambda_{\rm obs})}{33\ \rm hr}{-1}
    \lr{\delta \nu}{1.5625\ \rm MHz}{-1}
.
\label{eq:sT_chime}
\end{align}
$T_{\rm system}$ is the system temperature of the instrument, $N_{\rm pol}$ is the number of number of polarizations per antenna, and $\tau_{\rm p}$ is the observing time per pointing which we approximate as
\begin{equation}
    \frac{\tau_{\rm p}(\lambda_{\rm obs})}{33\ \rm hr}{}
   \approx 
    \lr{\tau_{\rm survey}}{1\ \rm yr}{}
    \lr{\Omega_{\rm FOV}(\lambda_{\rm obs})}{116.5\ \rm deg^2}{}
    \lr{\Omega_{\rm field}}{31000\ \rm deg^2}{-1},
\end{equation}
where $\Omega_{\rm FOV}$ is the instantaneous field of view of a single cylinder (\Reff{10.1117/12.2056962}):
\begin{equation}
    \frac{\Omega_{\rm FOV}(\lambda_{\rm obs})}{116.5\ \rm deg^2}\approx\biggr(\frac{\lambda_{\rm obs}}{45\ \rm cm}\biggr)\biggr(\frac{w_{\rm cyl}}{20\ \rm m}\biggr)^{-1}.
\end{equation}
$N_{\delta U}$ is the number of visibility measurements made by \textsf{CHIME} within the area element $\delta u^2$.
This is computed from the $(u,v)$-plane baseline number density, $n(u)$, for the experimental configuration of CHIME:
\begin{equation}
    N_{\delta U} = n(u,\lambda_{\rm obs}) \delta u^2(\lambda_{\rm obs}).
\end{equation}
We approximate $n(u)$ by assuming that the $(u,v)$-plane is uniformly sampled within the geometrically allowed annular region\footnote{{It is known that this approximation fails to precisely model \textsf{CHIME} (see App. D of \Reff{CosmicVisions21cm:2018rfq}) and one should use the code detailed in App.~C of \Reff{Bull:2014rha} or fitting formula from \Reff{CosmicVisions21cm:2018rfq} to model the $n(u)$ of \textsf{CHIME}. 
However, we decided not to do this for simplicity as the goal of this work is only to point out this direct correlation is feasible.
The median value of a realistically computed $n(u)\delta u^2$ for CHIME roughly matches our approximation.
}
}:
\begin{equation}
    n(u,\lambda_{\rm obs}) = \frac{N_d(N_d-1)}{2\pi (u_{\rm max}^2 - u_{\rm min}^2)},
\end{equation}
where the total number of antenna $N_d=N_{\rm ant}\times N_{\rm cyl}$ is the number of antenna per cylinder $N_{\rm ant}=256$ times the number of cylinders $N_{\rm cyl}=4$ and $u_{\rm min},u_{\rm max}$ are derived from the minimum and maximum baselines of \textsf{CHIME} (\Reff{CHIME:2022dwe}):
\begin{align}
    \frac{u_{\rm min}}{0.7\ \rm rad^{-1}} &\approx
    \lr{b_{\rm min}}{30.48\ \rm cm}{}
    \lr{\lambda_{\rm obs}}{45\ \rm cm}{-1}\\
    \frac{u_{\rm max}}{226\ \rm rad^{-1}} &\approx
    \lr{b_{\rm max}}{10200\ \rm cm}{}
    \lr{\lambda_{\rm obs}}{45\ \rm cm}{-1}
    .
\end{align}
In particular note that $b_{\rm min}=d_{\rm ant}$.
For future convinience we also define $\Delta b^2 \equiv b_{\rm max}^2 - b_{\rm min}^2$.
The power of fluctuations due to instrumental noise in the visibility measurement can be written in units of flux-densities using the Rayleigh-Jeans approximation:
\begin{equation}
\frac{P^{\epsilon_\V}_{\delta U}}{1\ \rm mJy^2} \approx
\lr{P^{\epsilon_T}_{\delta U}}{3 \ \rm \mu K^2}{}
\lr{\Omega_{\rm beam}}{157\ \rm deg^2}{2}
\lr{\nu_{\rm obs}}{663\ \rm MHz}{4}.
\label{eq:PeV}
\end{equation}
Since we assume the instrumental noise in each complex visibility measurement is Gaussian, we know that (1) the power of fluctuations due to instrumental noise in both the real and imaginary component of each complex visibility measurement is equal to $P^{\epsilon_\V}_{\delta U}/2$ and (2) fluctuations due to instrumental noise in the real and imaginary component of each complex visibility measurement is uncorrelated.

As it currently stands, we're in a awkward coordinate system.
Two axes of the intensity map have been Fourier transformed to $(u,v)$ space while one axis $\nu_{\rm obs}$ is still in real space.
To remove ourselves from this awkward position we can Fourier transform from $(u,v)$ space back to {real angular} $(x,y)$ space:
\begin{equation}
    \epsilon_I(\vx)
    = \int \frac{du}{2\pi}\int \frac{dv}{2\pi}\  \epsilon_\V(\vu)e^{i\vx \cdot \vu},
    \label{eq:eI1}
\end{equation}
where $\epsilon_\V$ is the random field of visibility fluctuations due to instrumental noise.
Note that by definition, $\epsilon_\V(\vu) = \epsilon_\V^*(-\vu)$ due to the realty of the intensity map.
Because we approximate the $(u,v)$-plane baseline number density as uniform in the geometrically allowed annular region, it is more natural to consider \Eq{eI1} in polar coordinates $(u,\phi)$:
\begin{equation}
\epsilon_I(\vx)
    = \int_{u_{\rm min}}^{u_{\rm max}} 
    \frac {u\times du} {(2\pi)^2} 
     \int_0^{2\pi} d\phi\ 
    \epsilon_\V(\vu)e^{i\vx\cdot \vu}.
\end{equation}
This can be naturally discretized to relate binned visibility fluctuations $\langle \epsilon _\V\rangle_{\delta U}$, whose statistics we derived above in \Eq{PeV}, to binned intensity fluctuations $\langle \epsilon_I\rangle _{\delta V}$:
\begin{equation}
\langle \epsilon_I \rangle_{\delta V} =  \sum_{i=1}^{N_u}\frac{u_i\times \delta u}{{(2\pi)^2}}\sum_{j=1}^{N_\phi(u_i)} \delta\phi(u_i)\langle \epsilon_\V\rangle_{\delta U_i}e^{i \vx\cdot \vu}
    ,
\end{equation}
where $N_u=(u_{\rm max}-u_{\rm min})/\delta u\approx 8$ is the number of radial bins we have, $u_i= u_{\rm min} + \delta u(i-1/2)$ is fixed to roughly the center of these radial bins, $\delta \phi(u_i)=\delta u / u_i$ is the angular size of a bin, and $N_\phi(u_i)=2\pi/\delta\phi(u_i)$ is the number of angular bins we have at radius $u_i$.
Since the instrumental noise in different bins in uncorrelated, with exception of the $\vu$ and $-\vu$ bins due to $\epsilon_\V(\vu) = \epsilon_\V^*(-\vu)$, the power of fluctuations due to instrumental noise in the intensity map can then be computed:
\begin{equation}
    P^{\epsilon_I}_{\delta V} = \frac {1} {(2\pi)^4}\left[\sum_{i=1}^{N_u} u_i^2\times (\delta u)^2 \sum_{j=1}^{N_\phi(u_i)} (\delta \phi(u_i))^2 \right] P^{\epsilon_\V}_{\delta U}
    .
\end{equation}
The double summation in the bracket can be evaluated analytically\footnote{While $N_u$ and $N_\phi$ are not integers, we evaluate the double summation pretending they are and apply the resulting expression to non-integer $N_u$ and $N_\phi$.}
\begin{equation}
    \left[\sum_{i=1}^{N_u} u_i^2\times (\delta u)^2 \sum_{j=1}^{N_\phi(u_i)} (\delta \phi(u_i))^2 \right] = \pi (\delta u)^2 (u_{\rm max}^2 - u_{\rm min}^2),
\end{equation}
thus yielding the following result for the power of fluctuations due to instrumental noise in {a voxel of the} 3-D intensity map:
\begin{equation}
    P^{\epsilon_I}_{\delta V} = \frac {1} {(2\pi)^4} \left[\pi (\delta u)^2 (u_{\rm max}^2 - u_{\rm min}^2)\right]  P^{\epsilon_\V}_{\delta U}
    .
\end{equation}
The simplicity of this result heavily implies that a greatly simplified derivation is possible.
This can then be related to the power of fluctuations in the intensity map itself with \Eq{sI_PeI}.
Thus our model of instrumental noise power for \textsf{CHIME} can be summarized as
\begin{align}
\nonumber
\frac{P^{\epsilon_I}_{\textsf{CHIME}}}{0.83\ \rm Mpc^3\ (kJy/sr)^2} \approx 
    \lr{\Omega_{\rm field}}{31000\ \rm deg^2}{}
    \lr{\tau_{\rm survey}}{1\ \rm yr}{-1}
    &
\\
\nonumber
\times
    \lr{T_{\rm system}}{55\ \rm K}{2} 
    \lr{N_{\rm ant}}{256}{-2}
    \lr{N_{\rm pol}}{2}{-1}
    \lr{N_{\rm cyl}}{4}{-2}
\\
\nonumber
\times
    \lr{\nu_{\rm obs}}{663\ \rm MHz}{}
    \lr{\fV(\nu_{\rm obs})}{32.1\ \rm Gpc^3}{}
    \lr{\Delta b^2}{10^4\ \rm m^2}{2}&
    \\
    \times
    \lr{d_{\rm ant}}{0.3048\ \rm m}{-3}
    \lr{w_{\rm cyl}}{20\ \rm m}{-2}
    \lr{\eta}{0.7}{-3}&
\label{eq:CHIME}
.
\end{align}

\subsection{\textsf{HETDEX}}
\label{app:HETDEX}
Our model of $P^{\epsilon_I}$ for \textsf{HETDEX} primarily follows \Reff{Cheng:2018hox} and is additionally informed by \Refs{2016ASPC..507..393H,Gebhardt:2021vfo,Schaan:2021hhy}.
Sensitivity metrics reported in \Reff{2016ASPC..507..393H} and \Reff{Gebhardt:2021vfo} are both consistent with the power of fluctuations from instrumental noise in a volume element of the 3-D flux map, $P^{\epsilon_F}_{\delta V}$, being roughly
\begin{equation}
    \sqrt{P^{\epsilon_F}_{\delta V}}\approx 10^{-17}\ \rm erg\ s^{-1}\ cm^{-2}.
\end{equation}
Relating flux to intensity yields the power of fluctuations from instrumental noise in a volume element of the intensity map:
\begin{align}
\nonumber   \frac{P^{\epsilon_I}_{\delta V}}{30\ \rm (kJy/sr)^2} \approx 
\biggr(\frac{P^{\epsilon_F}_{\delta V}}{10^{-34}\ \rm erg^2\ s^{-2}\ cm^{-4}}\biggr)&
\\
   \times
   \biggr( \frac{\mathcal R}{800}\biggr)^2
   \biggr(\frac{\nu_{\rm obs}}{700\ \rm THz}\biggr)^{-2}
   \biggr(\frac {\Omega_{\rm pixel}}{\ang{;;3}\times\ang{;;3}} \biggr)^{-2}&.
\end{align}
This can then be related with the power of fluctuations in the intensity map itself with \Eq{sI_PeI}.
Thus our model of instrumental noise power for \textsf{HETDEX} is
\begin{align}
\nonumber
\frac{P^{\epsilon_I}_{\textsf{HETDEX}}}{1.14\ \rm Mpc^3\ (kJy/sr)^2} \approx 
\biggr(\frac{P^{\epsilon_F}_{\delta V}}{10^{-34}\ \rm erg^2\ s^{-2}\ cm^{-4}}\biggr)&
\\
\times
\biggr( \frac{\Gamma(\nu_{\rm obs})}{42\ \rm Gpc^3}\biggr)
\biggr( \frac{\nu_{\rm obs}}{700\ \rm THz}\biggr)^{-3}
\biggr( \frac{\Omega_{\rm pixel}}{\ang{;;3}\times \ang{;;3}}\biggr)^{-1}
\biggr(\frac{\mathcal R}{800} \biggr)&
\label{eq:HETDEX}
.
\end{align}

\subsection{\textsf{COMAP}}
\label{app:COMAP}
Our model of $P^{\epsilon_I}$ for \textsf{COMAP} primarily follows \Reff{Li:2015gqa} with additional information from \Refs{Cleary:2021dsp,COMAP:2021qdn,COMAP:2021pxy,COMAP:2021sqw, COMAP:2021lae, Schaan:2021hhy}.
We assume that the instrumental noise is due to thermal fluctuations and thus has a white noise spectrum.
Using the radiometer equation (\Reff{2016era..book.....C}) we can then compute the power of fluctuation from instrumental noise in a volume element of temperature:
\begin{align}
\nonumber
\frac{P^{\epsilon_T}_{\delta V}}{1500\ \rm\mu K^2}
\approx \biggr(\frac{T_{\rm system}}{40\ {\rm K}} \biggr)^2\biggr(\frac{N_{\rm feeds}}{19} \biggr)^{-1}&
\\
\times
\biggr( \frac{\mathcal R}{800}\biggr)
\biggr( \frac{\nu_{\rm obs}}{30\ {\rm GHz}}\biggr)^{-1}
\biggr(\frac{\tau_{\rm pixel}}{25\ \rm min} \biggr)^{-1} 
&.
\end{align}
$T_{\rm system}$ is the system temperature of the instrument, $N_{\rm feeds}$ is the number of observing feeds, and $\tau_{\rm pixel}$ is the observing time per sky pixel which we approximate as
\begin{equation}
   \frac{\tau_{\rm pixel}}{25\ {\rm min}}  \approx \biggr( \frac{\Omega_{\rm pixel}}{\ang{;4.5;}\times\ang{;4.5;}/(8\ln 2)}\frac{12\ {\rm deg}^2}{\Omega_{\rm field}} \biggr)\biggr(\frac {\tau_{\rm survey}}{5000\ \rm hr} \biggr).
\end{equation}
We relate thermal fluctuations in a volume element of temperature to fluctuations due to instrumental noise in a volume element of intensity, $P^{\epsilon_I}_{\delta V}$, with the Rayleigh-Jeans approximation:
\begin{equation}
    \frac{P^{\epsilon_I}_{\delta V}}{1\ ({\rm kJy}/\rm sr)^2} \approx 
\frac{P^{\epsilon_T}_{\delta V}}{1500\ \rm\mu K^2}
\biggr(\frac{\nu_{\rm obs}}{30\ \rm GHz} \biggr)^4.
\end{equation}
This can then be related with the power of fluctuations in the intensity map itself with \Eq{sI_PeI}.
Thus our model of instrumental noise power for \textsf{COMAP} is
\begin{align}
\nonumber
\frac{P^{\epsilon_I}_{\sf COMAP}}{70\ \rm Mpc^3\ (kJy/sr)^2} \approx 
    \biggr(\frac{T_{\rm system}}{40\ {\rm K}} \biggr)^2
    \biggr(\frac {\tau_{\rm survey}}{5000\ \rm hr} \biggr)^{-1}&
    \\
    \times
    \biggr(\frac{\fV(\nu_{\rm obs})}{42\ \rm Gpc^3} \biggr)
    \biggr(\frac{\nu_{\rm obs}}{30\ \rm GHz} \biggr)^2
    \biggr(\frac{\Omega_{\rm field}}{12\ \rm deg^2}\biggr)
    \biggr(\frac{N_{\rm feeds}}{19} \biggr)^{-1}&
    \label{eq:COMAP}
    .
\end{align}

\subsection{\textsf{CCAT}}
\label{app:CCAT}
Our model of $P^{\epsilon_I}$ for \textsf{CCAT} comes directly from \Reff{Sato-Polito:2020cil}. 
In particular we assume that the power of fluctuations from instrumental noise in a volume element of the intensity map, $P^{\epsilon_I}_{\delta V}$, is that which is reported in Table 1. of \Reff{Sato-Polito:2020cil} for $z=3.5$:
\begin{equation}
    \textrm{(\Reff{Sato-Polito:2020cil})}:\quad P^{\epsilon_I}_{\delta V}(z=3.5) = (5.7\times 10^4\ \rm Jy/sr)^2.
\end{equation}
This can then be related with the power of fluctuations in the intensity map itself with \Eq{sI_PeI}.
Thus our model of instrumental noise power for \textsf{CCAT} is
\begin{align}
\nonumber    
\frac{P^{\epsilon_I}_{\textsf{CCAT}}}{2.22\times 10^{4}\ \rm Mpc^3\ (kJy/sr)^2} \approx  
\biggr(\frac{P^{\epsilon_I}_{\delta V}(z=3.5)}{(5.7\times 10^4\ \rm Jy/sr)^2} \biggr)&
\\
\times
    \biggr(\frac{\Omega_{\rm pixel}}{\frac{\ang{;;30}\times \ang{;;30}}{8\ln 2}} \biggr) 
    \biggr( \frac{\Gamma(z=3.5)}{40\ \rm Gpc^3} \biggr)
    \lr{\delta_\nu}{4.2\ \rm MHz}{}&
    \label{eq:CCAT}
\end{align}

\subsection{\textsf{SPHEREx}}
\label{app:SPHEREx}
Our model of $P^{\epsilon_I}$ for \textsf{SPHEREx} primarily follows \Reff{Cheng:2018hox} and is additionally informed by \Refs{SPHEREx:2014bgr, SPHEREx:2016vbo, SPHEREx:2018xfm, Schaan:2021hhy}.
\Reff{SPHEREx:2018xfm} reports that \textsf{SPHEREx} is able to make a 5-$\sigma$ detection a point source with an AB magnitude of 22:
\begin{equation}
    m_{AB} = 22.
\end{equation}
So, \textsf{SPHEREx} is able to make 5-$\sigma$ detections of point sources whose flux density is larger than \begin{equation}
    {f_*} =  10^{(8.9-m_{AB})/2.5}\ {\rm Jy} \approx 5.75\ \rm \mu Jy.
\end{equation}
This then defines the power of fluctuations for measurements of a point-source's flux density, $P^{\epsilon_{f}}_{\sf PS}$:
\begin{equation}
    \textsf{SNR}(f_*)=\frac{f_*}{\sqrt{P^{\epsilon_{f}}_{\sf PS}}}=5\Rightarrow 
    {\sqrt{P^{\epsilon_{f}}_{\sf PS}}}\approx 1.15\ \rm \mu Jy.
\end{equation}
As described in \Reff{SPHEREx:2014bgr}, \textsf{SPHEREx} combines flux measurements from different pixels to optimally extract the flux of a point source while accounting for smearing due to the point-spread function (PSF).
The parameter $N_{\tt eff}$ captures the effective number of pixels a point-source is smeared out into.
In other words, $N_{\tt eff}$ corresponds to the number of pixels whose fluxes we must combine to extract the flux of a point-source\footnote{We write this as a discrete sum for clarity but $N_{\tt eff}$ is not necessarily an integer.}:
\begin{equation}
    f_{\sf PS} \sim \sum_{i=1}^{N_{\tt eff}} f_{\tt i},
\end{equation}
where $f_{\tt i}$ is the flux from pixel $\tt i$. 
Assuming the fluctuations in $f_{\tt i}$ due to instrumental noise are independent but have the same power $P^{\epsilon_f}_{\delta V}$ yields
\begin{equation}
   P^{\epsilon_f}_{\sf PS} = N_{\tt eff}P^{\epsilon_f}_{\delta V}.
\end{equation}
Values of $N_{\tt eff}\gtrsim 2$ are reported in Fig. 9 of \Reff{SPHEREx:2014bgr} which give us
\begin{equation}
   \sqrt{P^{\epsilon_f}_{\delta V}} = 
   \sqrt{\frac{P^{\epsilon_f}_{\sf PS}}{N_{\tt eff}} }
    \lesssim 0.8\ \rm \mu Jy.
\end{equation}
We will conservitavely fix $P^{\epsilon_f}_{\delta V}$ to its maximum value.
Relating flux density to intensity yields the power of fluctuations from instrumental noise in a volume element of the intensity map:
\begin{align}
\nonumber     
\frac{P^{\epsilon_I}_{\delta V}}{1\ \rm (kJy/sr)^2} \approx
\biggr( \frac{P^{\epsilon_f}_{\delta V}}{(0.8\ \rm \mu Jy)^2}\biggr) \biggr(\frac{\Omega_{\rm pixel}}{\ang{;;6}\times \ang{;;6}}\biggr)^{-2}.
\end{align}
This can then be related with the power of fluctuations in the intensity map itself with \Eq{sI_PeI}.
Thus our model of instrumental noise power for \textsf{SPHEREx} is
\begin{align}
\nonumber  
\frac{P^{\epsilon_I}_{\textsf{SPHEREx}}}{4.1\ \rm Mpc^3 (kJy/sr)^2} \approx  
    \biggr( \frac{P^{\epsilon_{f}}_{\sf PS}}{(1.15\ \rm \mu Jy)^2}\biggr) 
    \biggr(\frac{\Gamma(\nu_{\rm obs})}{29.2\ \rm Gpc^3} \biggr)
    &
    \\
    \times
    \biggr(\frac{\nu_{\rm obs}}{335\ \rm THz} \biggr)^{-1}
    \biggr( \frac{\mathcal R}{41}\biggr)^{-1} 
    \biggr( \frac{\Omega_{\rm pixel}}{\ang{;;6}\times\ang{;;6}}\biggr)^{-1}
    \biggr(\frac{N_{\tt eff}}{2} \biggr)^{-1}&
    .
\label{eq:SPHEREx}
\end{align}

\section{Additional details on the toy model}
\label{app:toy}

The lightcone evolution kernel for LIM we choose to use in this toy model qualitatively mimics more realistic kernels that we considered in the remainder of this paper:
\begin{equation}
    \KL(\chi) = \begin{cases}
        (\chi - \chi_\star)^2 & \chi \leq \chi_\star\\
        0 & \chi > \chi_\star
    \end{cases}.
    \label{eq:KLtoy}
\end{equation}
In essence: ``more spectral line emission occurs as the universe gets older".

Lightcone evolution induces a convolution in the cross-spectrum:
\begin{align}
\nonumber    \textbf{This Work }(\textsc{Lightcone Evolution})\\
 \langle I(\kpar) \kappa\rangle = \int_{\qpar} K_{\rm LIM}(\kpar - \qpar) K_\kappa(\qpar) P(\qpar),
 \label{eq:Ik_toy}
\end{align}
meaning that even if we must discard long wavelength modes from our LIM (e.g. set $I(\kpar)=0$ for small $\kpar$) due to foreground contamination, the direct cross-spectrum does not lose it's dominant contribution from long wavelength modes in the weak lensing kernel (small $\qpar$ modes in $\Kk(\qpar)$).
However, if lightcone evolution is not included in the LIM kernel (\Eq{ItoynLC}) then the convolution in \Eq{Ik_toy} disappears, thus making the correlation hopeless:
\begin{align}
\nonumber    \textbf{Prev. Work }(\textsc{No Lightcone Evolution})&\\
 \langle I(\kpar) \kappa\rangle = \KL\Kk(\kparp)P(\kpar).&
        \label{eq:Ik_toy_nLC}
\end{align}
Namely, if lightcone evolution is not accounted for, discarding long wavelength modes from our LIM simultaneously discards the long wavelength modes in the weak lensing kernel $\Kk$ where power is concentrated.
In detail, we show in \Fig{toyIk} both analytically and through Monte Carlo simulation that neglecting lightcone evolution suppresses the cross-spectrum by almost four orders of magnitude for $\kpar=0.1 /$Mpc.

Additionally, lightcone evolution induces correlations between long and short wavelength modes in the LIM:
\begin{align}
\nonumber    \textbf{This Work }(\textsc{Lightcone Evolution})\\
     \langle I(\kpar)I(\kparp)^*\rangle = \int_\qpar \KL(\kpar-\qpar)K_{\rm LIM}^*(\kparp - \qpar ) P(\qpar).
    \label{eq:kkpcorr}   
\end{align}
These correlations are to be expected as lightcone evolution breaks translational symmetry in the radial direction.
In linear theory, there is no coupling between long and short wavelength modes when lightcone evolution is neglected:
\begin{align}
   \nonumber    \textbf{Prev. Work }(\textsc{No Lightcone Evolution})&\\
    \langle I(\kpar)I(\kparp)^*\rangle = (2\pi)\delta^{(D)}(\kpar - \kparp)P(\kpar).
    \label{eq:kkpcorr_nLC}
\end{align}
This is to be expected as the modes of the linear matter density field are equivalent to the modes of the LIM (\Eq{ItoynLC}).

{The computation of Eqs.~(\ref{eq:Ik_toy}), (\ref{eq:kkpcorr}), (\ref{eq:kkpcorr}), and (\ref{eq:kkpcorr_nLC}) needed for \Figs{toyIk}{toySNR} can be done through direct numerical evaluation (solid colored) but also with Monte Carlo (dashed black) as one only needs to generate a 1-D Gaussian random field with the appropriate power spectrum and average over realizations after appropriate transformations.
We publicly release both the theoretical and Monte Carlo computations of quantities in this toy model at \href{https://github.com/DelonShen/LIMxCMBL/blob/main/README.md}{\texttt{github.com/DelonShen/LIMxCMBL}} (\Reff{github}) along with numerical computations of the rest of the quantities in this paper.
}

The detectability calculation displayed in \Fig{toySNR} is a simplified version of the one described in \Sec{SNR}.
In the toy model, our observable of interest is $\langle I(k) \kappa\rangle$ (we will drop the $\parallel$ part of $\kpar$ moving forward for succinctness).
Just like in \Sec{SNR}, the simplest unbiased estimator for this quantity in a $k$-bin is
\begin{equation}
    \hat\O(k) = \frac 1 {N_k} \sum_{[k_\pm]} \frac{I(k')\kappa}{\chi_\star},
    \label{eq:toy_est}
\end{equation}
where we have taken the 1-D universe to have length $\chi_\star$.
A particular difference between this calculation and the one described in \Sec{SNR} is that $\langle I(k)\kappa\rangle$ is a complex quantity.
We will see shortly that working with $k$-bins in Fourier space simplifies varying the amount of foreground filtering when computing the detectability significantly in comparison to the primarily real space approach we took in \Sec{SNR} and \App{explicit}.
It would be interesting to explore if a Fourier space approach could simplify the burdensome detectability computation done in this paper.

The covariance between two $k$ bins, which we will index by integers ${\tt a,b}$, of our estimator \Eq{toy_est} for complex quantity $\langle I(k)\kappa\rangle$ is
\begin{equation}
    \C_{\tt a,b} = \frac 1 {N_k} \{\langle I_{\tt a}I_{\tt b}^*\rangle \langle \kappa^2 \rangle  +  \langle I_{\tt a }\kappa\rangle\langle I_{\tt b}^*\kappa\rangle\}.
\end{equation}
The presence of complex conjugates is to ensure the $\SNR{}^2$ is a real quantity:
\begin{equation}
    \SNR{}^2 = \sum_{\tt a,b} \langle I_{\tt a}\kappa \rangle (\C^{-1})_{\tt ab} \langle I_{\tt b}^*\kappa\rangle. \label{eq:toySNR}
\end{equation}
Indeed, we see that $(\SNR{}^2)^* = (\SNR{}^2)$.
This is the origin of the theory (cyan and magenta) curves in \Fig{toySNR}.
To vary $\Lambda$ is as simple as neglecting terms in the double summation of \Eq{toySNR} which correspond to momentum bins which contain contributions from  $k<\Lambda$.

\bibliography{main}


\end{document}